# Simulating super massive black hole mass measurements for a sample of ultra massive galaxies using ELT/HARMONI high spatial resolution integral-field stellar kinematics

Dieu D. Nguyen,[1] Michele Cappellari,[2] and Miguel Pereira-Santaella[3]
[1]*Université de Lyon 1, ENS de Lyon, CNRS, Centre de Recherche Astrophysique de Lyon (CRAL) UMR5574, F-69230 Saint-Genis-Laval, France*
[2]*Sub-department of Astrophysics, Department of Physics, University of Oxford, Denys Wilkinson Building, Keble Road, Oxford OX1 3RH, UK*
[3]*Instituto de Física Fundamental, CSIC, Calle Serrano 123, 28006 Madrid, Spain*

**ABSTRACT**

As the earliest relics of star formation episodes of the Universe, the most massive galaxies are the key to our understanding of the stellar population, cosmic structure, and SMBH evolution. However, the details of their formation histories remain uncertain. We address these problems by planning a large survey sample of 101 ultramassive galaxies ($z \leq 0.3$, $|\delta + 24°| < 45°$, $|b| > 8°$), including 76% ellipticals, 17% lenticulars, and 7% spirals brighter than $M_K \leq -27$ mag (stellar mass $2 \times 10^{12} \lesssim M_\star \lesssim 5 \times 10^{12}$ M$_\odot$) with ELT/HARMONI. Our sample comprises diverse galaxy environments ranging from isolated to dense-cluster galaxies. The primary goals of the project are to (1) explore the stellar dynamics inside galaxy nuclei and weigh SMBHs, (2) constrain the black hole scaling relations at the highest mass, and (3) probe the late-time assembly of these most massive galaxies through the stellar population and kinematical gradients. We describe the survey, discuss the distinct demographics and environmental properties of the sample, and simulate their HARMONI $I_z$, $I_z + J$, and $H + K$-band observations via combining the inferred stellar-mass models from Pan-STARRS observations, an assumed synthetic spectrum of stars, and SMBHs with masses estimated based on different black hole scaling relations. Our simulations produce excellent state-of-art IFS and stellar kinematics ($\Delta V_{\rm rms} \lesssim 1.5\%$) in a relatively short exposure time. We use these stellar kinematics in combination with the JAM to reconstruct the SMBH mass and its error using the MCMC simulation. Thus, these simulations and modelings can be benchmarks to evaluate the instrument models and pipelines dedicated to HARMONI to exploit the unprecedented capabilities of ELT.

## 1 INTRODUCTION

Supermassive black holes (SMBHs) discovered at the centers of massive galaxies ($10^{10} < M_\star \lesssim 2 \times 10^{11}$ M$_\odot$) have masses ($M_{\rm BH}$) correlating with the galaxy's macroscopic properties. These $M_{\rm BH}$–scaling relations include the galaxy luminosity (e.g. $L_K$ or $L_V$; Kormendy & Richstone 1995), the stellar mass of the galaxy-bulge component ($M_{\rm bulge}$) or the stellar-mass of the entire galaxy ($M_\star$; e.g. Magorrian et al. 1998), the stellar-bugle velocity dispersion ($\sigma_\star$; e.g. Gebhardt et al. 2000; Ferrarese & Merritt 2000), the circular velocity of the extended H I rotation curves ($v_c$; e.g. Ferrarese 2002; Sun et al. 2013), the velocity width of circumnuclear molecular gas discs ($\Delta V$; Smith et al. 2020), and the galaxy concentration (e.g. Graham et al. 2001, 2003). The tightness and dynamic range covering several orders of magnitude of these correlations indicate that the evolution of these objects is closely linked (e.g. Kormendy & Ho 2013; McConnell et al. 2013; Saglia et al. 2016; van den Bosch 2016; Sahu et al. 2019a,b; Greene et al. 2020). Thus, understanding the shapes, related scatters, and the universality of such correlations will reveal the physical processes involved in the growth of black holes and galaxies.

Investigations from the demographics of known galaxies host SMBHs have demonstrated the non-universality of the $M_{\rm BH}$–galaxy scaling relations (e.g. Sahu et al. 2019a; Greene et al. 2020). For instance, galaxies with active galactic nuclei (AGN; e.g. Chilingarian et al. 2018), masers (e.g. Greene et al. 2010), bars (e.g. Graham & Spitler 2009), pseudo-bulges (e.g. Gültekin et al. 2009), or late-type spirals (LTGs; e.g. Greene et al. 2016; Läsker et al. 2016) that are almost below the same relations of more massive counterparts interpolated towards the low-mass regimes of both SMBHs and host galaxies (e.g. Nguyen et al. 2014; Nguyen et al. 2017; Nguyen 2017; Nguyen et al. 2018, 2019; Nguyen 2019).

Additionally, examinations of the variation of $M_{\rm BH}$ function in the mass-size diagram of their stellar mass ($M_\star$) and effective radius ($R_e$, radius that encloses half-light of the galaxy) suggests various growth pathways (Cappellari 2016; Krajnović et al. 2018a, hereafter C16; K18). For example, numerous galaxy properties such as $\sigma_\star$, mass-to-light ratio ($M/L$), gas content, bulge fraction, stellar population, and morphology vary systematically along the lines with $R_e \propto M_\star$, where $\sigma_\star$ = constant (Fig. 23 of C16). The same variation along the lines of constant-$\sigma_\star$ happens for $M_{\rm BH}$ (Fig. 1 of K18). This $M_{\rm BH}$ transition occurs along the lines of constant-$\sigma_\star$ for galaxies that have $M_\star$ less than a critical mass $M_{\star,\rm crit} \approx 2 \times 10^{11}$ M$_\odot$, suggesting the primary growths of both SMBHs and host galaxies via cold gas accretion, gas-rich minor mergers, and

2     *Dieu D. Nguyen et al.*

secular evolution predicted by the current well-established $M_{\rm BH}$–galaxy scaling relations (e.g. Kormendy & Ho 2013).

However, one finds evidence for a change in the $M_{\rm BH}$ variation in galaxies that are more massive than this critical mass located at the opposite ends of the $M_{\rm BH}$–galaxy scaling relations (McConnell & Ma 2013). Their $M_{\rm BH}$ are positive outliers from the $M_{\rm BH}$–$L_{K,\rm bulge}$ and $M_{\rm BH}$–$M_{\rm bulge}$ relations (McConnell & Ma 2013; Walsh et al. 2015, 2016, 2017) or the $M_{\rm BH}$–$\sigma_\star$ relation (C16; K18; van den Bosch 2016, henceforth V16) or the correlation of $M_{\rm BH}$–host galaxy's core break radius ($r_b$) inferred from the galaxy's core-Sérsic surface brightness profile (aka the $M_{\rm BH}$–$r_b$ relation; Rusli et al. 2013; Dullo 2019), which used to describe the morphology of the most massive galaxy (see Section 4.3) approximately one order of magnitude of $M_{\rm BH}$. The $M_{\rm BH}$–$\sigma_\star$ (McConnell & Ma 2013) correlation and the $M_\star$–$R_e$ diagram (K18), which includes the four brightest cluster galaxies (BCGs; McConnell & Ma 2013), start to depart from their same correlations without having these four BCGs around the mass of $M_\star \approx 3 \times 10^{10}$ M$_\odot$ (C16). These suggest that massive galaxies assemble their matter, changing from a sequence of bulge growth to dry-merger growth (Krajnović et al. 2013), predominantly through dissipation-less equal-mass dry-mergers according to current numerical simulations that linearly increase $M_{\rm BH}$, $R_e$, $r_b$, and $M_\star$ but not $\sigma_\star$ (e.g. Boylan-Kolchin et al. 2006; Naab & Ostriker 2017).

Recent progress in looking for the best scaling relation and its universal indication of possible formation mechanisms starts with the two-channels-formation paradigm of galaxies, assuming that SMBHs and hosts evolve simultaneously affected by the galaxy-stellar mass and environment (e.g. Peng et al. 2010, K18). The idea was motivated by both theoretical (Oser et al. 2010) and observational (Cappellari 2013; van Dokkum et al. 2015) evidence (also see a review of C16). To test this hypothesis, we consider the distribution of galaxies with $M_{\rm BH}$ measurements in the $M_\star$–$R_e$ diagram (Cappellari et al. 2013b) to find the most massive galaxies ($M_\star > 10^{12}$ M$_\odot$) locate at the top of the galactic-mass ladder. Therefore, we search for the records of the growing dependence of $M_{\rm BH}$ with galaxy properties transit from $\sigma_\star$ to $M_\star$ in the highest-mass targets (Scott et al. 2013). In other words, to understand which of the correlations ($M_{\rm BH}$–$\sigma_\star$ versus $M_{\rm BH}$–$M_\star$) is more fundamental and a better predictor of $M_{\rm BH}$ at the highest-galaxy-mass regime, more systemic $M_{\rm BH}$ measurements are needed. However, these galaxies are extremely rare in the local universe ($D_A < 110$ Mpc), and to find them, one has to reach out to where the required spatial resolutions and sensitivities go below the limits of existing ground-based adaptive optics (AO) assisted telescopes (e.g. Gemini and Very Large Telescope, VLT). We thus employ the Extremely Large Telescope (ELT) integral field spectrograph (IFS) to investigate the physical conditions and dynamics deep inside galaxy nuclei.

In this work, we (1) utilize the available near-infrared (NIR) photometric surveys (Section 2.1) to define a volume-limited sample of the highest-mass galaxies accessible at the ELT site, then (2) investigate the potentials of using the High Angular Resolution Monolithic Optical and Near-infrared Integral field spectrograph (HARMONI; Thatte et al. 2016, 2020) on ELT in exploring the nuclear-stellar kinematics and dynamics within the sphere of influence radii (SOI, $r_{\rm SOI} = GM_{\rm BH}/\sigma_\star^2$, where $G$ is the gravitational constant) of SMBHs or more likely Most Massive Black Holes (MMBH), then weighing their $M_{\rm BH}$ at further distances (or $M_{\rm BH}$ at high redshift) than the ones which could be resolved by the current apparatuses (e.g. VLT and Gemini assisted with AO). We demonstrate the ELT capabilities in spatial and spectral resolution relative to the stringent technical requirements for direct $M_{\rm BH}$ measurements. In addition, we test the technically demanding nature of the required determinations and the limits of HARMONI and thus provide technical guidance for a wide range of studies to probe the underlying physics of galaxy and black hole co-evolution.

We describe the parent sample, define specific selected criteria to identify our MMBH survey sample and present their essential properties in Sections 2 and 3, respectively. We also describe the dynamical and photometric model we use for our simulations in the subsequent section in Section 4. In Section 5, we perform the NIR integral field spectroscopic (IFS) simulations using the HARMONI SIMulator (HSIM; Zieleniewski et al. 2015) software for ELT observations on the HARMONI instruments and demonstrate its simulated datacubes and stellar kinematics extractions. In Section 6, we discuss the potential of application for dynamical modeling to measure the masses of central black holes using these observations and their limits. We conclude our findings in Section 7.

Throughout this work, we quote all quantities using a foreground extinction correction (Schlafly & Finkbeiner 2011) and the Cardelli et al. (1989) interstellar extinction law, as well as assuming a standard flat Universe with the Hubble constant $H_0 \approx 70$ km s$^{-1}$ Mpc$^{-1}$, matter density $\Omega_{\rm m,0} \approx 0.3$ and dark energy density $\Omega_{\Lambda,0} \approx 0.7$, which is consistent with the latest constraints from Planck (Planck Collaboration et al. 2014) and WMAP (Calabrese et al. 2017). We use the AB-photometric magnitudes system (Oke 1974) throughout the analysis, otherwise will be indicated clearly in the text. All the maps presented in this article show the galaxy's major axis aligned along the horizontal direction and the galaxy's minor axis aligned along the vertical direction.

## 2 SAMPLE SELECTION

### 2.1 $K$s-band magnitude and distances

We utilized the photometric information provided in the NIR ($\approx 2.2$ $\mu$m) $K$s-band luminosity by the full-sky and homogeneous photometry of the Two Micron All Sky Survey (2MASS; Skrutskie et al. 2006) redshift survey (2MRS; Huchra et al. 2012) as our parent sample to search for the most massive galaxies. The $K$-band is $5 - 10$ times less sensitive to dust absorption than optical, and the $M/L_K$ varies within a factor $\approx 2$ or 3 times less than at optical (Bell & de Jong 2001; Maraston 2005). Ma et al. (2014) also tested the potentially underestimated luminosity of 2MASS $K$s-band magnitudes in galaxy selection caused by its relatively shallow photometry (Schombert & Smith 2012) and the relatively small size of the sources themselves, making it difficult to determine accurate Sérsic index for the light profiles. Ma et al. (2014) compared the 2MASS photometry against the *Hubble Space Telescope (HST)* photometry of 219 early-type galaxies (ETGs) from Lauer et al. (2007b) and found that $K$s-band selection does not appear to be much affected by potentially systematic underestimates in the 2MASS $K$s-band magnitude. In this work, we look for more massive targets than the MASSIVE sample (Ma et al. 2014), where this effect could be negligible. Thus, 2MRS is the best parent sample for selecting dust-poor, distant, bright candidates with robust stellar mass approximations.

However, it is necessary to have distances for deriving galaxy luminosities and stellar masses from the observed apparent magnitudes. For approximately 100,000 galaxies that have the NASA/IPAC Extragalactic Database (NED [1]) redshift-independent distances [2] (NED-D; Steer et al. 2017), we matched them with the 2MRS galaxies but adopted their NED-D distances obtained with $\approx (10-20)\%$ accuracy. Otherwise, for the targets from 2MRS that do not have independent distances available, we derived distances from redshift because, at the distances of our sample, peculiar motions due to the Virgo cluster, the Great Attractor, and the Shapley supercluster become negligible compared to the Hubble flow, making redshift distances accurate.

---

[1] https://ned.ipac.caltech.edu/

The 2MASS extended source (XSC) catalog (Skrutskie et al. 2006) provides the $K$s apparent magnitude (Vega system) measurements for approximately 1.6 million galaxies (k_m_ext XSC keyword). We converted these apparent magnitudes into absolute magnitudes, $M_K = K_T - 5\log D_L - 25 - 0.11 \times A_V$. Here, $K_T \equiv$ k_m_ext measured in an isophotal aperture of a single Sérsic surface-brightness profile extrapolated to the inner-unresolved profile (Jarrett et al. 2003); $A_V$ is the Galactic extinction in Landolt $V$-band from Schlafly & Finkbeiner (2011) and the reddening relation of Charlot & Fall (2000) with $R_V = A_V/E(B-V) = 3.1$; and $D_L$ is the luminosity distance.

To determine the selection criteria for galaxies in our MMBH IFS survey, we relied on (1) the nominal spatial resolution for the HARMONI's image quality and resolving their SMBHs' $r_{SOI}$, (2) the minimum luminosity in $K$-band ($M_K$), and (3) the availability of a tip-tilt star near the science target (or a natural-guide star, NGS) which serves as a reference in the sky to correct the atmospheric turbulence effect on the ground-based IFS.

For the first requirement, Thatte et al. (2016, 2020) argued that the intermediate spatial scale of $10 \times 10$ mas$^2$ is optimal because the instrument's long-exposure point spread function (PSF) with a full width at half maximum (FWHM) of 12 mas has an ensquared energy of $\geq 75\%$ within a $2 \times 2$ spaxel[2] box, i.e. one spaxel = 10 mas. In practice, we wish to detect the genuinely stellar-kinematic signature within the SMBH's SOI, which should stay within several spaxels at least. We started from the standard formula to estimate the black hole sphere of influence radius $r_{SOI} = GM_{BH}/\sigma_\star^2$. Given the units of Mpc for the angular-size distance ($D_A$[3]), $M_\odot$ for the black hole mass ($M_{BH}$), and km s$^{-1}$ for the velocity dispersion ($\sigma_\star$), we obtain $r_{SOI}$ in the unit of ″:

$$r_{SOI}('') \approx 8.87 \times 10^{-4} \left(\frac{M_{BH}}{M_\odot}\right)\left(\frac{Mpc}{D_A}\right)\left(\frac{km\ s^{-1}}{\sigma_\star}\right)^2 \quad (1)$$

Next, we conservatively adopt $\sigma_\star \approx 300$ km s$^{-1}$, which is a characteristic value for the most massive nearby ETGs (e.g. C16) and varies weakly with galaxy mass (Krajnović et al. 2013; Naab & Ostriker 2017). Using the $M_{BH}$–$\sigma_\star$ relation from equation (7) of Kormendy & Ho (2013), this $\sigma_\star$ corresponds to $M_{BH} = 1.8 \times 10^9$ $M_\odot$. Thus, the above equation becomes:

$$r_{SOI}('') \approx \frac{18.0}{D_A} \quad (2)$$

Finally, we require the spatial scale of $r_{SOI} \approx 20$ mas for our MMBH IFS survey, which is well sampled by two ten mas HARMONI spaxels in radius (discussed later in Section 5.2), or equivalently we have $\pi \times (20/10)^2 \approx 12$ spaxels inside the sphere of influence. Given this resolving scale and equation (2), the distance limit should be $D_A \approx 902$ Mpc. Since we define our selection based on observable redshift rather than $D_A$, we thus round our redshift selection to $z = 0.3$, which corresponds to a slightly larger $D_A \approx 950$ Mpc in the adopted standard cosmology.

We should note that this choice of $r_{SOI} \approx 20$ mas in our survey is a lower limit because of the following reasons: (i) we expect that the central black holes in our MMBH survey are MMBH, much larger than the predictions from the Kormendy & Ho (2013) $M_{BH}$–$\sigma_\star$ relations of the nearby and smaller-mass galaxies and (ii) we put the galaxies at the furthest $D_A$ of the sample.

For the second requirement, we chose candidates brighter than $M_K \leq -27.0$ mag to search for the ultramassive galaxies. This $M_K$ limit roughly corresponds to a $B$-band selection $M_B \lesssim -24.7$ mag for the typical $B - K$s $\approx 2.5$ mag color of at the faint end of our selected sample. To give a sense of the extreme masses of our selected galaxies, we note that the BCG NGC 4889 of the Coma cluster is the brightest galaxy within the local $D \lesssim 100$ Mpc volume, which has a magnitude $M_K = -26.6$ mag (Cappellari 2013).

Finally, we take into account the third requirement by examining the available images on the Panoramic Survey Telescope and Rapid Response System (Pan-STARRS), Sloan Digital Sky Survey (SDSS), and 2MASS archival database in multiple bands (e.g. $r$, $g$, and $i$) for target galaxies those passed the first and the second requirement above. Our main purpose is to search for tip-tilt stars that can be used as reference stars in the vicinity sky to correct the atmospheric disturbances on the ground-based IFS by applying the Laser Tomography Adaptive Optics (LTAO) mode when using the HARMONI instrument (see Section 5.2). We should note that 2MASS and Pan-STARRS lack the spatial resolution to identify such stars; we thus rely on SDSS only for finding tip-tilt stars. However, the sky coverage of SDSS does not match the ELT's observability. We, therefore, do not push this third observational requirement to be as strong as the first two conditions during the MMBH sample selection in this work.

Ideally, these NGS should be off the target galaxies' centers about $(12-60)''$ and have limiting magnitude in $H$-band of $< 19$ mag in the Vega magnitude system (or $< 20.4$ AB mag). For some NGS that do not have available $H$-band apparent magnitude but Sloan, we first made a correction to convert SDSS magnitude system (i.e. Asinh magnitude) to conventional magnitudes, although the difference between Asinh and conventional mag $< 1\%$ for objects brighter than Asinh mag $m(f/f_0) = 22.12, 22.60, 22.29, 21.85, 20.32$ for $ugriz$ (Lupton et al. 1999), where $f/f_0 =$ (counts/exposure time)$\times 10^{0.4\times(\text{photometric zeropoint + extinction coefficient}\times\text{airmass})}$. Second, we converted the SDSS magnitude system to AB magnitudes: $ugriz(AB) = 22.5 - 2.5 \times \log_{10} f_{ugriz} - q$, where $q = 0.042, 0.036, 0.015, 0.013, 0.002$ for $ugriz$. Third, we transformed the SDSS AB mag to the 2MASS (i.e. $JHK$s) AB mag using equation (4) and information in Table 5 of Bilir et al. (2008). We find that 90% of the galaxies in our MMBH sample have available SDSS imaging. Out of these, 80% satisfy this tip-tilt stars requirement for LTAO correction. We assume a similar fraction will apply to the whole MMBH sample.

### 2.2 Selection criteria

We thus enforced obvious observability criteria and described the following selection steps:

---

[2] http://ned.ipac.caltech.edu/Library/Distances/

[3] Because this MMBH survey targets the most massive galaxies that possibly host MMBH beyond the local universe, their redshifts become critical. It is thus necessary to distinguish the luminosity distance ($D_L$) and the angular-size distance ($D_A$). While we use $D_L$ to estimate $M_K$ and $M_\star$ only, $D_A$ has to be used to define the $r_{SOI}$ of central black holes. We thus quote them with careful indications throughout this article.

4    *Dieu D. Nguyen et al.*

**Table 1.** Targets selection criteria

| | |
|---:|:---|
| Redshift: | $z \leq 0.3$ |
| Galaxies total magnitude: | $M_K \leq -27.0$ mag |
| Observability: | $|\delta + 24°| < 45°$ |
| Galaxy zone of avoidance: | $|b| > 8°$ |
| Tip-tilt stars*: | (12–60)″ away from the target's centre |
| | $m_H < 19$ Vega mag (or < 20.4 AB mag) |

*Notes:* *: The Laser Tomography Adaptive Optics (LTAO) mode on the ELT/HARMONI instrument needs one Natural Guide Star (NGS) at least to work simultaneously with six other artificial off-axis Laser Guide Stars (LGS). The system brings the required NGS to be more than 10,000 times fainter than that from the classical AO used on Gemini and VLT. In this work, we use the first four criteria listed above for selecting the MMBH survey's members (Table 2). The Tip-tilt stars requirement is used to access a high likely fraction of the MMBH sample only where we can find a few suitable NGS using the available databases from SDSS (i.e. 2MASS and Pan-STARRS lack the spatial resolution to identify such stars) because the sky coverage of SDSS does not match the ELT's observability.

**Table 2.** Main characteristics of the MMBH IFS survey sample

| | |
|---:|:---|
| Angular-size distance: | $136 < D_A \leq 950$ Mpc |
| Luminosity distance: | $145 < D_L \leq 1,600$ Mpc |
| $K$s-band luminosity: | $L > 1.3 \times 10^{12} \, L_{\odot K}$ |
| Total stellar mass: | $2 \times 10^{12} \lesssim M_\star \lesssim 5 \times 10^{12} \, M_\odot$ |
| $B$-band absolute magnitude: | $M_B \lesssim -24.7$ mag |
| # Galaxies in the sample: | $N_{gal} = 101$ |
| # Ellipticals ($T \leq -3.5$): | 77 ($\approx 76\%$) |
| # Lenticulars ($-3.5 < T \leq -0.5$): | 17 ($\approx 17\%$) |
| # Spirals ($-0.5 < T \leq 8$): | 7 ($\approx 7\%$) |

*Notes:* The galaxy Hubble type ($T$) is defined from HyperLeda: `http://leda.univ-lyon1.fr/search.html`. The survey volume was removed the Milky Way exclusion zone and the declination selection.

(i) We expanded the explored volume out to $z \leq 0.3$, corresponding to $D_A \leq 950$ Mpc and $D_L \leq 1,600$ Mpc for the brightest targets that satisfy $M_K \leq -27.0$ mag. These selected targets are even more massive than the current sample of MUSE Most Massive Galaxies (M3G, $-26.7 \leq M_K < -25.7$ mag; Krajnović et al. 2018b).

(ii) We tightened the specific observability criterion based on the location of ELT on the top of Mount Cerro Armazones in the Atacama Desert of northern Chile and the limit on the zenith distance for a good AO correction: $|\delta + 24°| < 45°$, where $\delta$ is the sky declination (private communication with N. Thatte).

(iii) We excluded the Galactic equatorial plane and Galactic bulge regions, highly contaminated by dust: $|b| \leq 8°$, where $b$ is the Galactic latitude.

(iv) We checked for the existing NGS stars with $m_H < 19$ Vega mag (or < 20.4 AB mag) that should be at distances between (12–60)″ away from the science target's center. However, due to the lack of spatial resolutions and the difference in sky observability among the databases we used to find for NGS, we do not treat this criterion (iv) as the priority like the criteria (i), (ii), and (iii).

A brief summary of the selection criteria is given in Table 1, while some general properties of our MMBH-selected sample are in Table 2. Fig. 1 illustrates the parameters space of $D_A$ vs. $M_K$ of galaxies, showing a big jump in HARMONI resolving power now allows us to open the explored-comoving volume up to a Gpc$^3$ when ignoring the declination limit.

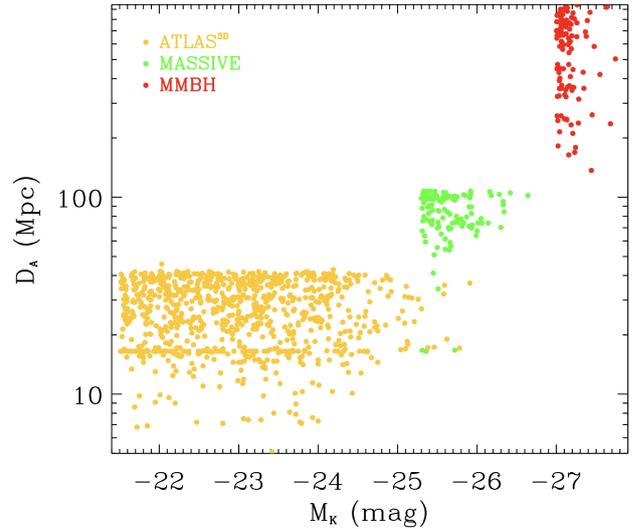

**Figure 1.** Angular-size distance ($D_A$) and absolute $K$s-band magnitude ($M_K$) of our Most Massive Black Holes (MMBH) survey (red; this work), MASSIVE survey (green; Ma et al. 2014), and ATLAS$^{3D}$ survey (yellow; Cappellari et al. 2011). There is no overlapping of our MMBH sample with the two others in these parameter spaces. We assumed $D_L \approx D_A$ (see footnote 3) for ATLAS$^{3D}$ and MASSIVE surveys due to their low redshifts ($z < 0.025$).

## 3  PROPERTIES OF THE SELECTED SAMPLE

### 3.1  Stellar mass and size

The galaxy photometric stellar masses are estimated from the total $K$s-band absolute magnitudes (extinction-corrected) using the relation in which both quantities related to each other as the prescription from equation (2) of Cappellari (2013) calibrated from 260 ATLAS$^{3D}$ ETGs: $\log(M_\star) = 10.58 - 0.44 \times (M_K + 23)$. On the other hand, the sources' sizes are defined by $R_e = 1.61 \times$ `j_r_eff`, where `j_r_eff` is the 2MASS XSC keyword for the semimajor axis of the isophote enclosing half of the total galaxy light in $J$-band (Cappellari 2013). The usage of `j_r_eff` instead of `k_r_eff` is because the $J$-band has a better signal-to-noise (S/N) than the equivalent $K$s-band (Cappellari et al. 2013a, K18).

Fig. 2 highlights the distinct parameters space in stellar masses and sizes occupied by our MMBH survey. The larger surveying volume from the highest redshift of MASSIVE ($z \approx 0.026$) to our adopted redshift ($z \leq 0.3$), allowing us to sample the galaxy mass function at $2 \times 10^{12} \lesssim M_\star \lesssim 5 \times 10^{12}$ M$_\odot$ for large galaxies with angular size $10 < R_e < 60$ kpc. Given no mass overlapping, about half of our MMBH-selected galaxies have similar $R_e$ with MASSIVE galaxies. Thus, our MMBH sample is the extreme mass and size complementary to ATLAS$^{3D}$ (Cappellari 2013), MaNGA (Bundy et al. 2015), SAMI (Croom et al. 2021, third and final data release), and MASSIVE (Ma et al. 2014) in the galaxy-hierarchical structure.

### 3.2  Shapes

It is well-established that low-mass elliptical galaxies appear to be fast rotators characterized by higher ellipticities, whereas giant ellipticals are slow rotators and round or mildly triaxial (Kormendy & Bender 1996, C16). These facts indicate strong correlations among the shapes, kinematics, and masses of massive ETGs. It is, there-fore, interesting to examine the distributions in galaxy shapes for our MMBH survey.

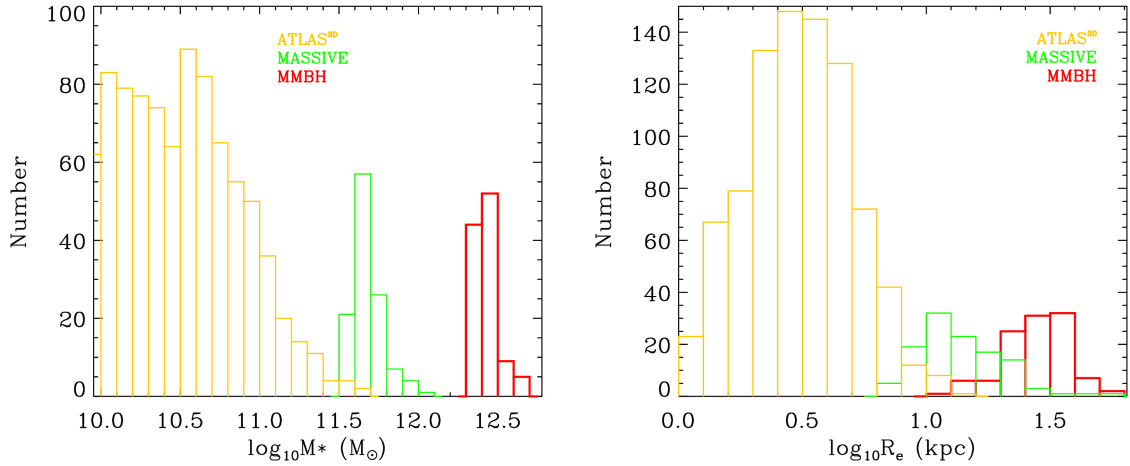

**Figure 2.** Distributions of of stellar masses estimated from 2MASS photometry (left) and half-light radii (right) for our MMBH (red, this work), MASSIVE (green; Ma et al. 2014) and ATLAS$^{3D}$ (yellow; Cappellari et al. 2011) galaxies.

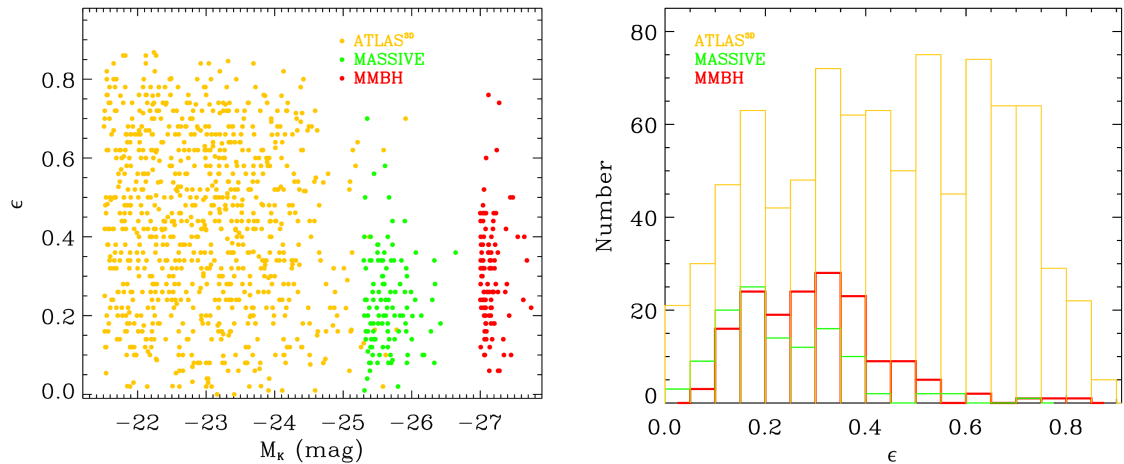

**Figure 3.** Ellipticity ($\epsilon$ = 1 − `sup_ba`) vs. $K$s-band luminosity (left) and ellipticity distribution (right) for our MMBH survey (red) in comparison with the MASSIVE survey (green; Ma et al. 2014) and ATLAS$^{3D}$ (yellow; Cappellari et al. 2011).

The left panel of Fig. 3 compares the ellipticity, $\epsilon = 1 - $ `sup_ba`, of our MMBH survey to that of MASSIVE and ATLAS$^{3D}$, where `sup_ba` is the 2MASS XSC parameter for the minor-to-major axis ratio fit to their "3$\sigma$ super-coadd isophote." Only five galaxies each in the MASSIVE and MMBH sample have $\epsilon > 0.5$, and that is about a quarter of ATLAS$^{3D}$ because the ellipticities are generally larger at larger radii, implying that `sup_ba` measured at the outermost isophote, is likely an upper limit to the effective ellipticity ($\epsilon_e$) used (Cappellari 2013). These galaxies are all in the fainter tails of $M_K < -25.7$ mag for MASSIVE and $M_K < -27.0$ mag for ours, respectively. There are depletions of high-$\epsilon$ galaxies in both samples, consistent with the fact that most of the galaxies are slow rotators for which C16 (Fig. 13) adopts an empirical separation at $\epsilon \approx 0.4$. Generally, some massive galaxies have ellipticity larger than this limit because 2MASS measures the ellipticity radii larger than the half-light radius used to define $\epsilon_e$. Moreover, the ellipticity of massive galaxies generally increases with radius. Our MMBH survey will provide direct measurements of the spatial profile of the nuclear-stellar kinematics (also ionized gas, if detected) of each galaxy and will allow us to quantify the distributions of galaxy rotations and shapes at the highest masses.

### 3.3 Supermassive black holes

None of the galaxies in our MMBH sample have published $M_{BH}$ as they are located at farther distances beyond the current telescopes' resolving powers. Considering within our sample's $D_A$ range only, ten smaller galaxies ($M_\star \lesssim 10^{12}$ M$_\odot$ and $R_e \leq 10$ kpc) have $M_{BH}$ measurements in the literature, mostly using the reverberation mapping emissions from the broad-line regions and the dynamics of maser spots (Table 2 of van den Bosch 2016), but they did not pass the selection criteria (Table 1). Amongst these measurements, only two BGCs were satisfied with our observability criteria but located highly in the Northern sky. These galaxies and their SMBHs occupy the high-mass ends and also have a large scatter on the $M_{BH}$–$\sigma_\star$ relation where there is a mix of very massive and lower-mass galaxies, raising concern that the black hole–galaxy scaling relation starts changing from $M_{BH}$–$\sigma_\star$ to $M_{BH}$–$M_\star$ if we would be able to push to the higher-mass regime (McConnell et al. 2013). The MASSIVE survey extended the ATLAS$^{3D}$ sample

6   *Dieu D. Nguyen et al.*

to the parameters space of $D < 110$ Mpc and $M_\star < 10^{12}$ M$_\odot$ (Ma et al. 2014). Our MMBH survey with ELT/HARMONI will thus provide a complimentary of $M_{\rm BH}$ in the highest mass galaxies ($2 \times 10^{12} \lesssim M_\star \lesssim 5 \times 10^{12}$ M$_\odot$) that locate at further distances ($136 < D_A \le 950$ Mpc or $145 < D_L \le 1,600$ Mpc) and very likely host MMBH predicted by the equation (3) of K18.

### 3.4 Black hole–galaxy scaling relations

Efforts of IFS-kinematic data and modeling developments in the past several years have substantially increased the number of dynamical measurements of $M_{\rm BH}$ in very high-mass galaxies (McConnell et al. 2011a,b; Rusli et al. 2011; McConnell et al. 2012; van den Bosch et al. 2012; McConnell & Ma 2013; Walsh et al. 2013), suggesting a possible offset above the current well-established scaling relations of lower-mass galaxies ($\approx 1$ order of magnitude of $M_{\rm BH}$; e.g. Walsh et al. 2016). Nevertheless, these adding IFS-kinematics and modelings also caused difficulty in discriminating the models for the galaxy–black hole coevolution (Peng 2007; Hirschmann et al. 2010; Jahnke et al. 2011; Anglés-Alcázar et al. 2013). Moreover, accurate determinations of the intrinsic scatter and the high-mass $M_{\rm BH}$ distribution is crucial for a tight constraint on $M_{\rm BH}$ function in quiescent galaxies (Lauer et al. 2007a; Lauer et al. 2007b), black hole demographics, the merging rate of SMBH binaries (e.g. van Haasteren et al. 2011), and the contribution of the gravitational wave background detected by the current pulsar timing experiments (Demorest et al. 2013; Shannon et al. 2013) or LISA (e.g. Gourgoulhon et al. 2019) in the future. Also, the hints of $M_{\rm BH}$-dependence in the scaling relation had changed from $\sigma_\star$ to $M_\star$ (e.g. McConnell et al. 2013; Kormendy & Ho 2013) in the highest mass galaxies. This change indicates that the most massive galaxies grow mainly from dry-mergers, distinguishing from a sequence of bulge growth of the lower-mass galaxies (e.g. C16; K18) predicted by current numerical simulations (Naab & Ostriker 2017). A systemic survey of dynamical-$M_{\rm BH}$ measurements in this MMBH survey without using the current scaling relations is necessary to progress our knowledge of black hole–galaxy coevolution.

Our MMBH sample comprises a significant fraction of core galaxies without central excess light profiles within a few central kpcs as a signature of black hole scouring (Begelman et al. 1980). They are the best laboratory to investigate the scaling relations between the core and the nuclear-galaxy structure, relating to the tangential stellar orbits (Kormendy & Bender 2009; Rusli et al. 2013; Thomas et al. 2014).

### 3.5 Uniqueness of our MMBH sample on the $M_\star$–$R_e$ diagram

We created the mass–size diagram ($M_\star$–$R_e$) from the 2MRS sources before applying the selection criteria and show it in Fig. 4. At the low mass ($M_\star < M_{\rm crit} \approx 2 \times 10^{11}$ M$_\odot$), the lines of constant-$\sigma_\star$ (also the lines of constant $M_{\rm BH}$) trace the mass concentration and the mass density (or bulge mass fraction) of galaxies, implying their central $M_{\rm BH}$ behave similarly with other galaxy properties. SMBHs evolution thus links to optical color, molecular-gas fraction, dynamical-$M/L$, initial mass function (IMF) normalization, age, metallicity, and $\alpha$-element abundance (Fig. 22 of C16), especially their $M_{\rm BH}$ growth following the same trend as galaxy properties arising from star formation within the host galaxies (Graham et al. 2018).

Next, we applied our target selection criteria in Section 2.2 (Table 1) for the 2MRS sources on the $M_\star$–$R_e$ diagram and plotted our MMBH-selected sample at the highest $M_\star$, $R_e$, and $\sigma_\star$ in Fig. 4. Given their highest $M_\star$ and $\sigma_\star$, the $M_{\rm BH}$ predicted by the $M_{\rm BH}$–$\sigma_\star$ relations (equation (2) of K18) differ by more than an order of magnitude from that predicted from the equation (3) of K18, where $M_{\rm BH}$ starts following the $M_{\rm BH}$–$M_\star$ relation when $M_\star > M_{\rm crit}$. We also added the ATLAS$^{\rm 3D}$ and MASSIVE galaxies in Fig. 4 to demonstrate the uniqueness of our highest-mass galaxies sample.

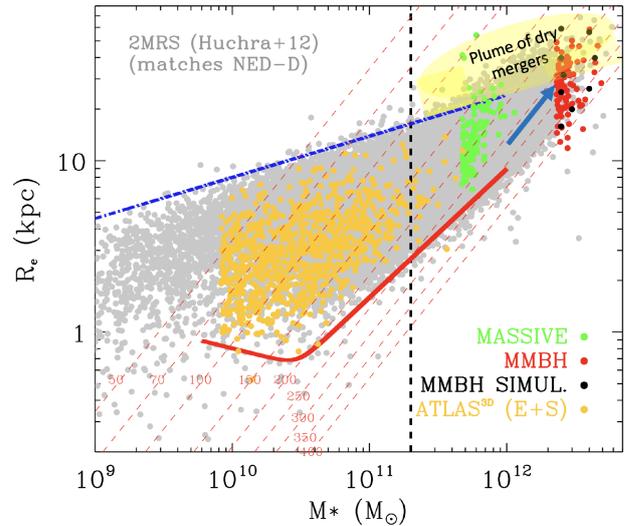

**Figure 4.** The Mass-size diagram of 2MRS sources (grey dots) shows in the stellar-mass range of $10^9 < M_\star < 7 \times 10^{12}$ M$_\odot$ only. The inclined-red-dashed lines are the constant-velocity dispersions (constant-$\sigma_\star$) ranging from $50 - 400$ km s$^{-1}$. The ATLAS$^{\rm 3D}$ (yellow dots; Cappellari et al. 2011, full sample of Ellipticals + Spirals), MASSIVE (green dots; Ma et al. 2014), and our MMBH (red and black dots; this work) galaxies surveys occupy different regions of the diagram. The thick-solid red curve defines the zone of exclusion (ZOE) described by equation (4) of Cappellari et al. (2013b) in the previously explored stellar-mass range of $6 \times 10^9 < M_\star < 1 \times 10^{12}$ M$_\odot$, while the thick blue arrow indicates the qualitative growth along constant-$\sigma_\star$ for dry mergers. The thick dash-dotted blue line shows the relation $(R_e/{\rm kpc}) = 8 \times [M_\star/(10^{10} \, {\rm M}_\odot)]^{0.24}$, which provides a convenient approximation for the lowest 99 percent contour for the distribution of ETGs (Cappellari et al. 2013b). The vertical dash-black line is the characteristic mass at $M_{\rm crit.} \approx 2 \times 10^{11}$ M$_\odot$. The yellow-shaded region at the top end of the highest mass regime (above the blue arrow) shows the signature of multiple dry mergers "plume" in the current data. This is a prediction from theoretical models (see Fig. 29 of C16), confirmed by these data. Black dots are our MMBH galaxies selected for HSIM IFS simulation in Section 5.2.

One can see in Fig. 4 a "plume" (indicated by the yellow shaded region on the top of the blue arrow) of galaxies with sizes larger than the extrapolation of the upper boundary of the ($M_\star$, $R_e$) distribution of lower mass galaxies (Cappellari et al. 2013a). The deviation starts appearing around $M_\star \approx M_{\rm crit.} \approx 2 \times 10^{11}$ M$_\odot$ and is particularly evident above $M_\star > 2 \times 10^{11}$ M$_\odot$. This deviation is the signature of multiple dry mergers expected to move galaxies along lines-of-nearly-constant sigma (or $R_e \propto M_\star$) on the ($M_\star$, $R_e$) plane (e.g. C16).

Although there is some evidence for a possible change in the black hole–galaxy scaling relations with the current existing-$M_{\rm BH}$, it is not clear beyond the high-mass regime due to the limited number of measurements above $M_{\rm crit}$. Consider at (1) a given $\sigma_\star$ with $M_\star > M_{\rm crit}$ and (2) the range of measured $M_{\rm BH}$ (varies more than $\approx$ an order of magnitude) and their uncertainties (a factor of $\approx 2$), which depends strongly on the type of data and type of models used. The observation of this $M_{\rm BH}$-dependence transition from the $\sigma_\star$ to the $M_\star$ is difficult to see with the current data. Furthermore, this effect is hampered by the increasing closeness of constant-$\sigma_\star$ lines and the lack of galaxies with masses $M_\star > 10^{12}$ M$_\odot$. Our proposed sample of the most massive galaxies utilizes the unprecedented advantages of ELT/HARMONI IFS in both angular and spectral resolutions and sensitivity, aiming to resolve all of these difficulties and discover a new physics in the previously untouchable regimes of SMBH and galaxy co-evolution.

### 3.6 Nine representative targets for our MMBH survey

Given our complete MMBH IFS survey sample of 101 ultramassive galaxies shown in red dots in Figs 1 and 4, it is not sensitive to the span of galaxy stellar-mass parameter (see Table 2); thus leaving the sample's properties is better to be examined in the $D_A - R_e$ plane because revealing the distributions of SMBH (or MMBH) population as a function of redshift ($D_A$) and effective radii ($R_e$) of the hosts will shed light on the underlying physics that drive the central massive black holes and the host galaxies obtained their masses and coevolution throughout the cosmic time. To make our ELT/HARMONI IFS simulations in subsequent sections (4, 5, and 6) representing the MMBH sample entirely, we select only nine targets from these 101 galaxies to perform HSIM to create IFS mock datacubes distributed over the full range of angular-size distance and size of our MMBH survey. Although these nine chosen targets are selected randomly from the $D_A - R_e$ plane as shown in Fig. 5, they must cover the full parameter ranges of the galaxy's $R_e$ and $D_A$. In this way, the reduced simulated sample minimally represents 101 ultramassive galaxies of the MMBH sample but optimally examines their SMBH/MMBH distributions at different cosmic times and galaxy densities.

### 3.7 Galaxy environments

As located at the highest galactic-mass ladder, galaxies in our MMBH-selected sample are commonly present in the centers of galaxy groups or clusters (Ma et al. 2014, C16). It is thus worth investigating the larger-scale environments where these $M_\star \gtrsim 2 \times 10^{12}$ M$_\odot$ galaxies reside. Only one galaxy in our sample lies within the full-sky volumes ($D_A < 150$ Mpc) of two galaxy-group catalogs: 2MRS (Crook et al. 2007, 2008) and galaxy-redshift 2M++ (Lavaux & Hudson 2011) but neither belong to these groups.

Dense-galaxy clusters significantly impact the galaxy growth and the evolution of the black hole–galaxy scaling relations because of supplied material supply and mergers. However, isolated galaxies live in low-density environments, likely surrounded by faint satellite galaxies (Jones et al. 2003), and might have stopped building up their masses a few billion years ago. Thus, our MMBH survey will provide an excellent sample for studying environmental effects on galaxy formation (Mulchaey & Jeltema 2010).

Fig. 6 shows the two-arcmin-squared FOV SDSS red-green-blue images of the subsample of nine galaxies (Section 3.6) shown in Fig. 5 and listed in Tables 3, 4, and 5. Such a large FOV reveals a diverse intergalactic vicinity of these nine galaxies, ranging from isolated (e.g. 2MASXJ22354078+0129053, 2MASXJ12052321+1022461) with some small satellites to dense-galaxy clusters (e.g. 2MASXJ00034964+0203594, 2MASXJ11480221+0237582). For other galaxies in the MMBH sample, we also carefully examined for such available NGS and found at least one to three of them in their surrounding neighborhood intergalactic environments, which satisfy our requirement (i.e. the selection criterion (iv)) but show in Fig. 6 the representative subsample only. The requirement of an NGS will shrink our MMBH sample size further because only 80% of the sample has such available NGS. As can be seen in Fig. 6, there are two galaxies in the reduced/simulated subsample (e.g. 2MASXJ12052321+1022461 and 2MASXJ11480221+0237582) that have no NGS, also giving ≈ 80% of the galaxies has NGS for LTAO performances.

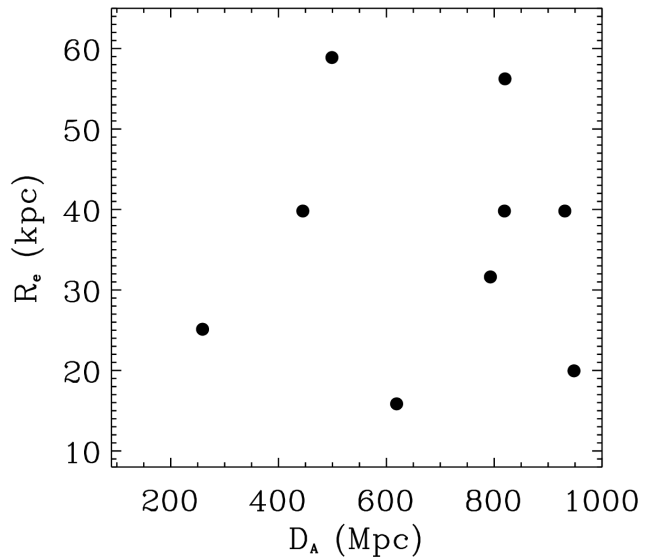

**Figure 5.** Nine galaxies in our MMBH sample span in the whole parameters space of angular-size distance and effective radius. Some specific properties of these nine galaxies are summarized in Tables 3 and 4. We perform HSIM IFS simulations for these all nine ETGs.

## 4 DYNAMICAL AND PHOTOMETRIC MODEL

We first describe the dynamical model and the synthetic library of stellar spectra we use in Sections 4.1 and 4.2, respectively. We next construct the mass models of all nine chosen galaxies (see Fig. 5) for HARMONI IFS simulation in Section 4.3 (only showing the galaxy 2MASXJ12052321+1022461 as an example) whose properties cover and represent our MMBH survey as a whole.

### 4.1 Jeans Anisotropic Model (JAM)

Our sample of galaxies consists of the most massive galaxies. These are generally close to spherical or weakly triaxial in their central regions (C16). For this reason, we constructed the mock kinematics using the dynamical model based on a solution of the Jeans equations, which assumes axisymmetry with a spherical aligned orientation of the velocity ellipsoid, which is likely to provide a better approximation to the dynamics of slow rotators (Cappellari 2020, henceforth C20) as implemented in the JAM software (which we call the JAM$_{\rm sph}$ model)[4]. To predict the mean velocity using JAM$_{\rm sph}$, we assumed a model with velocity ellipsoid axially symmetric around the radial direction, namely $\sigma_r \neq \sigma_\theta = \sigma_\phi$. This model converges to a non-rotating spherical model in the spherical limit and is an appropriate choice for modeling slow rotators expected to be not far from spherical in their central regions. This assumption corresponds to equation (55) of C20 achieved by setting `'gamma=beta'` in the `jam_axi_proj.py` procedure of JAM; see C20 for detailed descriptions of the model.

---

[4] jampy v6.4.0; available from https://pypi.org/project/jampy/

8  *Dieu D. Nguyen et al.*

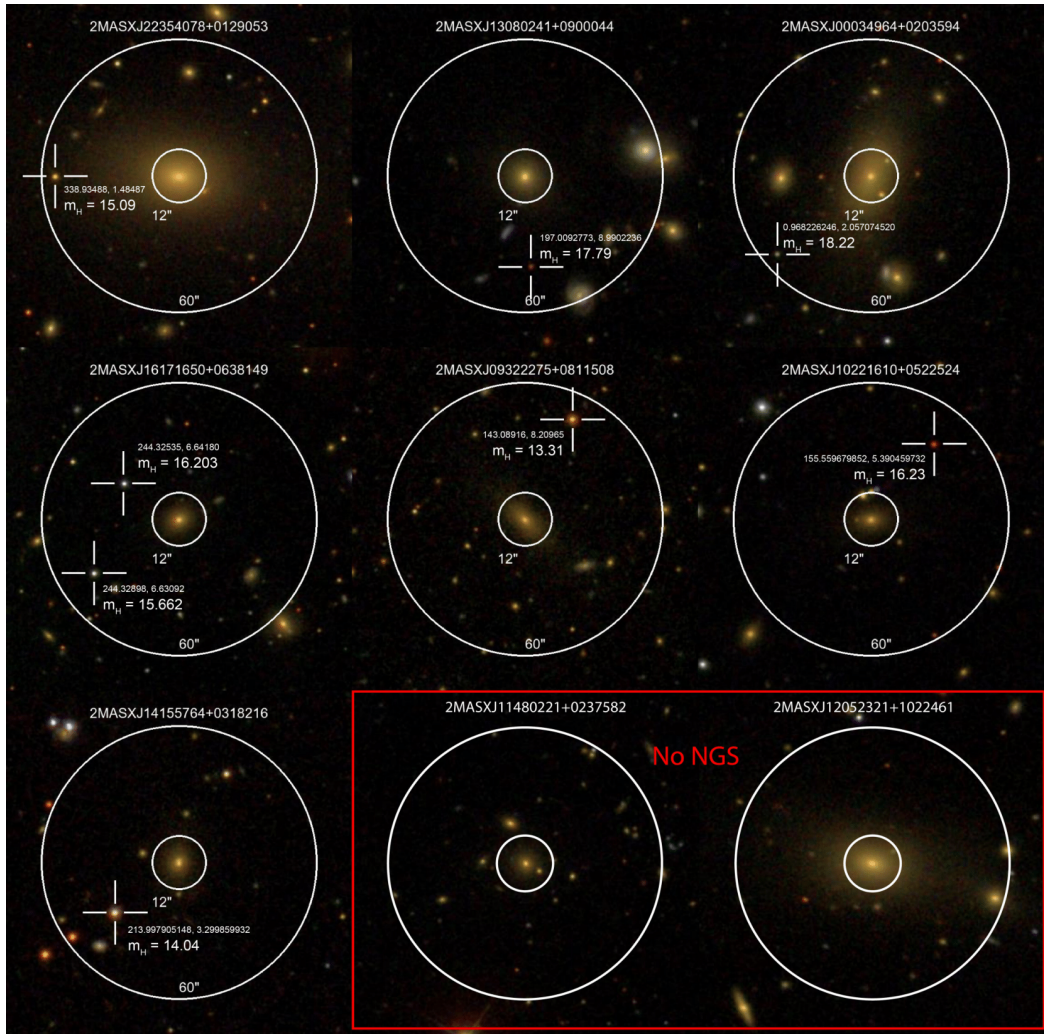

**Figure 6.** The red-green-blue SDSS images of the reduced subsample of nine galaxies which we chose to perform `HSIM` IFS simulations shown in Fig. 5. The large FOV ($> 2 \times 2$ arcmin$^2$) of these images reveals the intergalactic environments around these nine galaxies, ranging from isolated to dense-galaxy clusters. The name of each galaxy is shown in the top corner, while the two white circles define the galaxy's allowable vicinity ($12'' < r < 60''$ away from the galaxy centre) where to look for a faint NGS necessary for LTAO performance. The white crosses indicate the available NGS's locations, showing their R.A. and Decl. in degrees and the apparent AB magnitudes measured in the $H$ band. Among these nine selected galaxies for our simulations in this work, there are two of them, e.g. 2MASXJ12052321+1022461 and 2MASXJ11480221+0237582, which we cannot find NGS by our observational strategy, showing in the red rectangular at the bottom-right corner.

The detail of the adopted model is not critical for this work, as here we are interested in estimating the formal errors in the $M_{\rm BH}$ determinations due to the effect of noise and spatial resolution rather than the possible systematic biases of the modeling methods. We plan to study the latter in the future.

### 4.2  MARCS synthetic library of stellar spectra

We utilized the library of stellar population synthesis (SPS) spectra[5] by Maraston & Strömbäck (2011), based on the MARCS synthetic library of theoretical spectra by Gustafsson et al. (2008). Although

MARCS synthetic library spectra are not as reliable as an empirical stellar library, they are broad wavelength coverage (i.e. the vacuum wavelength covers from 1,300 Å to 20 $\mu$m) and high spectral resolution ($\sigma = 6.4$ km s$^{-1}$ or $R = \lambda/\Delta\lambda = 20,000$), sampling with 100,724 flux points ($\Delta\lambda \approx 0.065$ Å). We also assumed the Salpeter IMF, 10 Gyr, and Solar metallicity (`z002`) and truncated the SPS within the wavelength coverage of $0.7 - 2.5$ $\mu$m for HARMONI/$I_z$, $I_z + J$, and $H + K$-grating.

### 4.3  Galaxy mass models

#### 4.3.1  *The need for high-resolution imaging*

In order to obtain accurate constraints from dynamical modelings (e.g. $M_{\rm BH}$ and galaxy kinematics), the galaxy mass model must be precisely constructed at all components (i.e. stellar remnants, stars, dust, gas, dark matter) and scales (i.e. from a few ten parsecs away the galaxy center, where is comparable or within $r_{\rm SOI}$, to hundreds

---

[5] Available from: `https://marcs.astro.uu.se/`

**Table 3.** List of simulated targets and their core-Sérsic best-fit parameters from *i*-band PanSTARR images. The inner power-law's slope is fixed with $\gamma = 0.1$.

| Galaxy name | $\mu_b$ (mag/″²) | $n$ | $\alpha$ | $r_b$ (″) | $r_e$ (″) | $\mu_b$ (mag/″²) | $n$ | $\alpha$ | $r_b$ (″) | $r_e$ (″) |
|---|---|---|---|---|---|---|---|---|---|---|
| (1) | (2) | (3) | (4) | (5) | (6) | (7) | (8) | (9) | (10) | (11) |
| | MGE | MGE | MGE | MGE | MGE | IRAF | IRAF | IRAF | IRAF | IRAF |
| J22354078+0129053 | 15.34±0.04 | 2.80±0.03 | 2.98±0.02 | 0.83±0.02 | 5.03±0.03 | 15.21±0.03 | 2.52±0.04 | 3.58±0.03 | 0.98±0.03 | 5.95±0.03 |
| J13080241+0900044 | 15.17±0.03 | 2.97±0.04 | 3.72±0.04 | 0.79±0.02 | 5.34±0.04 | 15.14±0.02 | 2.34±0.02 | 3.75±0.04 | 0.94±0.02 | 6.04±0.03 |
| J12052321+1022461 | 15.36±0.03 | 2.86±0.05 | 2.79±0.04 | 0.74±0.05 | 5.40±0.03 | 15.22±0.04 | 2.43±0.02 | 3.95±0.01 | 0.95±0.05 | 5.98±0.04 |
| J00034964+0203594 | 15.05±0.04 | 2.92±0.02 | 3.27±0.03 | 0.81±0.03 | 5.48±0.05 | 14.97±0.03 | 2.22±0.03 | 3.89±0.04 | 0.92±0.04 | 6.12±0.05 |
| J16171650+0638149 | 15.30±0.02 | 2.90±0.02 | 3.85±0.05 | 0.85±0.04 | 5.52±0.02 | 15.20±0.05 | 2.21±0.05 | 3.84±0.05 | 0.91±0.03 | 5.97±0.05 |
| J09322275+0811508 | 15.15±0.05 | 2.88±0.02 | 3.44±0.03 | 0.90±0.04 | 5.85±0.05 | 15.16±0.05 | 2.45±0.03 | 3.69±0.02 | 0.97±0.02 | 5.83±0.03 |
| J10221610+0522524 | 15.13±0.05 | 2.91±0.03 | 3.81±0.03 | 0.95±0.03 | 5.57±0.04 | 15.05±0.06 | 2.26±0.03 | 3.83±0.01 | 0.95±0.03 | 5.79±0.05 |
| J14155764+0318216 | 15.05±0.05 | 2.90±0.04 | 3.15±0.02 | 0.93±0.02 | 5.96±0.03 | 14.93±0.04 | 2.31±0.05 | 3.52±0.05 | 0.93±0.03 | 5.82±0.03 |
| J11480221+0237582 | 15.31±0.05 | 2.85±0.03 | 3.60±0.04 | 0.87±0.02 | 5.82±0.03 | 15.25±0.05 | 2.57±0.04 | 3.78±0.01 | 0.97±0.03 | 6.21±0.05 |

*Notes:* Column 1: Galaxy name in which we assigned J ≡ 2MASXJ. The following five columns are the surface-brightness density $\mu_b$ at the break radius (column 2), the Sérsic index (column 3), the real galaxy profiles parameter (column 4), the break radius (column 5), and the effective radius of the outer core-Sérsic profile (column 6) derived from the MGE 1D profile. The last five columns (7, 8, 9, 10, 11) are the same as the former five but derived from the `IRAF ellipse` 1D profile. Details of these five parameters of the core-Sérsic profile are mentioned in the text (Section 4.3).

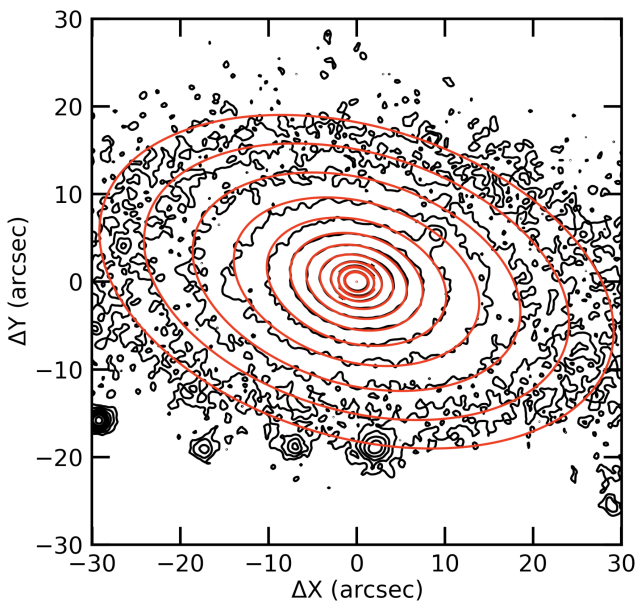

**Figure 7.** Comparison between the Pan-STARRS/*i*-band-image photometric data (black) versus its best-fit MGE model (red) of the galaxy 2MASXJ12052321+1022461 at the same radii illustrated in the form of 2D surface brightness density contours in the field-of-view (FOV) of 60″×60″. Contours are spaced by 0.5 mag arcsec$^{-2}$.

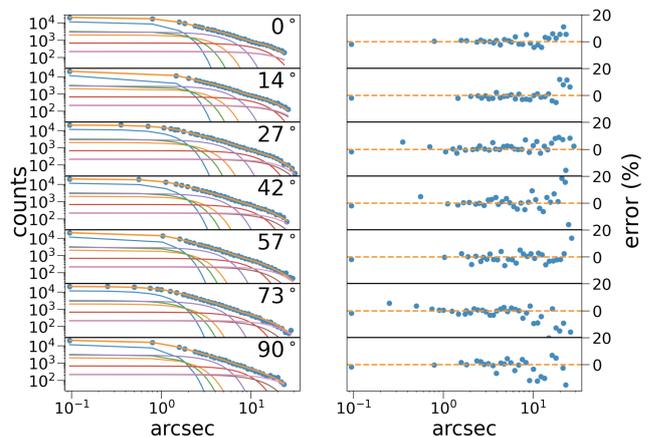

**Figure 8.** Same as Fig. 7, here is the comparison between the Pan-STARRS/*i*-band-image photometry of 2MASXJ12052321+1022461 (blue dots) and its corresponding best-fit MGE model with seven Gaussians at different position angles (color lines, left panels) within the FOV of 60″ × 60″, along with the correspondingly fractional errors (`data - model`)/`data` in right panels.

kpc at the dark-matter halo). Thus, wide-field images from broadband photometries at the same spatial resolutions (20–100 mas) and wavelength coverages (0.7–2.47 $\mu$m) of HARMONI IFS are highly demanded. In some cases, the optical images (0.45–0.76 $\mu$m) are also necessary to examine the galaxy's stellar variations and gas/dust extinction. Currently, such high-angular-resolution images for our MMBH IFS survey never exist in any archival databases except for their low-resolution ground-based images in optical and NIR surveys (e.g. Pan-STARRS, SDSS, 2MASS), which lack critical information of matter distribution at the scale of a few times of $r_{SOI}$. Using these available images without some appropriate assumptions to extrapolate the stellar distribution toward the galaxy center will bias the $M_{BH}$ estimate.

Imaging from space missions such as *HST* or *James Webb Space Telescope (JWST)* are possible alternatives as they probe deeper into the central regions of a few times of the MMBH's $r_{SOI}$ ($\approx 50 - 100$ mas), helping to reduce uncertainty on $M_{BH}$ estimate significantly. To obtain accurate measurements of the motions of stars (and gas) within $r_{SOI}$ – the key for reliable $M_{BH}$ estimates – a telescope must be able to at least marginally resolve it. Currently, the Enhanced Resolution Imager and Spectrograph (ERIS) on VLT provides the highest available spatial resolution ($\approx$50 mas), but even this is too low to probe the SOI of a typical SMBH beyond 100 Mpc, while our MMBH is detectable only out to $D_A \approx 200$ Mpc (see Fig. 1).

Ideally, possibly native ELT imaging obtained from the Multi-AO Imaging Camera for Deep Observations (MICADO) imager (FOV: 50″.5×50″.5, wavelength: 0.8−2.4 $\mu$m, filters: *IYJHK* broadband) at the same angular resolution (FWHM$_{PSF}$ ≈ 10 mas; Davies et al. 2010) with HARMONI is the best choice. MICADO takes advantage of the wide-field correction and uniform PSF offered by the multi-conjugate AO (MCAO) module to achieve almost a full

10  *Dieu D. Nguyen et al.*

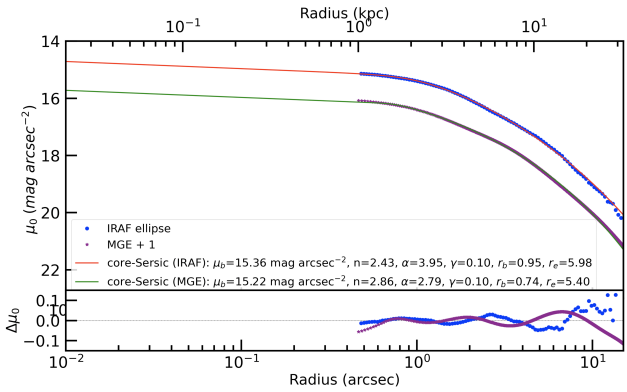

**Figure 9. Upper panel:** An example of the Pan-STARRS/*i*-band surface brightness profiles of the galaxy 2MASXJ12052321+1022461 constructed either directly from `IRAF ellipse` (blue dots) or indirectly from MGE model (purple stars). All magnitudes are corrected for foreground extinction. For better illustrations, we shifted the MGE surface brightness profile by +1 mag. The best-fitting core-Sérsic surface brightness profiles of these `IRAF ellipse` or MGE model are plotted in solid red and blue, respectively, and their best-fitting parameters are shown in the legend. We should note that within the radii of 0.″8 of these 1D profiles, we fixed $\gamma = 0.1$ (see text in Section 4.3 for details) and did not fit these parts to the data. **Lower panel:** The differences (or residuals (`data - model`)) between the `IRAF ellipse` (blue dots) and MGE (purple stars) surface brightness profiles and their corresponding best-fit core-Sérsic models illustrate the fit's goodness.

arcminute-squared FOV with 4 mas pixels to sample the diffraction limit, then fully resolves the MMBH's $r_{\rm SOI}$ to a few hundred Schwarzschild radius ($r_g = 2GM_{\rm BH}/c^2 \approx 0.001 - 400$ AU for black holes with masses in the range of $10^5 - 10^{10}$ M$_\odot$, 1 AU is the Sun-Earth distance = 150,000,000 km), thus provides excellent opportunities to observe the purely Keplerian motion of stars caused mainly by the central black hole's gravitational potential. Thus, the combination of HARMONI IFS and MICADO imager of ELT will be able to detect $10^4$ M$_\odot$ black holes at the distances of $D < 10$ Mpc (Nguyen et al. in preparation); heavier black holes ($M_{\rm BH} \approx 10^9$ M$_\odot$), will be detected up to the distances of $D_A \approx 1$ Gpc ($z \approx 0.3$; this work).

### 4.3.2 Building and extrapolating the galaxy-mass models

In this work, we used ground-based images and extrapolated the surface-brightness profiles following some assumptions sufficient for presenting the simulations and dynamical models. Among various ground-based images available (e.g. SDSS, Pan-STARRS, and 2MASS), Jensen et al. (2021) found the original Data Releases 1 (DR1) cutout images of Pan-STARRS are best even with their cosmetic defects for measuring large galaxies. In addition, the 2MASS, *HST*/Advanced Camera for Surveys (ACS), Pan-STARRS, and SDSS photometric magnitude systems are consistent near the centers of the galaxies. However, the Pan-STARRS profiles remain the same with ACS farther out, while the SDSS and 2MASS are not as deep, and the surface brightness measurements become inconsistent and noisy in the outer portion of the galaxy. We repeated this test carefully for nine chosen galaxies for HARMONI IFS simulation listed in Table 3 using both SDSS and Pan-STARRS images and found a consistent conclusion with Jensen et al. (2021). We, therefore, adopted the Pan-STARRS images for our photometric calibration. We thus performed the Pan-STARRS flux calibration to convert the imaging unit from counts/s to surface brightness in each pixel following the prescription described in Section 3.2 of Jensen et al. (2015). Note that the full-sky images and the cutout images do not have identical photometric calibration mentioned at the Pan-STARRS1 (PS1, including DR1 and DR2) Image Cutout Service webpage[6]. The cutout images are combined from several input images and divided by the imaging keyword `EXPTIME`. These images' individual photometric `ZERO POINTS` are in the imaging header, but we used the median value.

We adopt the multi-Gaussian expansion method (MGE; Emsellem et al. 1994; Cappellari 2002) to describe the galaxy surface brightness observed with Pan-STARRS/*i*-band (i.e. cutout image obtained via PS1[7]; Chambers et al. 2016) with the algorithm and software[8] of Cappellari (2002). During the fit, the model convolves (or analytically deconvolves the Pan-STARRS image) with an adopted *i*-band Gaussian PSF with a median full width at half maximum of FWHM $\approx 1.″25$ (Magnier et al. 2020; Waters et al. 2020).

We show the Pan-STARRS/*i*-band-image and the best-fitting MGE model of the galaxy 2MASXJ12052321+1022461 in Figs 7 and 8 as an example, illustrating the agreement/disagreement between the data and the model in the forms of radial profiles and 2D contours at the same radii and contour levels, respectively.

However, in this work, we did not use the above best-fitting MGE models of Pan-STARRS/*i*-band-images directly in our kinematic models because of their low-angular resolutions and coarse pixel sampling (1 pixel $\approx 0.″25$) compare to these desired scales of HARMONI in our simulations (pixel sampling of 10 mas). We used them to constrain the outer surface-brightness profiles of our chosen simulated galaxies, then extrapolating these profiles towards the regime of 10 mas surrounding their central black holes. We used the Trujillo et al. (2004) core-Sérsic profile (Sersic 1968) because our most massive galaxies sample comprised of core galaxies:

$$I(r) = I' \left[ 1 + \left(\frac{r_b}{r}\right)^\alpha \right]^{\frac{\gamma}{\alpha}} \exp\left\{-b\left[\frac{(r^\alpha + r_b^\alpha)}{r_e^\alpha}\right]^{\frac{1}{n\alpha}}\right\}, \quad (3)$$

where $I' = I_b 2^{-\frac{\gamma}{\alpha}} \exp\left[b 2^{\frac{1}{n\alpha}} \left(\frac{r_b}{r_e}\right)^{\frac{1}{n}}\right]$ and *b* is a function of the various parameters ($n, \alpha, \gamma, r_b,$ and $r_e$) that can be determined by solving the relation (A10) of Trujillo et al. (2004) when the enclosed luminosity at $r_e$ equals to half of the total luminosity, $2L(r_e) = L_T$. In numerical practice, Ciotti & Bertin (1999) use the asymptotic expansion theorem for the $1/r^n$ law to solve for *b* as an analytical function of Sérsic index *n* as $b = 2n - \frac{1}{3} + \frac{4}{405n} + \frac{46}{25515n^2} + O(\frac{1}{n^3})$. Here, *n* is the Sérsic index, which controls the shape of the outer Sérsic part. $r_e$ is the effective radius of the profile. $r_b$ is the break radius, which is the point at which the surface brightness changes from the outer Sérsic part to the inner power-law regime of the profile. $I_b$ is the intensity at the break radius(converted to surface-brightness density $\mu_b$ in Table 3) that controls the sharpness of the transition between the cusp and the outer Sérsic profile. $\alpha$ is the sharpness parameter, which describes the transition between outer Sérsic and inner power-law regimes.

Firstly, we converted the best-fitting deconvolved MGE models into one-dimensional (1D) surface brightness profiles, then fit them

---

[6] https://panstarrs.stsci.edu/
[7] http://ps1images.stsci.edu/cgi-bin/ps1cutouts
[8] v5.0.14 available from https://pypi.org/project/mgefit/

**Table 4.** List of simulated targets and their essential properties used for `HSIM` to produce their mock IFS simulated datacubes and kinematics.

| Galaxy name | $z$ | $D_A$ (Mpc) | $\log R_e$ (kpc) | $M/L_i$ ($M_\odot/L_\odot$) | $\sigma_\star$ (km s$^{-1}$) | $M_\star$ ($10^{12}M_\odot$) | $M_{\rm BH},\sigma_\star$ ($10^9 M_\odot$) | $M_{\rm BH},M_\star$ ($10^{10}M_\odot$) | $R_{\rm SOI},\sigma_\star$ ($''$) | $R_{\rm SOI},M_\star$ ($''$) | NGS ($''$, $m_H$) |
|---|---|---|---|---|---|---|---|---|---|---|---|
| (1) | (2) | (3) | (4) | (5) | (6) | (7) | (8) | (9) | (10) | (11) | (12) |
| J22354078+0129053 | 0.05798 | 259 | 1.41 | 2.0 | 295 | 2.5 | 1.3 | 1.6 | 0.050 | 0.63 | 53, 15.09 |
| J13080241+0900044 | 0.09340 | 499 | 1.77 | 2.5 | 211 | 2.3 | 0.2 | 0.3 | 0.010 | 0.12 | 39, 17.79 |
| J12052321+1022461 | 0.09502 | 445 | 1.62 | 3.0 | 262 | 2.9 | 0.7 | 1.0 | 0.020 | 0.29 | –, – |
| J00034964+0203594 | 0.11812 | 420 | 1.75 | 2.0 | 243 | 4.0 | 0.5 | 0.9 | 0.018 | 0.32 | 53, 18.22 |
| J16171650+0638149 | 0.15357 | 619 | 1.23 | 3.0 | 405 | 2.5 | 6.6 | 8.2 | 0.057 | 0.72 | 43, 16.20 |
| J09322275+0811508 | 0.19251 | 793 | 1.50 | 2.5 | 272 | 2.6 | 0.8 | 1.0 | 0.015 | 0.12 | 48, 13.31 |
| J10221610+0522524 | 0.25531 | 819 | 1.61 | 2.5 | 232 | 2.5 | 0.4 | 0.4 | 0.008 | 0.08 | 43, 16.20 |
| J14155764+0318216 | 0.30212 | 931 | 1.63 | 3.0 | 320 | 4.4 | 2.0 | 4.0 | 0.019 | 0.37 | 35, 14.04 |
| J11480221+0237582 | 0.31399 | 948 | 1.33 | 2.5 | 346 | 3.0 | 3.0 | 4.4 | 0.025 | 0.35 | –, – |

*Notes:* Column 1: Galaxy name which is assigned to J ≡ 2MASXJ (Skrutskie et al. 2006). Column 2: galaxy's redshift (Huchra et al. 2012). Column 3: angular-size distance to the galaxy obtained from NED (double check with `https://www.astro.ucla.edu/~wright/CosmoCalc.html` and redshift; Wright 2006, gives a little different in anngular-size distance probably because of the unclear indication either $D_A$ or $D_L$ in NED). Column 4: galaxy's effective radius (or half-light radius $R_e$ = 1.61 × j_r_eff; Skrutskie et al. 2006; Cappellari 2013; Cappellari et al. 2013a, K18). Column 5: assuming mass-to-light ratio (estimated in Section 5.2). Column 6: stellar velocity dispersion from the galaxy's bulge component $\sigma_\star^2 = G \times M_\star/(5 \times R_e)$ (Krajnović et al. 2018a). Column 7: galaxy's stellar mass (equation (2) of Cappellari et al. 2013a, $\log(M_\star) = -0.44 \times (M_K + 23) + 10.58$). Column 8: central SMBH mass estimated based on equation (2) from K18. Column 9: central MMBH mass estimated based on equation (3) from K18. Column 10: $R_{\rm SOI}$ of SMBH calculated from $\sigma_\star$ and central SMBH mass estimated based on equation (2) from K18. Column 11: $R_{\rm SOI}$ of MMBH calculated from $\sigma_\star$ and central MMBH mass estimated based on equation (3) from K18. We calculate these two $R_{\rm SOI}$ using equation (1). Column 12: Natural-guide star's distance from the galaxy centre and its apparent $H$-band magnitude used in the LTAO mode for the atmospheric turbulence correction.

**Table 5.** Mock `HSIM` IFS of the nine chosen targets (DIT = 900 s = 15 min.)

| Galaxy name | HSIM band | Exptime DIT × NDIT (min.) | Sensitivity – (min.) |
|---|---|---|---|
| (1) | (2) | (3) | (4) |
| J22354078+0129053 | $I_z$, $H+K$ | 30 = DIT × 2 | 10 |
| J13080241+0900044 | $I_z$, $H+K$ | 45 = DIT × 3 | 15 |
| J12052321+1022461 | $I_z$, $H+K$ | 45 = DIT × 3 | 15 |
| J00034964+0203594 | $I_z$, $H+K$ | 60 = DIT × 4 | 15 |
| J16171650+0638149 | $I_z + J$, $H+K$ | 60 = DIT × 4 | 15 |
| J09322275+0811508 | $I_z + J$, $H+K$ | 75 = DIT × 5 | 20 |
| J10221610+0522524 | $I_z + J$, $H+K$ | 90 = DIT × 6 | 30 |
| J14155764+0318216 | $I_z + J$, $H+K$ | 120 = DIT × 8 | 45 |
| J11480221+0237582 | $I_z + J$, $H+K$ | 120 = DIT × 8 | 45 |

*Notes:* Column 1: Galaxy name which is assigned to J ≡ 2MASXJ (Skrutskie et al. 2006). Column 2: HSIM band chosen to perform IFS simulation for observational mock datacubes and their kinematic measurements. The choice of these `HSIM` band is optimal and trade-off between redshift and spectral resolution. Column 3: real exposure time we put in `HSIM` for our simulated kinematics maps presented in Figs 10 and 11 and Figs A1, A5, A9, A13, A17, A21, A25, A29 in Apendix A. Column 4: Sensitivity in terms of exposure time at which we test the lowest limit of S/N from the simulated IFS so that our pPXF still extract accurate kinematics (will be discussed later in Section 6.2). We should note that the estimated time show in Columns 3 and 4 are the science time on targets without accounting for the target acquisition, overhead, and AO setup time.

with the core-Sérsic function above. Here, we fixed the inner power-laws slope of the core-Sérsic profiles with the typical $\gamma = 0.1$ for core galaxies (e.g. Lauer et al. 2007b) expected in the brightest galaxies (e.g. Faber et al. 1997), while allowing the other five parameters to vary. Table 3 shows the best-fit values of these five free parameters of the surface-brightness profiles of nine chosen simulated galaxies.

As a sanity check, we used the `Image Reduction and Analysis Facility (IRAF) ellipse` task (Jedrzejewski 1987) to extract radial surface-brightness profiles of the stars in concentric annuli with varying position angles and ellipticities, although keeping both fixed does not change our results (Nguyen et al. 2022). We then fit these stellar radial light surface-brightness profiles with a core-Sérsic function. The fits were carried out using a non-linear least-squares algorithm (`IDL MPFIT function`[9]; Markwardt 2009). To compare the model and data, before extracting the 1D spatially deconvolved (i.e. intrinsic) `IRAF` profile, we firstly made a two-dimensional (2D) Gaussian PSF adopted from the Pan-STARRS $i$-band image above, then convolved it with the image secondly. Thirdly, we iterated the core-Sérsic function to the spatially deconvolved profile for its best-fit parameters. We showed the consistency of these two approaches in determining the best-fitting parameters of the core-Sérsic profiles using the case of galaxy 2MASXJ12052321+1022461 in Fig. 9 as an example and listed these parameters for all nine simulated galaxies in Table 3.

Secondly, we used these derived parameters to reconstruct the interpolated-MGE model towards the central 10 mas for each galaxy via the `mge_fit_1d.py` routine (Cappellari 2002, see footnote 9) to fit the analytic core-Sérsic with a constant ellipticity $\epsilon = 0$ and 10 Gaussians across the radii of from ≈14$''$ to ≈30$''$, depending on the apparent size of galaxies.

Finally, we created a mass-follow-light surface density by assuming a constant $M/L_i$. This stellar-mass component will be added in an SMBH or an MMBH with a specific mass as a point source. In this work, we ignored (1) the possible variation in $M/L$ inferred from stellar population variation (McConnell et al. 2013; Mitzkus et al. 2017; Li et al. 2017; Nguyen et al. 2017, 2018, 2019; Nguyen et al. 2020, 2021, 2022; Thater et al. 2022, 2023) due to the lack of spatial resolution of Pan-STARRS observations and (2) the distribution of dark matter halo (Navarro et al. 1996) as we concern the stellar kinematics within the FOV of $0.''4 \times 0.''4$ of HARMONI only where the central black hole's potential dominates. In this nuclear region, any possible $M/L$ gradient due to the stellar population or dark matter is insignificant for our tests.

---

[9] Available from: `http://purl.com/net/mpfit`.

## 5 HARMONI IFS SIMULATION

We first describe the HARMONI instrument on ELT and HSIM simulator in Section 5.1. Next, we combine the mass-MGE models of all nine chosen galaxies constructed in Section 4.3 with the HSIM simulator to simulate their $I_z$ (0.83 − 1.05 $\mu$m), $I_z + J$ (0.81 − 1.37 $\mu$m), and $H+K$ (1.45 − 2.45 $\mu$m) mock datacubes in Section 5.2. Finally, we present the extracted kinematics of all nine galaxies with different redshifts and sizes (aka $D_A$ vs. $R_e$, Fig. 5) in Section 5.3.

### 5.1 HARMONI instrument and HSIM simulator

HARMONI is an optical and NIR instrument on ELT, which will provide IFS at four different spatial scales (i.e. $4 \times 4$, $10 \times 10$, $20 \times 20$, and $30 \times 60$ mas$^2$) and three spectral resolving powers (i.e. $\lambda/\Delta\lambda \approx 3,355$, $\approx 7,104$, and $\approx 17,385$). Given a 39 m single-field in design with 798 hexagonal segments (each $\approx 1.4$ m across), ELT can collect spectra of $152 \times 214$ ($\approx 32,530$) spaxels equipped with laser guide star AO. This technical design is at best to perform a wide range of science programs from diffraction-limited to ultra-sensitive such as morphology, spatially resolved populations and kinematics, abundances, and line ratios of distant sources (Thatte et al. 2016), allowing it to achieve a particular S/N in relatively short exposure time, even in faint surface brightness targets. In particular, such unprecedentedly powerful technicals will revolutionize our understanding of the physics of mass assembly in high-redshift galaxies and hunting for the missing intermediate-mass black holes (IMBHs, $M_{BH} \lesssim 10^5$ M$_\odot$) population in nearby dwarf galaxies or stellar clusters (Zieleniewski et al. 2015; García-Lorenzo et al. 2019). A full description of the instrument is presented in Thatte et al. (2020) and on the HARMONI webpage[10].

HARMONI Simulator (HSIM[11]) is the pipeline for simulating observations with the HARMONI instrument on ELT (Zieleniewski et al. 2015). It uses high spectral and spatial resolution IFS cubes without random noise generated in Section 5.2 as inputs, encodes with the celestial target's physical properties, and then creates simulated cubes. The simulations incorporate detailed models of the atmospheric effects and realistic detector statistics to mimic realistically mock data. This paper concentrates in-depth on the simulations of the ELT AO observations' quality by measuring the nuclear stellar kinematics in distant galaxies and estimating their $M_{BH}$. From those, we will explore the limits at which HARMONI can produce feasible observables.

### 5.2 Simulations of the mock IFS datacubes

To understand the effects of high redshifts and galaxy sizes on the stellar kinematic measurements and sensitivities, we chose to simulate the HARMONI IFS observations in nine galaxies with different redshifts and sizes using the dedicated HSIM pipeline (see footnote 11). Due to our survey's wide range of redshifts (0.028 < $z \lesssim 0.3$), specific stellar features are used to estimate the nuclear-stellar kinematics shift along the spectral dimension differently for each target. The CO-bandheads absorptions (2.29 − 2.47 $\mu$m; e.g. CO(2–0) $\lambda 2.293$ $\mu$m and CO(3–1) $\lambda 2.312$ $\mu$m bands) fall off the $H + K$- and $K$-band and cannot be used for galaxies with $z > 0.04$. Nevertheless, the CaT stellar absorption (0.86 − 0.88 $\mu$m) features stay safely within the $I_z$-band for galaxies with $z < 0.12$ and within the $I_z + J$-band for our other selected galaxies with higher redshift. Additionally, to account for the spectral resolutions and to test the feasibility of different stellar features, we performed simulations for the $I_z$-band (0.83−1.05 $\mu$m) and $I_z+J$-band (0.81−1.37 $\mu$m), which have $\sigma_{instr} \approx 18$ km s$^{-1}$, $\lambda/\Delta\lambda \approx 7,104$ and $\sigma_{instr} \approx 38$ km s$^{-1}$, $\lambda/\Delta\lambda \approx 3,355$, respectively. It is also necessary to test the capacity of using some stellar features in $H + K$-band (1.45 − 2.45 $\mu$m) to measure the stellar kinematics, which are not used widely in the current works (but see Crespo Gómez et al. 2021).

Depending on redshift, some strong absorption lines from atomic species at the blue part of $K$-band shifted to its red part (e.g. Na I $\lambda 2.207$ $\mu$m, Ca I $\lambda 2.263$ $\mu$m, and, Mg I $\lambda 2.282$ $\mu$m) for $z < 0.1$ galaxies in the MMBH sample. And a larger number of atomic absorption lines in $H$-band (e.g. Mg I $\lambda 1.487$, 1.503, 1.575, 1.711 $\mu$m, Fe I $\lambda 1.583$ $\mu$m and S I $\lambda 1.589$ $\mu$m) remains on the $H + K$-band for the rest higher redshift galaxies. In addition, there are several strong CO absorption features that are very sensitive to the star surface gravity and effective temperature in $H$-band (Silge & Gebhardt 2003; Crespo Gómez et al. 2021) mainly produced at the atmospheres of evolved giant stars, with a non-negligible contribution of cool asymptotic giant branch (AGB) stars (Kleinmann & Hall 1986; Dallier et al. 1996; Wallace & Hinkle 1997; Förster Schreiber 2000; Böker et al. 2008; Kotilainen et al. 2012; Dametto et al. 2014) such as CO(3–0) $\lambda 1.540$ $\mu$m, CO(4–1) $\lambda 1.561$ $\mu$m, CO(5–2) $\lambda 1.577$ $\mu$m, CO(6–3) $\lambda 1.602$ $\mu$m, CO(7–4) $\lambda 1.622$ $\mu$m, and CO(8–5) $\lambda 1.641$ $\mu$m (see Fig. 11). We relied on these atomic absorption lines and CO absorption features as the most significant carriers of the kinematic information – whenever they are on the redshifted wavelength coverages of the $H + K$ grating – to build the inputs and to extract stellar kinematics from the outputs of HSIM in this work. They will be the benchmark for future usage in deriving stellar kinematics from IFS data.

All essential properties of the nine chosen galaxies needed for the modelings are presented in Table 4, while the chosen grating IFS and HSIM simulations are shown in Table 5. However, regarding the AO performance during HSIM simulations, we did not use the appropriate NGS listed in Table 4 for each galaxy (i.e. we cannot find the realistic NGS for two galaxies in this nine simulated sample) but supplied these simulations with the LTAO mode with a star of 17.5 mags in $H$ band within the distance of 30$''$, the standard zenith seeing of FWHM = 0$''$64 and airmass of 1.3. These parameters are defaulted in HSIM to perform median observational conditions but can be changed from target to target as long as the selected criterion (iv) mentioned in Section 2.2 is satisfied. The necessary NGS information listed in Column 10 of Table 4 and showed in Fig. 6 is a reduced version of 101 selected galaxies, showing the robustness of LTAO performances for our MMBH IFS survey. The LTAO mode combines six off-axis LGS with a faint NGS to deliver diffraction-limited image quality over a large fraction of the sky. It also implements several off-axis wavefront sensors, but optimises to analyze the center while better sampling the on-axis turbulence cylinder of the FOV in detail, resulting in high performance across a small FOV, limited by tomographic error and low order residuals and increased (medium) sky coverage.

We simulated the IFS within the FOV of $400 \times 400$ mas$^2$ and sampled the pixel size to $10 \times 10$ mas$^2$. This choice of a 10 mas pixel size ensures that we sample the ELT PSF FWHM of 12–18 mas with 1–2 spaxels (Thatte et al. 2016, 2020), resulting in precisely kinematic measurements at the galaxy center at the scale of a factor 2× smaller than the resolving power in radius (i.e. our given proposed survey spatial resolution of $20 \times 20$ mas$^2$ gives $\approx$12 pixels within the black holes' SOI with the simulated pixel sampling

---

[10] http://harmoni-web.physics.ox.ac.uk/
[11] v3.10; available from https://github.com/HARMONI-ELT/HSIM

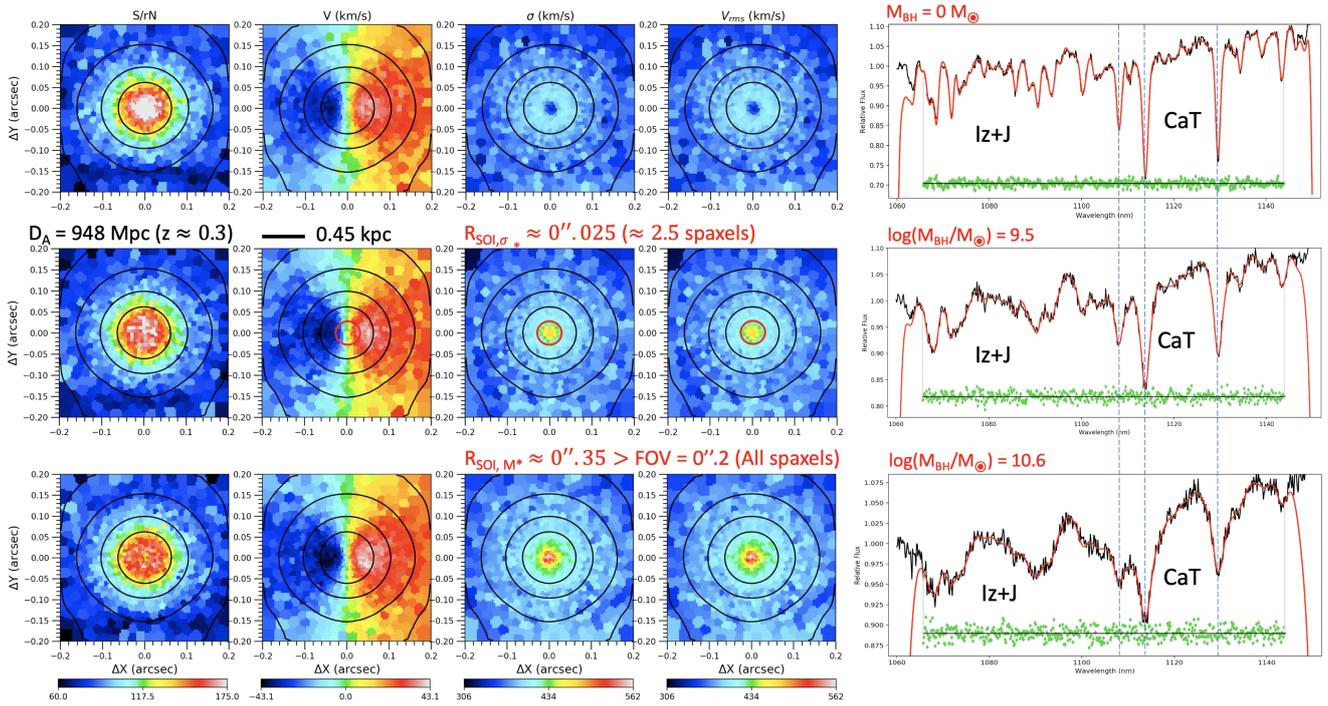

**Figure 10.** The stellar kinematic maps of 2MASXJ11480221+0237582 extracted from a spectral part (1.065 $\mu$m < $\lambda$ < 1.145 $\mu$m, $z \approx 0.3$) – which contains the CaT stellar-absorption line ($\lambda$0.86–0.87 $\mu$m) – of its mock $I_z + J$ HSIM IFS cubes produced from JAM$_{sph}$ (Section 5.2) using pPXF. These maps are present with three different black hole masses: $M_{BH} = 0$ M$_\odot$ (top-row plots), $3 \times 10^9$ M$_\odot$ (middle-row plots), and $4.4 \times 10^{10}$ M$_\odot$ (bottom-row plots). On each row, these maps are listed from left to right with (1) the signal-to-residual noise (S/rN) measuring the standard deviation of the residuals between the galaxy spectrum and the best-fitting pPXF model to define a residual noise (rN) for each Voronoi bin, (2) relative velocity ($V$), (3) velocity dispersion ($\sigma_\star$), (4) root-mean-squared velocity ($V_{rms}$); the black contours in all four maps indicate the isophotes from the collapsed HSIM IFS cubes spaced by 0.5 mag arcsec$^{-2}$, and (5) part of the simulated spectrum, showing the CaT-absorption features indicated by vertically thin-dashed lines) of the stellar component extracted from one bin (black line) and its best-fit model produced by pPXF (red line). Two gray-vertical lines limit the wavelength range where the spectrum was fit, and green dots show the residual between the galaxy spectrum and the best-fitting model (data-model). Color bars at the bottom of the corresponding maps are fixed at the same scale for all three black hole masses to illustrate the kinematic effects of the central black holes and also indicate the robustness of our proposed kinematic measurements having at the centers of these highest mass galaxies as the kinematic signatures for SMBHs/MMBHs. The red circle at the centers of middle-row maps demonstrates the size of the SMBH's SOI radius ($R_{SOI}$)

.

of $10 \times 10$ mas$^2$). Thus, the stellar kinematics dominated mainly by the central black holes will be robustly detected. The exposure time of each simulation will change substantially depending on redshifts and gratings carefully to ensure the S/N in every spaxel at the measured-stellar features $\gtrsim 5$, but we will bin pixels together later for higher S/N. However, to mimic the actual observations on ELT, we applied multi-exposure frames and dithering by setting DIT = 900 s (15 min.) for each; the total exposure time will be counted by the number of exposure NDIT = an interger in the HSIM pipeline.

For simplicity, we assumed the light-of-sight velocity distribution (LOSVD) to be Gaussian (e.g. $V_{rms}^2 = V^2 + \sigma_\star^2$). Thus, we created the 2D intrinsic first and second velocity moments (i.e. $V$ and $\sigma_\star$) in terms of a Gaussian using the JAM$_{sph}$ modeling (C20) and the galaxy-mass model mentioned in Sections 4.1 and 4.3, respectively.

We also assumed a constant $M/L_i$ (Table 4) to convert the axisymmetric stellar-light MGEs inferred from the core-Sérsic profile (Table 3) into the galaxy-mass model for each galaxy. This $M/L_i$ was estimated using the Pan-STARRS ($g - i$) color and the Roediger et al. (2015) color–$M/L$ scaling relation assumed the Charlot & Fall (2000) prescription for dust + interstellar medium (ISM) attenuation correction. Details of these color–$M/L$ conversion and dust +

ISM correction processes followed descriptions from Nguyen et al. (2018). In order to estimate the $M/L_i$ accurately, we calculated the background level of each image in small regions as far away from the galaxy center as possible (radial range of $20'' - 35''$ depending on the apparent size of each galaxy) and subtracted it off. The $5'' \times 5''$ central regions of these nine chosen simulated galaxies show mostly constant ($g - i$) color for each nucleus with the values ranging from 1.17 mag to 1.35 mag, resulting in constants $M/L_i$ changing from 2.0 (M$_\odot$/L$_\odot$) to 3.0 (M$_\odot$/L$_\odot$) (see Fig. 7 and Table 1 of Roediger et al. 2015, for estimating $M/L$ based color).

In the JAM$_{sph}$ modelings (C20), we assumed an average inclination ($i \approx 60°$) and chose to model three kinematics of three different $M_{BH}$, including $M_{BH} = 0$ M$_\odot$ and two other black holes, either (i) following the $M_{BH}$–$\sigma_\star$ relation or (ii) assuming the $M_{BH}$ follows the $M_{BH}$–$\sigma_\star$ for $M_\star < M_{crit}$ and switches to being proportional to $M_\star$ for $M_\star > M_{crit}$. And for this, we used equations (2) and (3) of K18, respectively. These two black hole masses for each simulated galaxy are presented in Table 4. The kinematic maps were computed with JAM$_{sph}$ on a regular grid with the FOV of $200 \times 200$ mas$^2$ and pixel size of $5 \times 5$ mas$^2$. This scale will be convolved with the HARMONI PSF, rebinned, and interpolated to the specific pixel-sampling scale of $10 \times 10$ mas$^2$ by HSIM. Also, previous dynamical analysis with the IFS (e.g. WHT/OASIS and VLT/SINFONI) and

14  *Dieu D. Nguyen et al.*

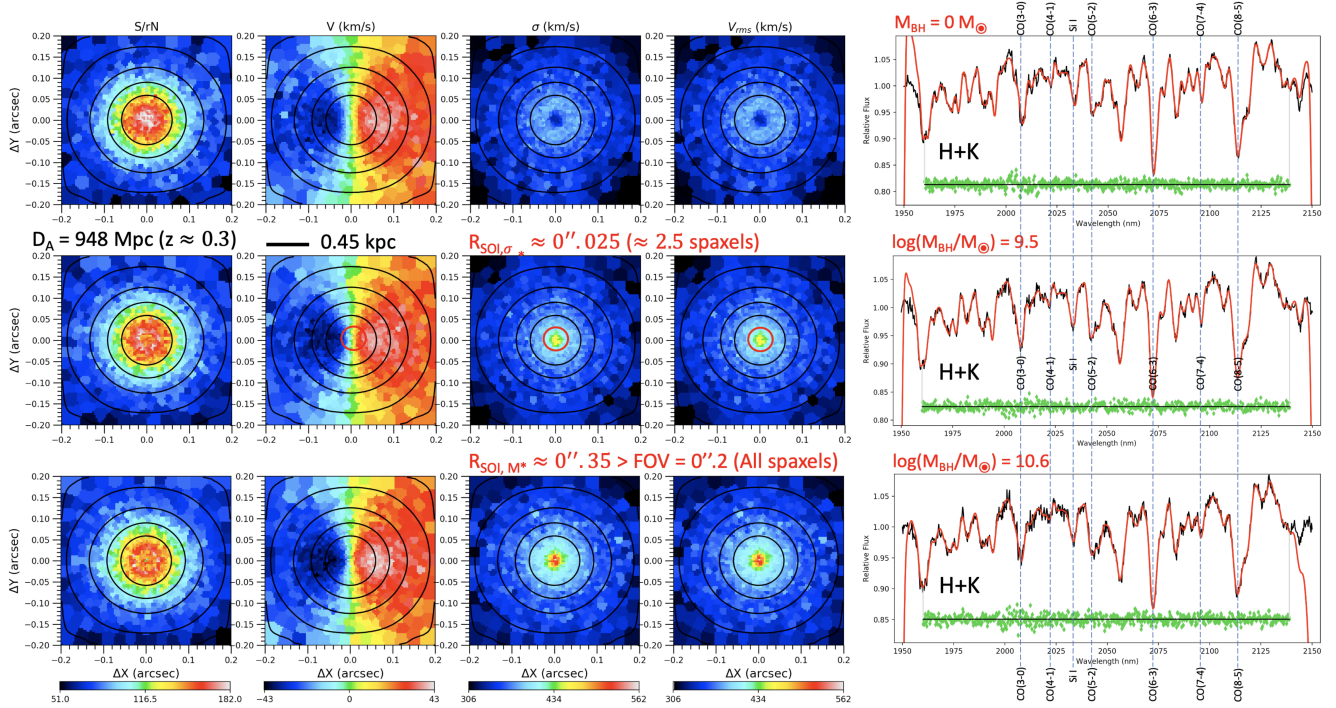

**Figure 11.** Same as Fig. 10 but the stellar-kinematic maps of the galaxy 2MASXJ11480221+0237582 extracted from part of the strongest stellar-absorption features (1.955 $\mu$m < $\lambda$ < 2.135 $\mu$m, $z \approx 0.3$; indicated by vertically thin-dashed lines and labels) of its mock $H + K$ HSIM IFS cubes produced from JAM$_{sph}$ using pPXF. This spectrum part contains a strong atomic absorption line S I $\lambda$1.589 $\mu$m) and some CO-absorptions lines from the atmospheres of evolved giant stars and cool AGB stars: CO(3–0) $\lambda$1.540 $\mu$m, CO(4–1) $\lambda$1.561 $\mu$m, CO(5–2) $\lambda$1.577 $\mu$m, CO(6–3) $\lambda$1.602 $\mu$m, CO(7–4) $\lambda$1.622 $\mu$m, and CO(8–5) $\lambda$1.641 $\mu$m.

the Schwarzschild orbit-based model (Schwarzschild 1979) that included the effects of a central SMBH, the mass distribution of the stars, and a dark matter halo for massive (core) slow-rotator galaxies found the tangential anisotropy ($\beta_r < 0$) in the cores and radial anisotropy ($\beta_r > 0$) at larger radii among the population of stellar orbits (Cappellari et al. 2008; Thomas et al. 2014). We thus accounted for this fact in our simulations by adopting $\beta_r = -0.2$ for several innermost-MGE components inferred mainly from the power-law part, which describes the core of the surface brightness profile separated from the outer Sérsic one by the break radius $r_b$, and $\beta_r = +0.2$ for the rest, while setting the tangential anisotropy of the individual kinematic-tracer MGE Gaussians, $\sigma_\theta = \sigma_\phi$ as assumed in equation (55) of C20.

Given all those assumptions, we created an input noiseless-IFS cube for HSIM by employing the following steps:

(i) We accounted for the targets' redshifts on the MARCS SPS spectra by shifting the spectral range (i.e. for specific HARMONI grating bands) by a factor of $(1 + z)$.

(ii) We did logarithmically-rebin the synthetic-stellar spectrum of the chosen population (SPS; Section 4.2) to a scale at which the velocity scale is set as velscale = 2 km s$^{-1}$ so that the spectrum has a constant $\Delta \log \lambda$ intervals.

(iii) For each spatial position in the cube, we constructed kinematics Gaussian kernel, sampled at steps $\Delta V = 2$ km s$^{-1}$, with the mean velocity and velocity dispersion ($V, \sigma_\star$) computed by the JAM$_{sph}$ model for that position.

(iv) We convolved the logarithmically-rebinned spectrum created from step one with the Gaussian kernel generated from step two, then linearly interpolated this logarithmically-rebinned spectrum to the constant wavelength step $\Delta \lambda \approx 0.02$ Å, which is small enough so that no information is lost by the interpolation.

(v) We rebinned the spectrum by co-adding an integer number of adjacent spectral pixels to reach a step in wavelength at a minimum 2× smaller than the smallest HARMONI instrumental resolution in terms of FWHM (e.g. for the *JHK* gratings, $\Delta \lambda \approx 0.2$ nm = 2 Å). This is a rigorous integral over the pixels, and no information is lost.

(vi) We stored the resulting redshifted, linearly-sampled, LOSVD-convolved, and noiseless spectrum in a cube.

(vii) We estimated the surface brightness of each galaxy (i.e. integrated intensities) using its core-Sérsic MGE model inferred from Table 3, then assigned every spaxel's intensity to its corresponding linearly-sampled, LOSVD-convolved, and noiseless spectrum in the cube. Since the core-Sérsic profile only describes the galaxy's surface brightness along the major axis, our galaxies are not all spherical. To deal with the galaxy shapes, we computed the elliptical radius of every pixel using the relation: $r_{\text{ellipse}} = \sqrt{x^2 + \left(\frac{y}{q}\right)^2}$, where $x$, $y$, and $q$ are the positions of pixels $(x, y)$ and the axis ratio from the stellar-light MGE model (Section 4.3), respectively, then assigned the flux by the core-Sérsic profile $I(r_{\text{ellipse}})$ in equation (3).

(viii) We scaled the template spectrum in each 5×5 mas$^2$ spaxels in such a way that its mean flux in the *i*-band, in erg s$^{-1}$ cm$^{-2}$ Å$^{-1}$, is equal to the surface brightness, in erg s$^{-1}$ cm$^{-2}$ Å$^{-1}$ arcsec$^{-2}$, times the 5 × 5 mas$^2$ spaxels area. The robustness of our surface brightness estimations is critical to the HARMONI sensitivity (and thus the required S/N within a reasonable exposure time), which should be high enough for measuring the stellar (or ionized gas emission, if detected) kinematics accurately. In Section 4.3, we tested the consistency of the surface brightness derived from the Pan-STARRS and SDSS images at the galaxy centers.

### 5.3 `HSIM` mock datacubes and extracted kinematics

Figs 10 and 11 show the kinematic maps of the galaxy 2MASXJ11480221+0237582, the furthest target in our simulated sample ($z \approx 0.3$) as an example, extracted from the $I_z + J$ and $H + K$ HSIM mock datacubes with JAM$_{\rm sph}$ model (C20), respectively. In each Fig., three different black holes that have different masses are assumed to reside at the galaxy center, including zero black holes ($M_{\rm BH} = 0$ M$_\odot$), $M_{{\rm BH},\sigma_\star} = 3 \times 10^9$ M$_\odot$ (equation (2) of K18), and $M_{{\rm BH},M_\star} = 4.4 \times 10^{10}$ M$_\odot$ (equation (3) of K18), are shown in each row plot (also see Table 4).

In order to create these maps, we adopt the adaptively `Voronoi Binning` method (`vorbin`[12]; Cappellari & Copin 2003) to spatial bin 2D data to the threshold-adopted S/N ≳ 75 per bin. This technique increases the S/N by adding up the signals of many spaxels within one bin and reduces the uncertainty of the kinematic measurement of that bin. We also took into account both the quality of the simulated data and the quality of the spectral fit by using the signal-to-residual noise (S/rN) measured as the standard deviation of the residuals between the galaxy spectrum and the best-fitting `Penalized PiXel-Fitting` (pPXF [13]; Cappellari 2022) model to define a residual noise (rN) for each `Voronoi` bin. Due to the high S/rN of the mock datacubes as seen in the left panels of these Figs, we obtained a small root-mean-squared velocity scatter with typical values of $\Delta V_{\rm rms} \lesssim 2.5\%$ (i.e. $\Delta \sigma_\star \lesssim 2\%$ and $\Delta V \lesssim 1\%$). In the successive panels of each row plot, the kinematic maps show orderly with rotation subtracted for the systemic $V_{\rm sys}$, velocity dispersion $\sigma_\star$, and root-mean-squared velocity $V_{\rm rms} = \sqrt{V^2 + \sigma_\star^2}$.

We also demonstrate in these Figs the pPXF fits for these kinematic maps using the stellar CaT-absorption features (1.06–1.14 $\mu$m) in $I_z + J$ band and some strongly stellar features (e.g. 1.95–2.15 $\mu$m) in $H + K$ band for 2MASXJ11480221+0237582 at the redshift $z \approx 0.3$. During this pPXF fit between the mock simulated spectra and stellar model, we used the MARCS (Gustafsson et al. 2008) version of the (Maraston & Strömbäck 2011) SPS models and default Legendre polynomials for correcting the template continuum shape (i.e. by setting `mdegree = 0, degree = 4`), and fit only for $V$ and $\sigma_\star$ (i.e. by setting `moments = 2`). In addition, we also accounted for the HARMONI IFS instrumental broadening by broadening the stellar templates with the constant instrumental dispersion adopted by HSIM, which must be done before log-rebinning the spectra. And, to make our fit more realistic, we included 13 templates with ages from 3 to 15 Gyrs and Solar metallicities (`z002`). The best-fit SPS template overlaid on the simulated spectra. Their residuals (`data-model`) are also shown simultaneously in the same panel to illustrate the quality of the fits.

We tested the usage of the $H + K$-band wavelength region, which is rarely used for kinematics studies (but see Crespo Gómez et al. 2021) (i.e. using some strongly stellar features but not using the CO-absorption bandheads because they fall out of the grating wavelength). Thus, we first tested with different chunks of wavelength ranges in the $H + K$ band, for example, the $H + K$ short (blue, 1.70–1.97 $\mu$m, where contains the atomic absorption: Mg I $\lambda$1.487 $\mu$m) and $H + K$ long (red, 1.97–2.15 $\mu$m, where contains the atomic absorption: S I $\lambda$1.589 $\mu$m and the CO absorptions: CO(3–0) $\lambda$1.540 $\mu$m, CO(4–1) $\lambda$1.561 $\mu$m, CO(5–2) $\lambda$1.577 $\mu$m, CO(6–3) $\lambda$1.602 $\mu$m, CO(7–4) $\lambda$1.622 $\mu$m, and CO(8–5) $\lambda$1.641 $\mu$m), and found they provide consistent kinematic maps. Note that these ranges of $H + K$ short and $H + K$ long are subject to change substantially depending on the redshift of the target. Next, we compared the kinematic results extracted from the $H + K$ band to those extracted from the stellar CaT features in the $I_z + J$ band. They prove that some of these strong-stellar features in $H + K$ grating to measure stellar kinematics are robust and feasible to measure SMBH masses with minimum uncertainty.

The stellar kinematic maps shown in Figs 10 and 11 have the central drops of $\sigma_\star$ and $V_{\rm rms}$ for the case of zero black holes that consistent with our core-Sérsic most massive galaxies. The central drop in velocity dispersion is a general feature of the predicted stellar kinematics of galaxies without central SMBHs, for a range of assumed density and anisotropy profiles (e.g. Tremaine et al. 1994). Instead, models with an SMBH with the mass either following equation (2) or (3) of K18 create centrally raising peaks towards the galaxy center in both velocity dispersion and root-mean-squared velocity map. This fact agrees with the general expectation that the central velocity dispersion should be increased in a Keplerian way generally where the central SMBH's potential dominates (e.g. Tremaine et al. 1994). The difference between these kinematic maps at the galaxy center is visible very clearly, especially for the cases with and without a central SMBH in 2MASXJ11480221+0237582, one of the farthest targets of our most massive survey sample, demonstrating the unprecedented spectral and spatial resolving powers of ELT/HARMONI in detecting stellar-kinematic signatures of central SMBHs at a large distance (e.g. out to the redshift of $z \leq 0.3$) and measuring their mass accurately and dynamically.

The edge effect is visible clearly on the kinematic maps, which always produces higher velocity dispersions (and thus higher root-mean-squared velocity) for the top- and bottom-bins (see the $\sigma_\star$ and $V_{\rm rms}$ maps in Figs 11 and 10) than their actual predictions of decreasing values because these spaxel-bins are away from the center. This effect also resulted in the squared shape for several outermost surface brightness contours. To avoid any uncertainty due to this instrumental issue in our dynamical modelings, we exclude all these high $V_{\rm rms}$ bins in our recovering models for the $M_{\rm BH}$ (Section 6.1).

Other eight chosen simulated galaxies (listed in Tables 3 and 4) with their kinematic results extracted from the $I_z/I_z + J$ and $H + K$ HSIM mock datacubes are shown in Figs A1, A5, A9, A13, A17, A21, A25, and A29 of Appendix A (available as supplementary material) side by side for comparison but excluded the pPXF fitting plots. We should note that the first four of these eight galaxies have low redshifts and their CaT features still stay in the wavelength range of the $I_z$ grating (0.83–1.05 $\mu$m). We thus simulated their $I_z$ observations instead of $I_z + J$ for a higher spectral resolution.

## 6 RESULTS

### 6.1 Black hole mass recovering

In Section 5, we used the JAM$_{\rm sph}$ (C20) modeling to generate the 2D intrinsic first and second-order velocity distributions of stellar kinematics (i.e. $V$ and $\sigma_\star$), which then used to convolve with the Maraston SPS models based on MARCS library (Gustafsson et al. 2008, Section 4.2) to simulate the IFS datacubes and extracted their corresponding kinematics maps ($V$, $\sigma_\star$, $V_{\rm rms}$, Section 5.3).

---

[12] v3.1.5 available from https://pypi.org/project/vorbin/
[13] v8.2.1 available from https://pypi.org/project/ppxf/

16  *Dieu D. Nguyen et al.*

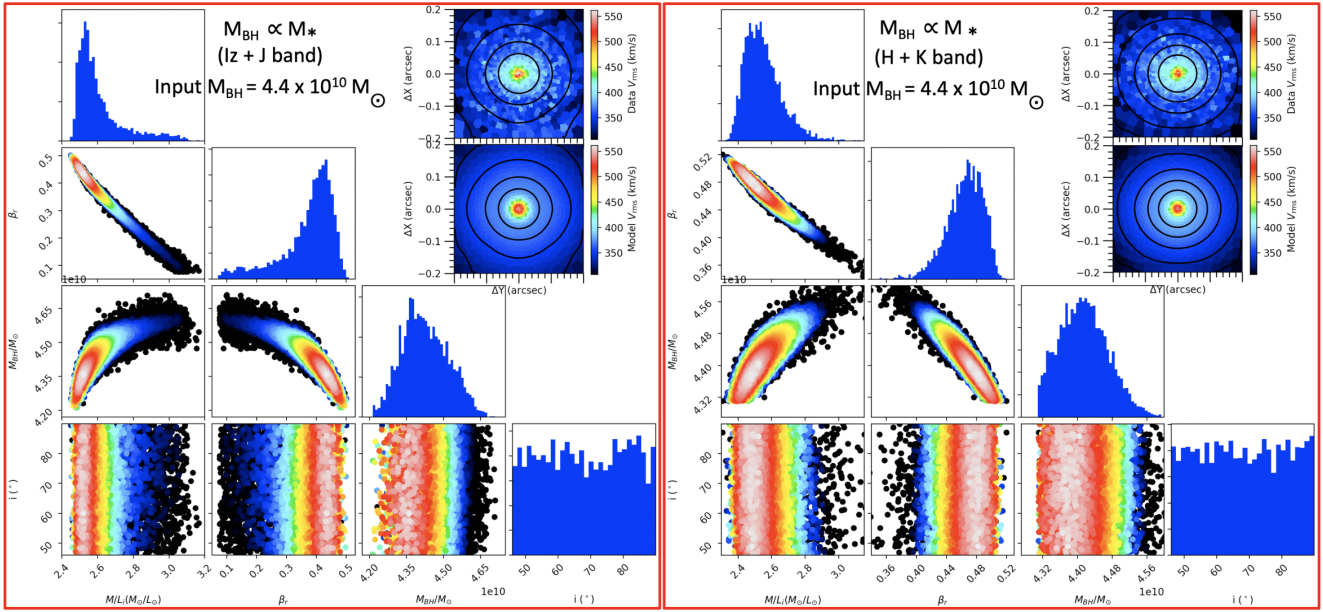

**Figure 12.** The `adamet` MCMC post-burn-in phase posterior distributions for our best-fit JAM$_{sph}$ models with the assuming central black hole with a mass that follows the $M_{BH}$–$M_\star$ relation for galaxies with masses above $M_{crit.}$ predicted by equation (3) of K18 (see text for details). These posterior distributions were obtained when optimizing the JAM$_{sph}$ models to the `HSIM` simulated kinematics of the galaxy 2MASXJ11480221+0237582 created using the JAM$_{sph}$ models (Section 5.2). The scatter plots show the projected 2D distributions for each parameter. The histograms show the projected 1D distribution. From the top left to bottom right, the panels show the inclination $i$, $M_{BH}$, $M/L_i$, and $\beta_r$ for JAM$_{sph}$. The insect $V_{rms}$ maps are the simulated kinematic maps extracted from the simulated datacubes (top), while the maps recovered from the best-fit JAM$_{sph}$ models (bottom) are shown to visually illustrate the agreements/disagreements at every spaxel between the simulated data and our adopted best-fit model. These posteriors are produced using the $I_z + J$ (left) and $H + K$ (right) band HARMONI simulated kinematics with $M_{BH} = 4.4 \times 10^{10}$ M$_\odot$; other input parameters are listed in Table 4.

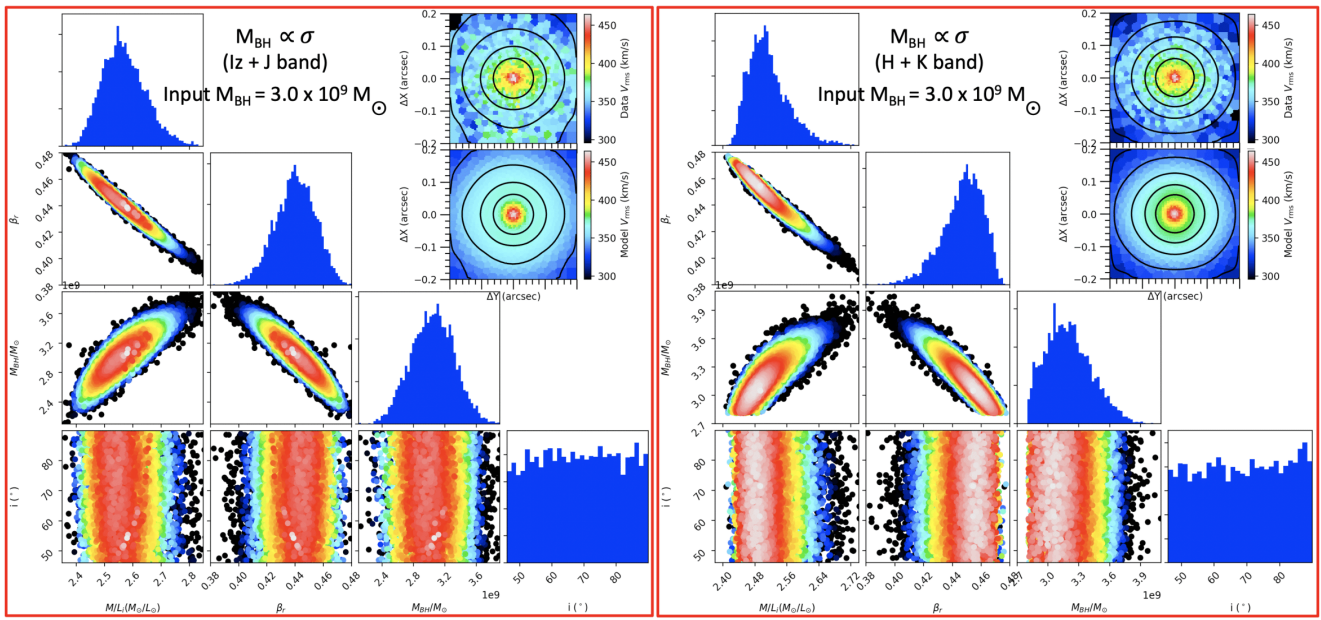

**Figure 13.** Same as Fig. 12 but for the HARMONI simulated kinematics with $M_{BH} = 3 \times 10^9$ M$_\odot$ that follows the $M_{BH}$–$\sigma_\star$ relation for galaxies with masses below $M_{crit.}$ predicted by equation (2) of K18 for the galaxy 2MASXJ11480221+0237582.

During this process, we assumed some dynamical parameters and hypotheses for a central compact dark mass object ($M_{BH}$), stellar orbitals ($\beta_r$), stellar mass ($M/L_i$), and inclination angle ($i$). In this section, we do a reversed process using the JAM$_{sph}$ modeling itself and assume the available stellar-kinematic measurements from our mock HARMONI IFS cubes (Section 5.3) to infer (or recovery) these dynamical parameters, especially $M_{BH}$.

The JAM$_{sph}$ model fits the simulated kinematics data with the following parameters: (1) inclination angle ($i$), (2) the mass of a point-like SMBH $M_{BH}$, (3) $M/L_i$, which parameterizes the $M/L$ relative to the best-fit stellar population estimated in $i$-band, and

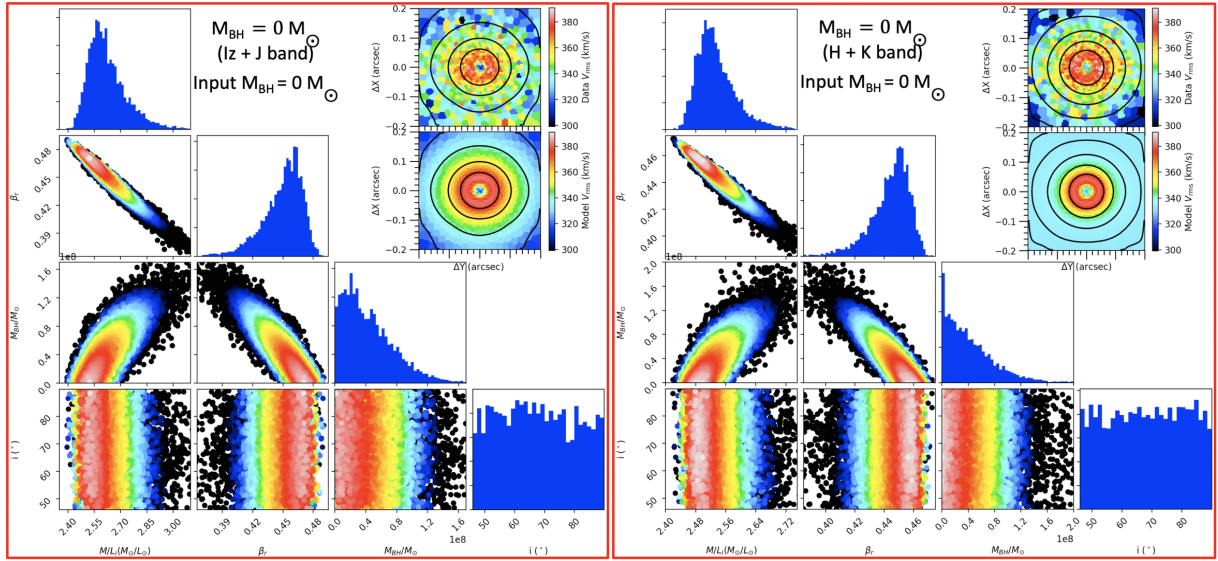

**Figure 14.** Same as Fig. 12 but for the HARMONI simulated kinematics with $M_{BH} = 0$ M$_\odot$ for the galaxy 2MASXJ11480221+0237582.

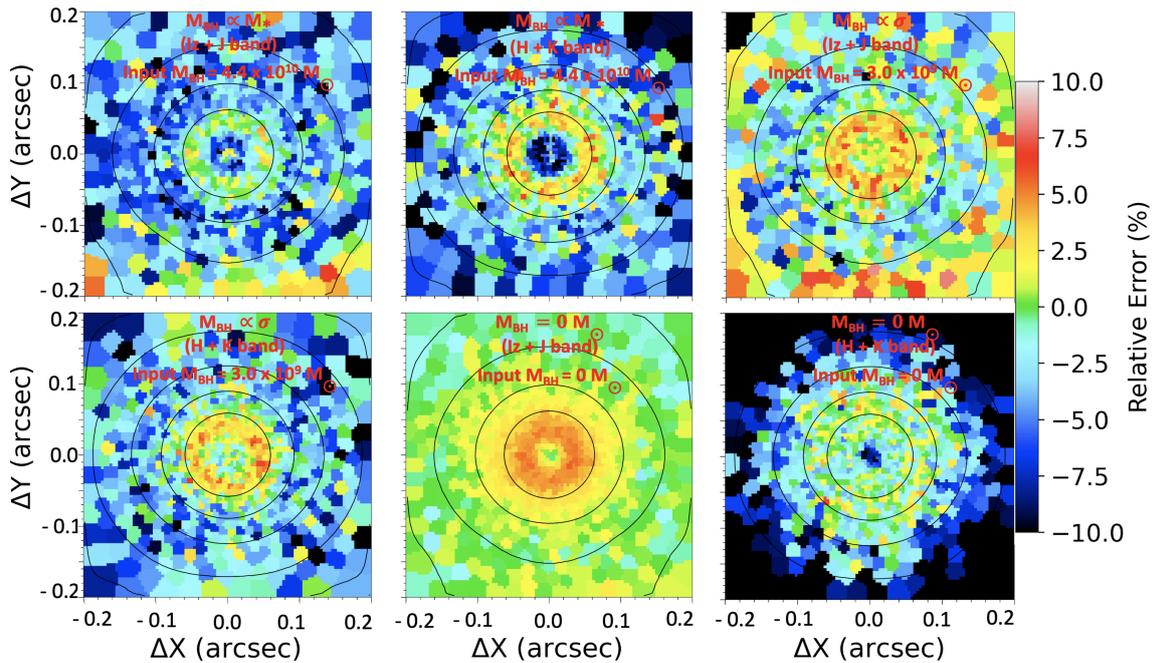

**Figure 15.** $V_{rms}$ residual maps demonstrating the spatially relative agreements/disagreements between the HSIM simulated kinematic data and the best-fit JAM$_{sph}$ model, (data−model)/data, of the galaxy 2MASXJ11480221+0237582, which showed in the insert plots of Figs 12, 13, and 14. The corresponding information of input black holes and HSIM IFS bands are listed in the legend of each panel. All maps share a common color bar on the right. Our recovering best-fit JAM$_{sph}$ models fit well to data with the relative error <10% for all cases across the simulated FOV of $0''.4 \times 0''.4$.

(4) the global anisotropic parameter $\beta_r$ among the population of stellar orbits ($\gamma = \beta_r$; implemented in Section 5.2). All four parameters are space in the linear scales. Note that although we create synthetic models with an anisotropy profile that varies slightly with radius, the fitted model assumes a constant anisotropy for simplicity. JAM$_{sph}$ generates kinematic models that can be compared with their corresponding simulated values ($V_{rms}$) within their errors. We also tested with accurately known HSIM LTAO PSF of ELT/HARMONI, which has $\sigma_{PSF} \approx 5$ mas (or FWHM$_{PSF} \approx 12$ mas).

In JAM$_{sph}$ modeling, we created a Markov Chain Monte Carlo (MCMC) simulation to fully sample parameter space of $i$, $M/L_i$, $M_{BH}$, and $\beta_r$. The model is used to fit the simulated kinematic data to find their best-fit values and statistics uncertainties using the adaptive Metropolis algorithm (Haario et al. 2001) in the Bayesian framework (adamet[14] package; Cappellari et al. 2013a). We ran our MCMC chains for the JAM$_{sph}$ models with a total of $3 \times 10^4$ calculations. We excluded the first 20% of calculations as the burn-in phase to produce the full probability distribution function (PDF)

---

[14] v2.0.9 available from https://pypi.org/project/adamet/

18  *Dieu D. Nguyen et al.*

**Table 6.** Best-fitting JAM$_{sph}$ parameters and their statistical uncertainties for the $I_z + J$ and $H + K$ band simulated kinematics of the 2MASXJ11480221+0237582 galaxies, the farthest one in our nine simulated targets.

| Parameter name (1) | Search range of parameters (2) | Input value for HSIM (3) | Best-fit value (4) $I_z + J$ | $1\sigma$ error (16–84%) (5) $I_z + J$ | $3\sigma$ error (0.14–99.86%) (6) $I_z + J$ | Best-fit value (7) $H + K$ | $1\sigma$ error (16–84%) (8) $H + K$ | $3\sigma$ error (0.14–99.86%) (9) $H + K$ |
|---|---|---|---|---|---|---|---|---|
| **Assuming no central SMBH ($M_{BH} = 0$ M$_\odot$)** | | | | | | | | |
| $M_{BH}/M_\odot$ | $(0 \longrightarrow 10^{12})$ | 0 | $1.2 \times 10^5$ | $\pm 3.5 \times 10^7$ | $\pm 9 \times 10^7$ | $1.1 \times 10^5$ | $< 3.8 \times 10^7$ | $< 1.1 \times 10^8$ |
| $M/L_i$ (M$_\odot$/L$_\odot$) | $(0 \longrightarrow 10)$ | 2.5 | 2.52 | $\pm 0.11$ | $\pm 0.43$ | 2.50 | $\pm 0.05$ | $\pm 0.26$ |
| $i$ (°) | $(45 \longrightarrow 90)$ | 60.0 | 88.4 | $\pm 15.6$ | $\pm 21.9$ | 65.0 | $\pm 15.8$ | $\pm 21.9$ |
| $\beta_r$ | $(-15 \longrightarrow 1)$ | $\pm 0.2$ | 0.47 | $\pm 0.02$ | $\pm 0.07$ | 0.46 | $\pm 0.01$ | $\pm 0.06$ |
| **Assuming a central SMBH with mass $M_{BH} = 3 \times 10^9$ M$_\odot$, derived from the $M_{BH}$–$\sigma_\star$ relation (equation (2) from K18)** | | | | | | | | |
| $M_{BH}/M_\odot$ | $(0 \longrightarrow 10^{12})$ | $3.0 \times 10^9$ | $3.1 \times 10^9$ | $\pm 3.0 \times 10^8$ | $\pm 7.7 \times 10^8$ | $3.0 \times 10^9$ | $\pm 2.5 \times 10^8$ | $\pm 5.5 \times 10^8$ |
| $M/L_i$ (M$_\odot$/L$_\odot$) | $(0 \longrightarrow 10)$ | 2.5 | 2.58 | $\pm 0.08$ | $\pm 0.24$ | 2.50 | $\pm 0.05$ | $\pm 0.15$ |
| $i$ (°) | $(45 \longrightarrow 90)$ | 60.0 | 54.3 | $\pm 15.7$ | $\pm 21.9$ | 71.4 | $\pm 16.1$ | $\pm 21.9$ |
| $\beta_r$ | $(-15 \longrightarrow 1)$ | $\pm 0.2$ | 0.44 | $\pm 0.02$ | $\pm 0.04$ | 0.45 | $\pm 0.02$ | $\pm 0.04$ |
| **Assuming a central SMBH with mass $M_{BH} = 4.4 \times 10^{10}$ M$_\odot$, derived from the $M_{BH}$–$M_\star$ relation (equation (3) from K18)** | | | | | | | | |
| $M_{BH}/M_\odot$ | $(0 \longrightarrow 10^{12})$ | $4.4 \times 10^{10}$ | $4.4 \times 10^{10}$ | $\pm 1.0 \times 10^9$ | $\pm 2.3 \times 10^9$ | $4.4 \times 10^{10}$ | $\pm 6.3 \times 10^8$ | $\pm 1.4 \times 10^9$ |
| $M/L_i$ (M$_\odot$/L$_\odot$) | $(0 \longrightarrow 10)$ | 2.5 | 2.52 | $\pm 0.13$ | $\pm 0.33$ | 2.52 | $\pm 0.11$ | $\pm 0.37$ |
| $i$ (°) | $(45 \longrightarrow 90)$ | 60.0 | 59.9 | $\pm 15.9$ | $\pm 21.9$ | 85.8 | $\pm 16.0$ | $\pm 21.9$ |
| $\beta_r$ | $(-15 \longrightarrow 1)$ | $\pm 0.2$ | 0.43 | $\pm 0.09$ | $\pm 0.21$ | 0.47 | $\pm 0.03$ | $\pm 0.07$ |

*Notes:* The table columns list each parameter name (column 1), search range (column 2), input value (column 3; input value of $\beta_r$ was discussed in Section 5.2), best-fit (or upper limit), and uncertainty at the a$1\sigma$ (16–84% of the PDF) and $3\sigma$ (0.14−99.86% of the PDF) confidence levels (columns 4, 5, 6 are the results obtained from the $I_z + J$ simulated kinematics; and columns 7, 8, 9 are the results obtained from the $H + K$ simulated kinematics. Also, see Fig. 16 for a graphically short summary of this Table. The number 9 is following the target order in Tables 4 and 3.

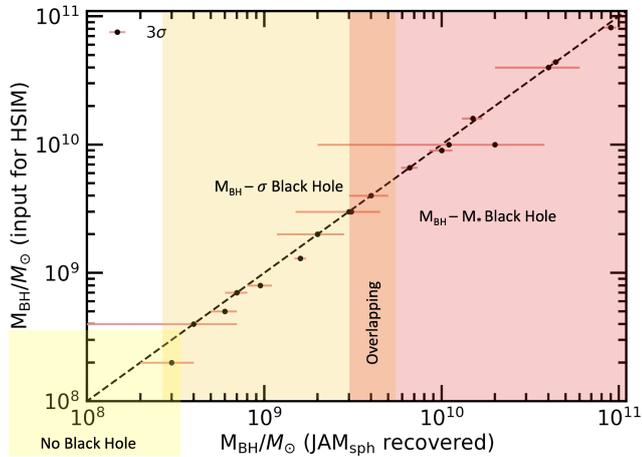

**Figure 16.** A summarised comparison of input black holes for HSIM and our recovered black holes (including statistic error bars at the $3\sigma$ confidential levels) using JAM$_{sph}$ and IFS mock data cubes ($I_z/I_z + J$ and $H + K$) for all nine simulated galaxies (Table 4), which are listed in Tables 6, A1, A2, A3, A4, A5, A6, A7, and A8. The black dashed line is the equal mass line between the input black holes masses for HSIM and our recovered black hole masses. The error bars of the recovered black hole masses are $3\sigma$ uncertainties (at 0.14–99.86 percentiles) found from the MCMC fits of the JAM$_{sph}$ modellings to the correspondingly simulated HSIM kinematic measurements (also listed in mentioned above Tables).

from the 80% remaining measures. The best-fit parameters are the highest likelihood of the PDF.

As examples, we show in Figs 12–14 our best-fit JAM$_{sph}$ (Fig. 12 for input $M_{BH} = 4.4 \times 10^{10}$ M$_\odot$ predicted from equation (3) of K18, Fig. 13 for input $M_{BH} = 3 \times 10^9$ M$_\odot$ predicted from equation (2) of K18, and Fig. 14 for input $M_{BH} = 0$ M$_\odot$) parameters and their associated statistic uncertainties, respectively, that describe the corresponding simulated HARMONI IFS and the derived stellar kinematics of three different $M_{BH}$ for the galaxy 2MASXJ11480221+0237582 in two bands, $I_z + J$ and $H + K$, accordingly. Here, we adopt the 2D distributions scatter plots for each parameter, with colored points indicating their likelihood (white color corresponds to the maximum likelihood and black to a confidence level smaller than $3\sigma$). The histograms show the 1D distributions for each parameter. We used the 1D distributions to calculate the best-fit values and their corresponding uncertainties listed in Tables 6.

To highlight the differences between the model and the data, we demonstrate in Fig. 15 the $V_{rms}$ residual maps ((data-model)/data) produced from the inserted top and bottom $V_{rms}$ shown in Figs 12, 13, and 14) of the galaxy 2MASXJ11480221+0237582 (i.e. as an example represents for all nine simulate), which show the relative agreements/disagreements from pixel to pixel of the $V_{rms}$ maps. The considerable disagreements are all mostly minor within 10% clearly seen in two regions with clear causes. The very central region ($r \lesssim 0.''06$) where we assumed an input anisotropy $\beta_r = -0.2$ (tangential) in the simulated IFS that is different from the main body's anisotropy ($\beta_r = +0.2$, radial), while in the recovering JAM$_{sph}$ models we performed with a fixed anisotropy ($\beta_r$ = constant). The purpose of using a common anisotropy among the population of stellar orbits is for simplicity but can obtain as best as the fit and speed up the calculations, although it is not the thing we will do with the actual data. At larger radii ($r > 0.''06$), even though we approximated the dynamics of core-Sérsic galaxies to be slow rotators and simulated the IFS with the JAM$_{sph}$ model, some amount of rotation seems to have a significant contribution. The existence of a large fraction rotation in the kinematics (i.e. $V/\sigma \gtrsim 0.3$) can be seen with other galaxies in Appendix (available as supplementary material) figures: 2MASXJ22354078+0129053 (Fig. A1), 2MASXJ12052321+1022461 (Fig. A9), and 2MASXJ00034964+0203594 (Fig. A13). Although the differences are not sufficiently large to rule out the JAM$_{sph}$ models, a large fraction of $V/\sigma$ in some galaxies suggests that the using of JAM$_{sph}$ instead of the Jeans equations, which assumes axisymmetry with a cylindrical aligned orientation of the velocity ellipsoid (JAM$_{cyl}$; Cappellari 2008) is a relatively poor assumption.

Our models recovered the $M_{BH}$ and $M/L_i$ values very well and are close to the input values used when creating the input-noiseless cubes, which were supplied to the HSIM simulations for the HARMONI IFS. Specifically, these differences are $\lesssim 5\%$ for both $M_{BH}$ and $M/L_i$. For the uncertainties, we caution that these statistic errors found from the MCMC routines are formal and small (e.g. $3\sigma \approx 3\%$) obtained because (1) our simulated kinematics are high quality and (2) the proposed black hole's SOI (i.e. $r_{SOI} \approx 20$ mas; Section 2.1) is totally resolvable with our HARMONI $10 \times 10$ mas$^2$ simulated angular scale. In fact, $r_{SOI}$ depends on both $M_{BH}$ and $\sigma_\star$, and therefore is different from galaxy to galaxy, as listed in Table 4 and indicated by the red circles in Figs 13 and 14 in the main text and Figs A1, A5, A9, A13, A17, A21, A25, A29 in Appendix A (available as supplementary material), or by the text on the Figs' legends if the $r_{SOI}$ are larger than the simulated FOV. Graphically, a short summary of these black hole masses comparisons between the input values for HSIM and their corresponding recovered values using the adamet MCMC algorithm and JAM$_{sph}$ modelings also demonstrate in Fig. 16. We also included in this Fig the recovered black hole masses for the other eight galaxies listed in Table 4 for sample completeness because we considered these nine galaxies as representative of our MMBH survey sample.

In cases of inputs $M_{BH} = 3 \times 10^9$ M$_\odot$ and $M_{BH} = 4.4 \times 10^{10}$ M$_\odot$, it seems to have a "covariance" between $M_{BH}$ and $M/L_i$ due to the degeneracy between the potentials of the central black holes and the galaxy itself, resulting in a "banana shape" of the $3\sigma$ confidence levels in the 2D PDF found between these two parameters. However, this should not be the case because our simulated observational scale of $10 \times 10$ mas$^2$ is high enough to resolve within the central black hole's SOI even though this galaxy is at the upper limit of our MMBH survey sample's redshift range ($z \approx 0.3$; and remind that our proposed survey is at $r_{SOI} \approx 20$ mas). In the meantime, we observe the "banana shapes" in these 2D PDFs of $M_{BH}$ and $M/L_i$ correlations (Figs 12 and 13), which follow a purely positive trend. This purely positive banana shape does also appear for the case of input $M_{BH} = 0$ M$_\odot$ (Fig. 14) resulting in an upper limit for $M_{BH}$. These $M_{BH}$ are smaller than the statistic errors of the two former cases due to the high-angular and spectral resolutions of the simulated kinematic data.

In addition, we observed a variety distribution trends of this "banana shapes" in Appendix A (available as supplementary material), including purely positive (e.g. 2MASXJ13080241+0900044 in Figs A6, A7, and A8), purely close to zero (e.g. galaxy 2MASXJ09322275+0811508 in Figs A22, A23, and A24 and galaxy 2MASXJ10221610+0522524 in Figs A26, A27, and A28), or a mixture between positive, negative (or anti-correlation, which is usually expected in the covariance between $M_{BH}$ and $M/L_i$ in dynamical modellings), and close to zero (e.g. 2MASXJ22354078+0129053 in Figs A2, A3, and A4, 2MASXJ12052321+1022461 in Figs A10, A11, and A12, 2MASXJ00034964+0203594 in Figs A14, A15, and A16, 2MASXJ16171650+0638149 in Figs A18, A19, and A20,

2MASXJ14155764+0318216 in Figs A30, A31, and A32) for kinematics extracted from the $I_z + J$ and $H + K$ IFS simulated cubes. A possible explanation for the lack of consistent anti-correlation between $M_{BH}$ and $M/L_i$ – which is the most expectation – is that we used a spatially varying anisotropy, with tangential anisotropy near the central black holes, to generate the mock data with JAM$_{sph}$, but then we applied a constant anisotropy to fit the data. This resulted in significant residuals and some unexpected positive correlations. This approach thus may not have been the best choice in retrospect. Another alternative reason for the existence of plenty of 2D PDF "banana shape" perhaps because the $M/L_i$ parameter is not well constrained due to the small FOV or perhaps due to the systemic difference between models based on $I_z + J$ and $H + K$ kinematic extractions.

That recoveries of $\beta_r$ are also tightly constrained by the models, resulting in tiny uncertainties (see Figs 12, 13, 14 and Tables 6 for the galaxy 2MASXJ11480221+0237582). The preferred values of $\beta_r \approx (0.1 - 0.5)$ for $I_z + J$ and $\beta_r \approx (0.3 - 0.5)$ for $H + K$ (i.e. recovered $\beta_r$ are dominated with $\beta_r > 0$ comparing with their input $\beta_r = \pm 0.2$, see Section 5.2), respectively, suggesting that the radial stellar orbits ($\beta_r > 0$) dominate. Similarity happens for the other eight galaxies with the constrained range of $\beta_r$ can be seen in the aforementioned Figs in Appendix A (available as supplementary material).

As expected, we find that the inclination ($i$) is nearly unconstrained by the data. This is because, for all assumed inclinations, the modeled galaxies are by construction quite close to spherical, and in the spherical limit, a galaxy looks the same from any inclination.

In Appendix A (available as supplementary material), we will mention the other eight simulated targets with kinematic results and their black hole masses recovering and associated statistical uncertainties shortly.

## 6.2 Sensitivity limits

Since we considered the nine most massive galaxies as the representative of our MMBH survey sample (Table 3 and 4), their HARMONI IFS simulations at the pixel-sampling scale of $10 \times 10$ mas$^2$ in accurately determining stellar kinematics and dynamical-$M_{BH}$ measurements have demonstrated that the angular-size distances to the targets can be extended further than previous measurements ($\approx 100$ Mpc; van den Bosch 2016) up to a factor of $\approx 9$ and $\approx 90$ for $M_{BH}$ predicted from equation (2) and (3) of K18, respectively. Thus, our proposed survey could push the current spatial-resolution limit of dynamical $M_{BH}$-measurements and $M_{BH}$-scaling relation evolution probe to redshift $z \leq 0.3$. In principle, this angular-size distance can be extended further up to a factor of 5 (i.e. accounting for the fact that we propose a survey at $r_{SOI} = 20$ mas vs. the highest spatial resolution of ELT is 4 mas). However, since the cosmological dimming effect limits the dynamical detections and measurements of SMBHs at further distances, these tasks must be ceased at a specific redshift. We will explore this limit in future work.

It is also worth testing the ultra sensitivity in terms of exposure time of the instruments for the same purposes. We repeated the same simulations for nine representative galaxies, decreasing the exposure times (and thus decreasing the spectra S/N) until the simulated HARMONI IFS cubes marginally provide meaningful kinematic maps after we bin the spaxels together via Vorbin with a specific bin-S/N of $\approx 25$. This handful test gives us the required exposure time for each galaxy in Column 4 of Table 5. Note that the galaxy's surface brightness was measured from the Pan-STARRS image and interpolated towards the regime of 10 mas using the

20     *Dieu D. Nguyen et al.*

core-Sérsic function in Section 4.3. And we also considered the spread of flux along the grating's wavelengths (Section 5.2). These required exposure times are relatively short and smaller than 45 minutes; they are thus suitable for ELT. Here, because of lacking high spatial-resolution imaging for precise surface brightness measurements, we should caution that our sensitivity estimates in terms of exposure time only provide insight values for the MMBH survey. The real measurements of sensitivities are probably a bit higher (or longer exposure time) because higher spatial-resolution imaging (e.g. $\lesssim 0''.05$ from *HST*/ACS and *JWST*/NIRCam or a few mas from ELT/MICADO) will resolve out some flux. We are thus demonstrating the possibility of our proposed science with ELT/HARMONI.

## 7 CONCLUSIONS

Given the purposes of exploring the stellar dynamics deep inside galaxy nuclei and weighing the central SMBHs, we investigated the potential applications of the unprecedented high-spatial-resolution and ultra-sensitivity observations offered by the ELT/HARMONI instrument. These are best for hunting for the most fundamental $M_{\rm BH}$–galaxy scaling relation (whether $M_{\rm BH}$–$\sigma_\star$ or $M_{\rm BH}$–$M_\star$) and shed light on physical processes that derive the evolution picture between SMBHs and galaxies in a large sample of the most massive galaxies.

We defined such a complete sample ($z \leq 0.3$ and $M_K \leq -27.0$ mag) that can be accessible at the location of ELT ($|\delta + 24°| < 45°$, $|b| > 8°$). Our selection criteria base on the mass-selection ($K$-band) of 2MRS assisted by NED-D, resulting in a sample of 101 highest-mass galaxies ($2 \times 10^{12} < M_\star \lesssim 5 \times 10^{12}$ M$_\odot$, statistically with 77% ellipticals, 17% lenticulars, and 7% spirals). This sample extends to the locally largest-mass galaxies ($D_A \leq 950$ Mpc) beyonds the currently well-known and large surveys of galaxies like ATLAS$^{\rm 3D}$ (Cappellari et al. 2011), MASSIVE (Ma et al. 2014), and MaNGA (Graham et al. 2018), and similar to the M3G (Krajnović et al. 2018b) sample but includes a wide range of environments from isolations to dense galaxy clusters. Our extensive survey of MMBHs is crucial for gaining insights into the mass buildup of the most massive galaxies. We achieve this through a comprehensive analysis of stellar and, if detectable, gas kinematics, photometric profiles, and dynamical masses, all within the context of their respective environments. The HARMONI IFS observations of this MMBH sample will be compared against the modeling predictions to test formation scenarios and to develop the models at the top end of the galaxy mass function. Thus, our limited redshift range ($z \approx 0.02 - 0.3$) survey is essential to investigate the evolution of galaxy-global parameters with redshift and trace galaxy evolution back in time in combination and comparison with the availability of lower redshift sample, i.e. ATLAS$^{\rm 3D}$, MaNGA, MASSIVE.

We tested the capacity of HARMONI IFS observations in measuring $M_{\rm BH}$ by doing the HSIM simulation for the $I_z$, $I_z + J$, and $H + K$ band IFU. For the $I_z$ and $I_z + J$ gratings, we made use of the stellar absorption features of CaT (0.86–0.88 $\mu$m) to extract the simulated stellar kinematics. We also provided a guideline for using the $H + K$ IFS to obtain the stellar kinematic measurements in the future. There are many strong stellar-atomic absorptions (Mg I $\lambda 1.487$ $\mu$m and S I $\lambda 1.589$ $\mu$m) and CO absorptions (CO(3–0) $\lambda 1.540$ $\mu$m, CO(4–1) $\lambda 1.561$ $\mu$m, CO(5–2) $\lambda 1.577$ $\mu$m, CO(6–3) $\lambda 1.602$ $\mu$m, CO(7–4) $\lambda 1.622$ $\mu$m, and CO(8–5) $\lambda 1.641$ $\mu$m). We found consistent kinematic maps extracted from the listed absorption features above within the instrument resolution ($\Delta V \lesssim 40$ km s$^{-1}$). We then use these data to estimate SMBHs masses in combination with the interpolated stellar-mass model from the Pan-STARRS image and JAM$_{\rm sph}$ (C20) modelings. In this paper, we only tested the capacity of using the JAM$_{\rm sph}$ model to produce simulated HARMONI IFS cubes and recover the $M_{\rm BH}$ correspondingly from the simulated kinematic maps.

We found that the recovering $M_{\rm BH}$ and $M/L_i$ from the simulated data are totally consistent with our input values (uncertainties $\lesssim 5\%$) during the simulated process, although we made some different assumptions on the input and output anisotropy (input varying $\beta_r$ vs. output constant $\beta_r$). However, we should note that we did not compare the different JAM versions of the coordinate-aligned orientation of the velocity ellipsoid (spherical vs. cylindrical, JAM$_{\rm cyl}$) in estimating the $M_{\rm BH}$ directly. Our simulations thus demonstrated that ELT/HARMONI will be able a unique facility for measuring SMBH mass and exploring the black hole mass–galaxy scaling relations evolution.

Our canonically proposed angular-resolution survey of $r_{\rm SOI} = 20 \times 20$ mas$^2$ with the simulated pixel scales of $10 \times 10$ mas$^2$ is high enough to resolve the stellar kinematics within the central black hole's SOI even though the galaxy is at the upper limit of our MMBH sample's redshift range ($z \approx 0.3$). The covariance between $M_{\rm BH}$ and $M/L_i$ due to the degeneracy between the potentials of the central black holes and the galaxy itself should not be seen in the $3\sigma$ confidence levels in the 2D PDF. However, we observed the "banana shape" in the posterior PDF of these two parameters with a variety of shapes, including the purely positive, negative (anti-correlation), or the mixture between positive, negative, and close to zero for kinematics extracted from the $I_z + J$ and $H + K$ IFS simulated cubes. The reasons for the existence of plenty of 2D PDF "banana shapes" and lack of anti-correlation, perhaps we used a spatially varying anisotropy (with tangential anisotropy near the SMBH) to generate the mock data with JAM$_{\rm sph}$, but then we applied a constant anisotropy to fit the data, which may be not have been the best choice in retrospect. Or perhaps the $M/L_i$ parameter is not well constrained due to the small FOV or the systemic difference between models based on $I_z + J$ and $H + K$ kinematic extractions.

Our simulations predict that within a relatively short observing time with ELT/HARMONI (i.e. less than one hour) one can obtain high-quality IFS and stellar kinematics data, demonstrating that HARMONI will be a cutting-edge instrument for investigating the above science goals.

## ACKNOWLEDGEMENTS

The authors would like to thank the anonymous referee for his/her careful reading and useful comments, which helped to improve the paper greatly. DDN is grateful to the LABEX Lyon Institute of Origins (ANR-10-LABX-0066) Lyon for its financial support within the program "Investissements d'Avenir" of the French government operated by the National Research Agency (ANR). MPS acknowledges funding support from the Ramón y Cajal program of the Spanish Ministerio de Ciencia e Innovación (RYC2021-033094-I).

The authors would like to thank Professor Joseph Jensen of the Department of Physics, Utah Valley University for enlightening discussions on Pan-STARRS photometric calibration. We also thank some students of Vietnam National University in Ho Chi Minh City, Vietnam: The University of Natural Science (Ngo Ngoc Hai and Tong Gia Huy), The University of Technology (Le Nguyen Tuan), and The International University (Le Thong Quoc Tinh and On Tuan Phong) for their partially helps on computations.

This research has made use of the NASA/IPAC Extragalactic

Database (NED) which is operated by the Jet Propulsion Laboratory, California Institute of Technology, under contract with the National Aeronautics and Space Administration.

The Pan-STARRS1 Surveys (PS1) and the PS1 public science archive have been made possible through contributions by the Institute for Astronomy, the University of Hawaii, the Pan-STARRS Project Office, the Max-Planck Society and its participating institutes, the Max Planck Institute for Astronomy, Heidelberg and the Max Planck Institute for Extraterrestrial Physics, Garching, The Johns Hopkins University, Durham University, the University of Edinburgh, the Queen's University Belfast, the Harvard-Smithsonian Center for Astrophysics, the Las Cumbres Observatory Global Telescope Network Incorporated, the National Central University of Taiwan, the Space Telescope Science Institute, the National Aeronautics and Space Administration under Grant No. NNX08AR22G issued through the Planetary Science Division of the NASA Science Mission Directorate, the National Science Foundation Grant No. AST-1238877, the University of Maryland, Eotvos Lorand University (ELTE), the Los Alamos National Laboratory, and the Gordon and Betty Moore Foundation.

Funding for SDSS-III has been provided by the Alfred P. Sloan Foundation, the Participating Institutions, the National Science Foundation, and the U.S. Department of Energy Office of Science. The SDSS-III website is www.sdss3.org/. SDSS-III is managed by the Astrophysical Research Consortium for the Participating Institutions of the SDSS-III Collaboration including the University of Arizona, the Brazilian Participation Group, Brookhaven National Laboratory, Carnegie Mellon University, University of Florida, the French Participation Group, the German Participation Group, Harvard University, the Instituto de Astrofisica de Canarias, the Michigan State/Notre Dame/JINA Participation Group, Johns Hopkins University, Lawrence Berkeley National Laboratory, Max Planck Institute for Astrophysics, Max Planck Institute for Extraterrestrial Physics, New Mexico State University, New York University, Ohio State University, Pennsylvania State University, University of Portsmouth, Princeton University, the Spanish Participation Group, University of Tokyo, University of Utah, Vanderbilt University, University of Virginia, University of Washington, and Yale University. This publication makes use of data products from the Two Micron All Sky Survey (https://old.ipac.caltech.edu/2mass/), which is a joint project of the University of Massachusetts and the Infrared Processing and Analysis Center/California Institute of Technology, funded by the National Aeronautics and Space Administration and the National Science Foundation.

*Facilities:* Pan-STARRS DR1 and DR2, SDSS DR12, 2MASS, and *HST*

*Software:* python 3.10 (https://www.python.org/), Matplotlib 3.6.0 (https://matplotlib.org/), numpy 1.22 (https://www.scipy.org/install.html), scipy 1.3.1 (https://www.scipy.org/install.html), photutils 0.7 (https://photutils.readthedocs.io/en/stable//), MPFIT (http://purl.com/net/mpfit), plotbin 3.1.3 (https://pypi.org/project/plotbin/), astropy 5.1 (Astropy Collaboration et al. 2022), adamet 2.0.9 (Cappellari et al. 2013a), jampy 6.4.0 (Cappellari 2008, 2020), pPXF 8.2.1 (Cappellari 2022), vorbin 3.1.5 (Cappellari & Copin 2003), MgeFit 5.0.14 (Cappellari 2002) and HSIM 3.10 (Zieleniewski et al. 2015).

## DATA AVAIBILITY

All data and software used in this paper are public. We provided their links in the text when discussed. The data produced underlying this article will be shared on reasonable request to the corresponding author.

22    *Dieu D. Nguyen et al.*

## APPENDIX A: PPXF KINEMATICS MEASUREMENTS AND BLACK HOLE MASS RECOVERING FROM THE HARMONI IFS OF EIGHT OTHERS SIMULATED GALAXIES

We extracted stellar kinematic maps of the rest eight galaxies from the CaT part (the first four-panel plots) and the significant stellar features (the second four-panel plots) of its mock $I_z/I_z+J$ and $H+K$ HSIM IFS cubes, respectively, produced from JAM$_{sph}$ (Section 5.2) using pPXF are shown in Figs A1, A5, A9, A13, A17, A21, A25, and A29. These maps are present with three different black hole masses: $M_{BH} = 0$ M$_\odot$ (top-row plots), $M_{BH,\sigma_\star}$ (equation (2) of K18, middle-row plots), and $M_{BH,M_\star}$ (equation (3) of K18, bottom-row plots) in the same fashion with the left parts of Figs 10 and 11 but excluding their right parts of showing the galaxy spectrum and its best-fitting pPXF model. The creation of these maps, their uncertainties ($\Delta V$, $\Delta \sigma_\star$, $\Delta V_{rms}$), and testing them with different wavelength ranges are following the same procedure but accounting for the galaxy redshift described in Section 5.3.

The kinematic signature of the central black holes is visible in these maps. In particular, the velocity dispersion ($\sigma_\star$) and root-mean-squared velocity ($V_{rms}$) of $\approx 12$ central spaxels within 20 mas are increasing substantially and distinguishably for three of the above SMBH masses. These central drops of $\sigma_\star$ and $V_{rms}$ for the case of zero black holes are consistent with our core-Sérsic most massive galaxies, while that seen in the case of existing an SMBH with the mass of $M_{BH,\sigma_\star}$ because the black hole mass is small. Its stellar kinematic effects are smaller than the surrounding kinematics produced by the stellar component. However, models with an SMBH with the mass of $M_{BH,M_\star}$ create centrally raising Keplerian peaks towards the galaxy center in both $\sigma_\star$ and $V_{rms}$ map where the central SMBH's potential dominates.

We used these simulated kinematic maps, which are presented in Figs A1, A5, A9, A13, A17, A21, A25, and A29 and the JAM$_{sph}$ modelling to recover their central SMBH masses, which were incorporated into these HARMONI IFS data cubes during their creations using HSIM (Section 5.2), following the same procedure discussed in Section 6.1. We demonstrate these SMBH recovering results for all six simulated IFS kinematics (e.g. three black holes: $M_{BH} = 0$ M$_\odot$, $M_{BH,\sigma_\star}$, and $M_{BH,M_\star}$ for two HARMONI grating $I_z/I_z + J$ and $H + K$; see Table 4) in Figs A2, A3, A4 and summarize their results in Table A1 for 2MASXJ22354078+0129053, in Figs A6, A7, A8 and Table A2 for 2MASXJ13080241+0900044, in Figs A10, A11, A12 and Table A3 for 2MASXJ12052321+1022461, in Figs A14, A15, A16 and Table A4 for 2MASXJ00034964+0203594, in Figs A18, A19, A20 and Table A5 for 2MASXJ16171650+0638149, in Figs A22, A23, A24 and Table A6 for 2MASXJ09322275+0811508, in Figs A26, A27, A28 and Table A7 for 2MASXJ10221610+0522524, and in Figs A30, A31, A32 and Table A8 for 2MASXJ14155764+0318216.

As demonstrated in these Figs, the inclination ($i$) is unconstrained by the data. However, both $M_{BH}$ and $M/L_i$ are well constrained and are close to the input values used when creating the input-noiseless cubes. Specifically, these differences are $\lesssim 25\%$ for $M_{BH,\sigma_\star}$ and $\lesssim 15\%$ for $M_{BH,M_\star}$, and $\lesssim 25\%$ for $M/L_i$. The statistic errors found from the MCMC routines are formal and reasonable (e.g. $3\sigma \approx 35\%$) obtained because (1) our simulated kinematics are high quality and (2) the black hole's SOI is resolvable with HARMONI $20 \times 20$ mas$^2$ angular scale.

There is a covariance between $M_{BH}$ and $M/L_i$ in all cases of black hole masses. The models also tightly constrain those recoveries of the anisotropy parameter ($\beta_r$). There is also covariance with $M_{BH}$. Nevertheless, anisotropy parameter ($\beta_r$) is mainly negative for 2MASXJ22354078+0129053, 2MASXJ12052321+1022461, 2MASXJ00034964+0203594, 2MASXJ09322275+0811508, and 2MASXJ10221610+0522524, where tangential stellar orbits ($\beta_r < 0$) dominate in the models; and mainly positive for 2MASXJ13080241+0900044, 2MASXJ16171650+0638149, and 2MASXJ14155764+031826 where radial stellar orbits ($\beta_r > 0$) dominate in the models.

This paper has been typeset from a T$_E$X/L$^A$T$_E$X file prepared by the author.

24　*Dieu D. Nguyen et al.*

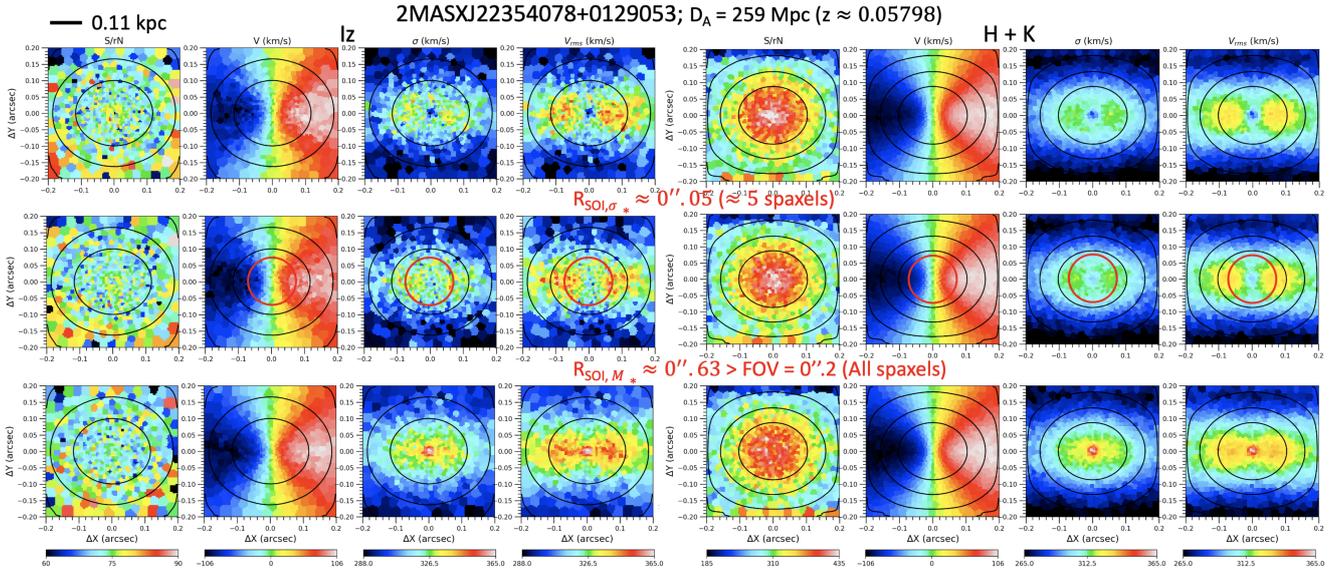

**Figure A1.** Same as Figs 10 and 11, but stellar-kinematic maps of the galaxy 2MASXJ22354078+0129053 on each row extracted from its mock $I_z$ (the left four-panel plots) and $H + K$ (the right four-panel plots) HSIM IFS cube, respectively.

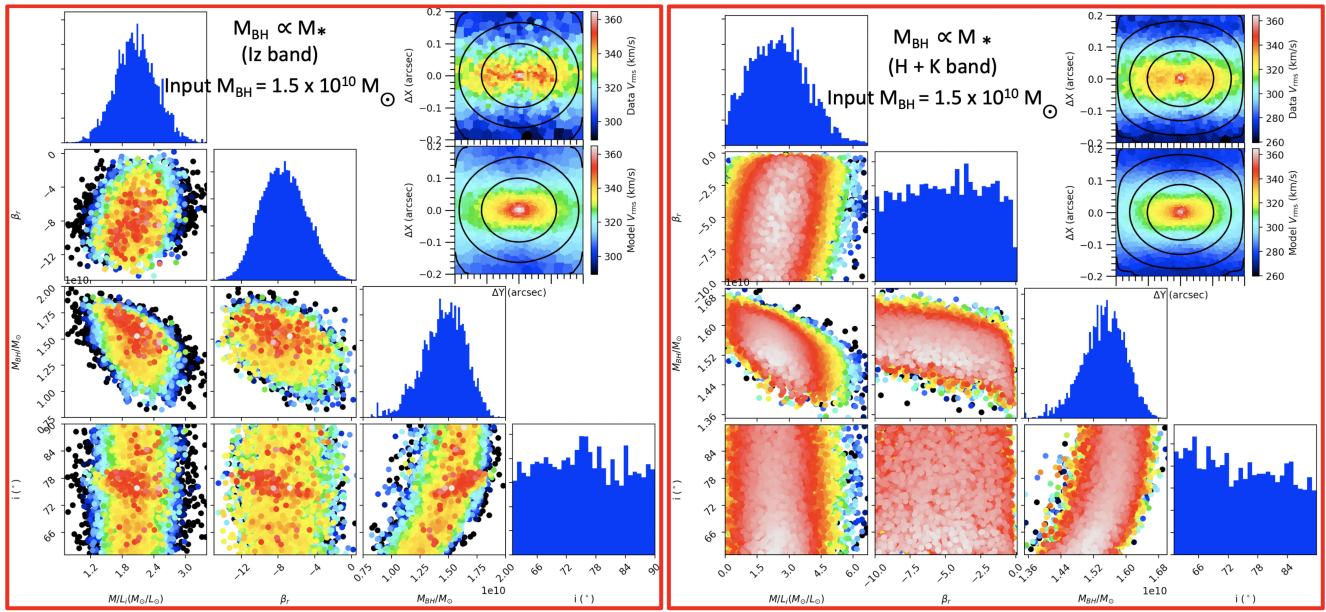

**Figure A2.** Same as Fig. 12 but for the HARMONI simulated kinematics with $M_{BH} = 1.5 \times 10^{10}$ M$_\odot$ that follows the $M_{BH}$–$M_\star$ relation for galaxies with masses above $M_{crit.}$ predicted by equation (3) of K18 for the galaxy 2MASXJ22354078+0129053. The input parameters, their JAM$_{sph}$ best-fit models, and statistics uncertainties are listed in Table A1.

*Simulating SMBH mass measurements with HARMONI IFS* 25

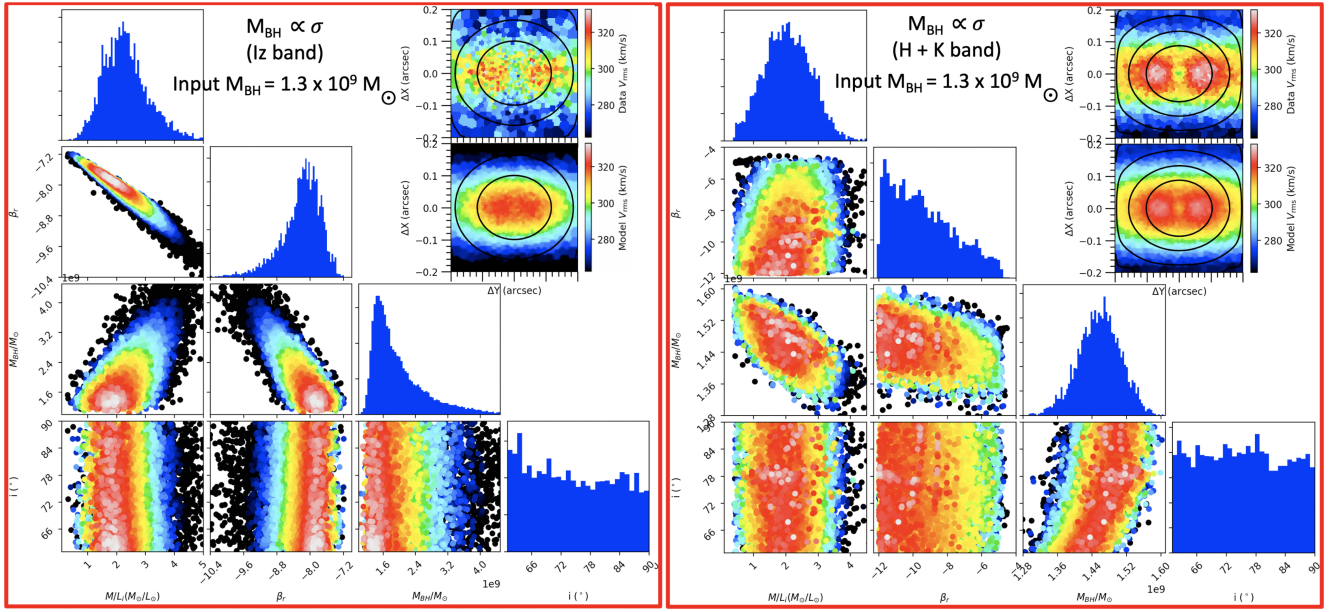

**Figure A3.** Same as Fig. 12 but for the HARMONI simulated kinematics with $M_{BH} = 1.3 \times 10^9$ M$_\odot$ that follows the $M_{BH}-\sigma_\star$ relation for galaxies with masses below $M_{crit.}$ predicted by equation (2) of K18 for the galaxy 2MASXJ22354078+0129053. The input parameters, their JAM$_{sph}$ best-fit models, and statistics uncertainties are listed in Table A1.

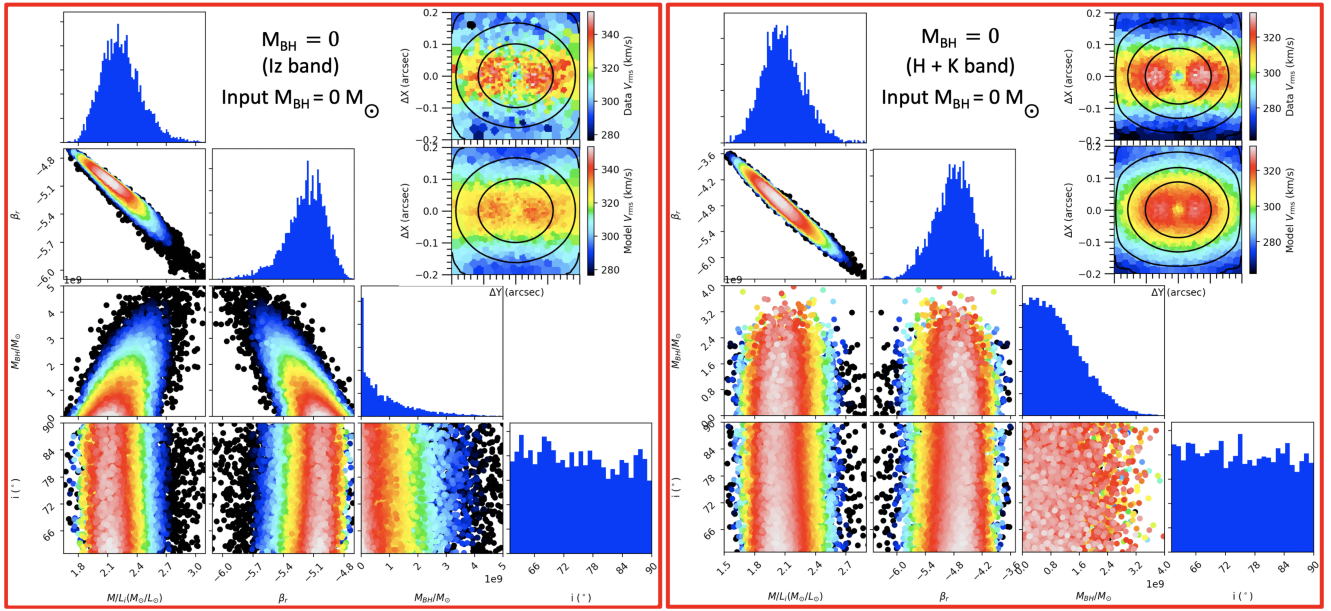

**Figure A4.** Same as Fig. 12 but for the HARMONI simulated kinematics with $M_{BH} = 0$ M$_\odot$ for the galaxy 2MASXJ22354078+0129053. The input parameters, their JAM$_{sph}$ best-fit models and statistics uncertainties are listed in Table A1.

26    *Dieu D. Nguyen et al.*

**Table A1.** Best-fitting JAM$_{sph}$ parameters and their statistical uncertainties for the $I_z$ and $H+K$ band simulated kinematics of 2MASXJ22354078+0129053.

| Parameter name (1) | Search range of parameters (2) | Input value for HSIM (3) | Best-fit value (4) $I_z$ | $1\sigma$ error (16–84%) (5) $I_z$ | $3\sigma$ error (0.14–99.86%) (6) $I_z$ | Best-fit value (7) $H+K$ | $1\sigma$ error (16–84%) (8) $H+K$ | $3\sigma$ error (0.14–99.86%) (9) $H+K$ |
|---|---|---|---|---|---|---|---|---|
| **Assuming no a central SMBH ($M_{BH} = 0\ M_\odot$)** | | | | | | | | |
| $M_{BH}/M_\odot$ | ($0 \longrightarrow 10^{12}$) | 0 | $0.2 \times 10^9$ | $< 2.0 \times 10^9$ | $< 4.0 \times 10^9$ | $0.5 \times 10^9$ | $< 2.0 \times 10^9$ | $< 3.5 \times 10^9$ |
| $M/L_i$ (M$_\odot$/L$_\odot$) | ($0 \longrightarrow 10$) | 2.0 | 2.10 | $\pm 0.03$ | $\pm 0.08$ | 2.10 | $\pm 0.03$ | $\pm 0.09$ |
| $i$ (°) | ($60 \longrightarrow 90$) | 60 | 75 | $\pm 10$ | $\pm 25$ | 75 | $\pm 10$ | $\pm 25$ |
| $\beta_r$ | ($-15 \longrightarrow 1$) | $\pm 0.2$ | $-0.50$ | $\pm 0.10$ | $\pm 0.30$ | $-0.47$ | $\pm 0.04$ | $\pm 0.10$ |
| **Assuming a central SMBH with mass $M_{BH} = 1.3 \times 10^9\ M_\odot$, derived from the $M_{BH}$–$\sigma_\star$ relation (equation (2) from K18)** | | | | | | | | |
| $M_{BH}/M_\odot$ | ($0 \longrightarrow 10^{12}$) | $1.3 \times 10^9$ | $1.5 \times 10^9$ | $\pm 0.7 \times 10^9$ | $< 3.0 \times 10^9$ | $1.4 \times 10^9$ | $\pm 0.3 \times 10^8$ | $\pm 1.2 \times 10^8$ |
| $M/L_i$ (M$_\odot$/L$_\odot$) | ($0 \longrightarrow 10$) | 2.0 | 2.0 | $\pm 1.0$ | $\pm 2.0$ | 2.0 | $\pm 1.0$ | $\pm 2.0$ |
| $i$ (°) | ($60 \longrightarrow 90$) | 60 | 75 | $\pm 10$ | $\pm 25$ | 75 | $\pm 10$ | $\pm 25$ |
| $\beta_r$ | ($-15 \longrightarrow 1$) | $\pm 0.2$ | $-8.0$ | $\pm 0.8$ | $\pm 1.6$ | $-10.0$ | $\pm 1.5$ | $\pm 5.0$ |
| **Assuming a central SMBH with mass $M_{BH} = 1.5 \times 10^{10}\ M_\odot$, derived from the $M_{BH}$–$M_\star$ relation (equation (3) from K18)** | | | | | | | | |
| $M_{BH}/M_\odot$ | ($0 \longrightarrow 10^{12}$) | $1.5 \times 10^{10}$ | $1.5 \times 10^{10}$ | $\pm 0.2 \times 10^{10}$ | $\pm 0.5 \times 10^{10}$ | $1.5 \times 10^{10}$ | $\pm 0.1 \times 10^{10}$ | $\pm 0.2 \times 10^{10}$ |
| $M/L_i$ (M$_\odot$/L$_\odot$) | ($0 \longrightarrow 10$) | 2.0 | 2.0 | $\pm 0.4$ | $\pm 1.0$ | 2.5 | $\pm 1.5$ | $\pm 2.5$ |
| $i$ (°) | ($60 \longrightarrow 90$) | 60 | 75 | $\pm 10$ | $\pm 25$ | 75 | $\pm 10$ | $\pm 25$ |
| $\beta_r$ | ($-15 \longrightarrow 1$) | $\pm 0.2$ | $-8.0$ | $\pm 2.8$ | $\pm 8.0$ | $-5.0$ | $\pm 2.5$ | $\pm 5.0$ |

*Notes:* Same as Table 6 but are the best-fit JAM$_{sph}$ modellings for the galaxy 2MASXJ22354078+0129053 optimized to its HARMONI simulated kinematics with different $M_{BH}$ which are shown in Figs A2, A3, and A4. Also, see Fig. 16 for a graphically short summary of this Table A1.

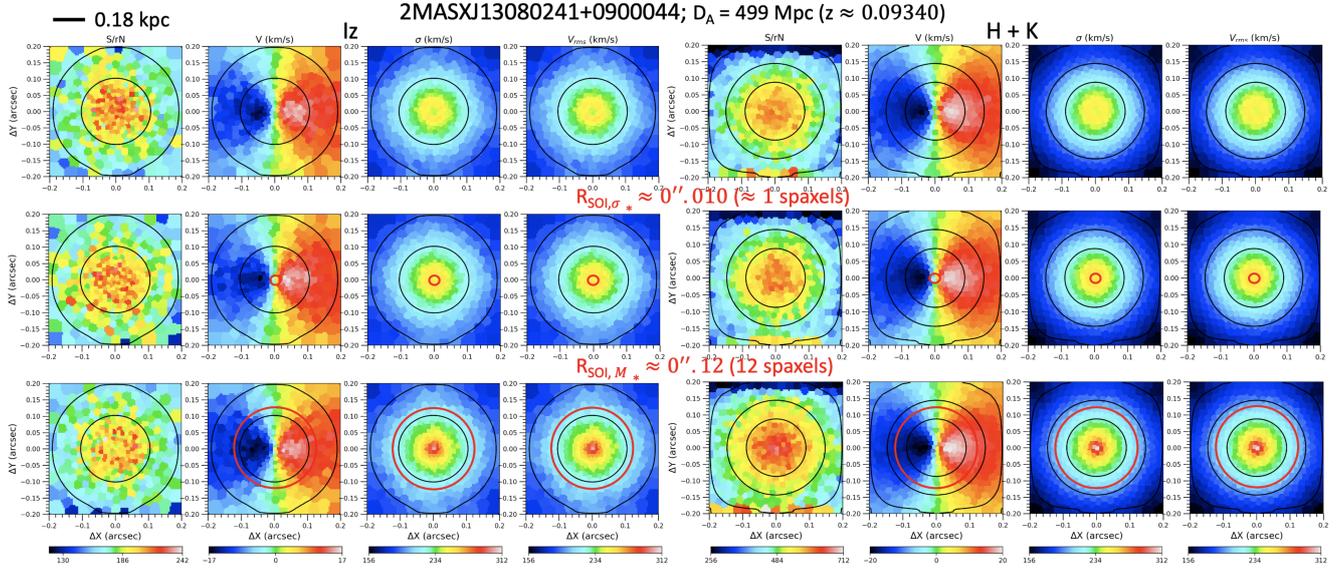

**Figure A5.** Same as Fig. A1, but stellar-kinematic maps of the galaxy 2MASXJ13080241+0900044 on each row extracted from its mock $Iz$ (the left four-panel plots) and $H+K$ (the right four-panel plots) HSIM IFS cube, respectively.

*Simulating SMBH mass measurements with HARMONI IFS* 27

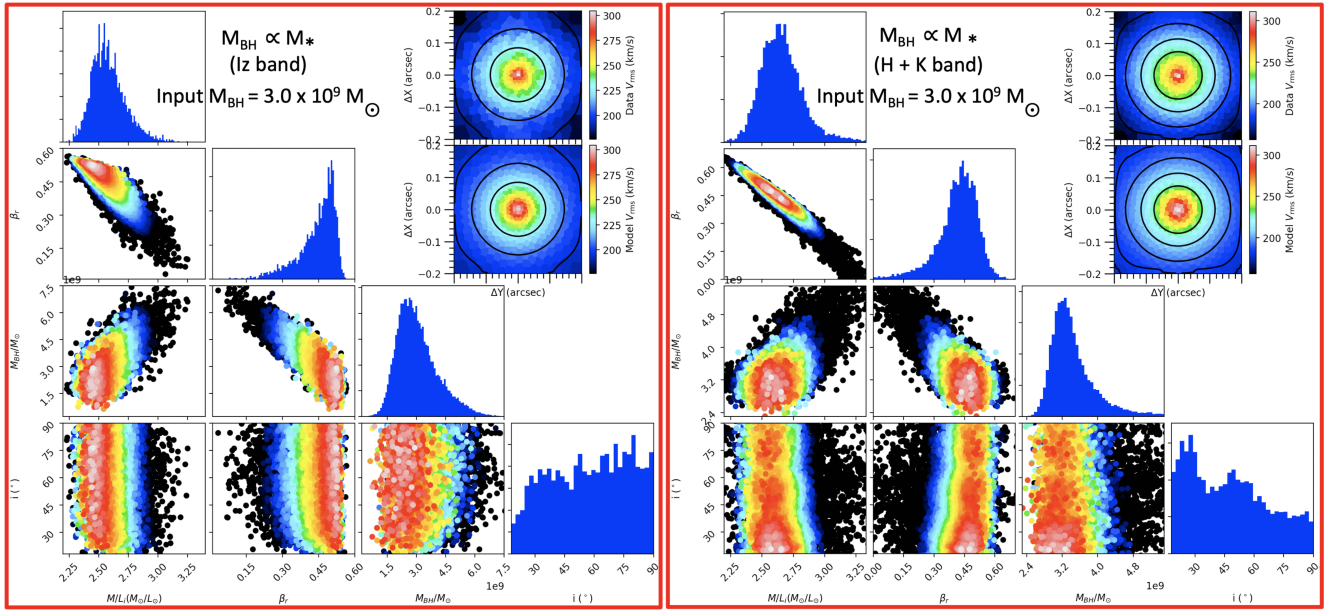

**Figure A6.** Same as Fig. 12 but for the HARMONI simulated kinematics with $M_{BH} = 3.0 \times 10^9$ M$_\odot$ that follows the $M_{BH}-M_\star$ relation for galaxies with masses above $M_{crit.}$ predicted by equation (3) of K18 for the galaxy 2MASXJ13080241+0900044. The input parameters, their JAM$_{sph}$ best-fit models, and statistics uncertainties are listed in Table A2.

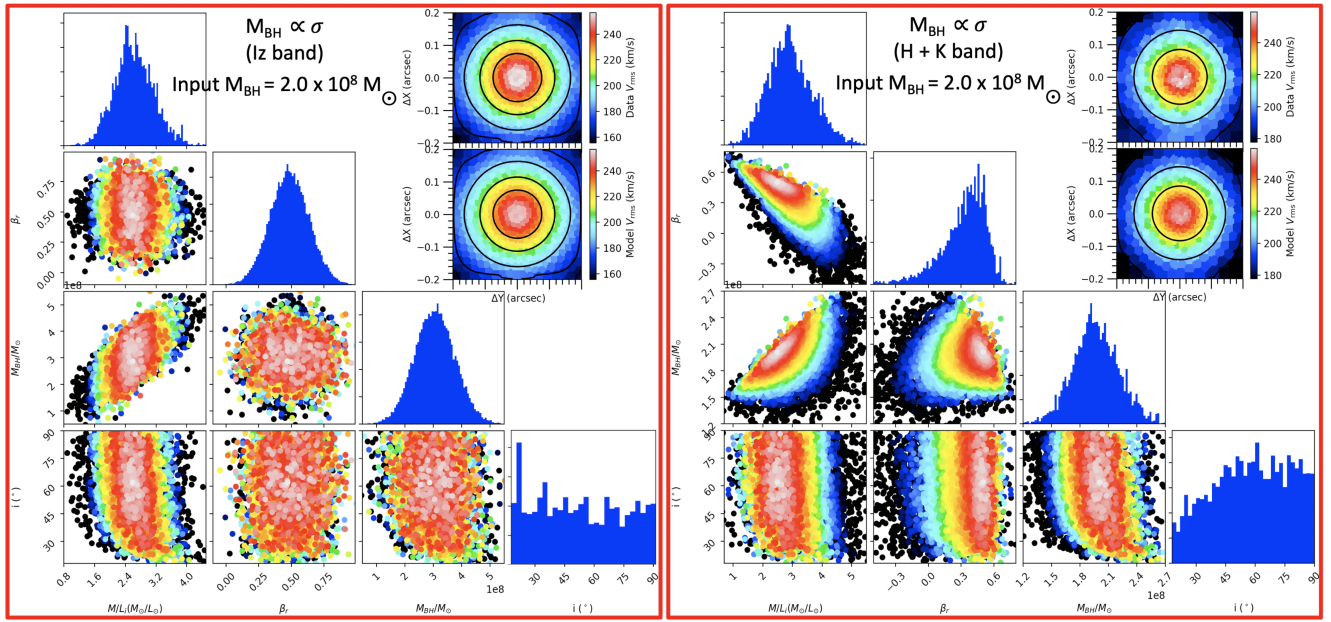

**Figure A7.** Same as Fig. 12 but for the HARMONI simulated kinematics with $M_{BH} = 2.0 \times 10^8$ M$_\odot$ that follows the $M_{BH}-\sigma_\star$ relation for galaxies with masses below $M_{crit.}$ predicted by equation (2) of K18 for the galaxy 2MASXJ13080241+0900044. The input parameters, their JAM$_{sph}$ best-fit models, and statistics uncertainties are listed in Table A2.

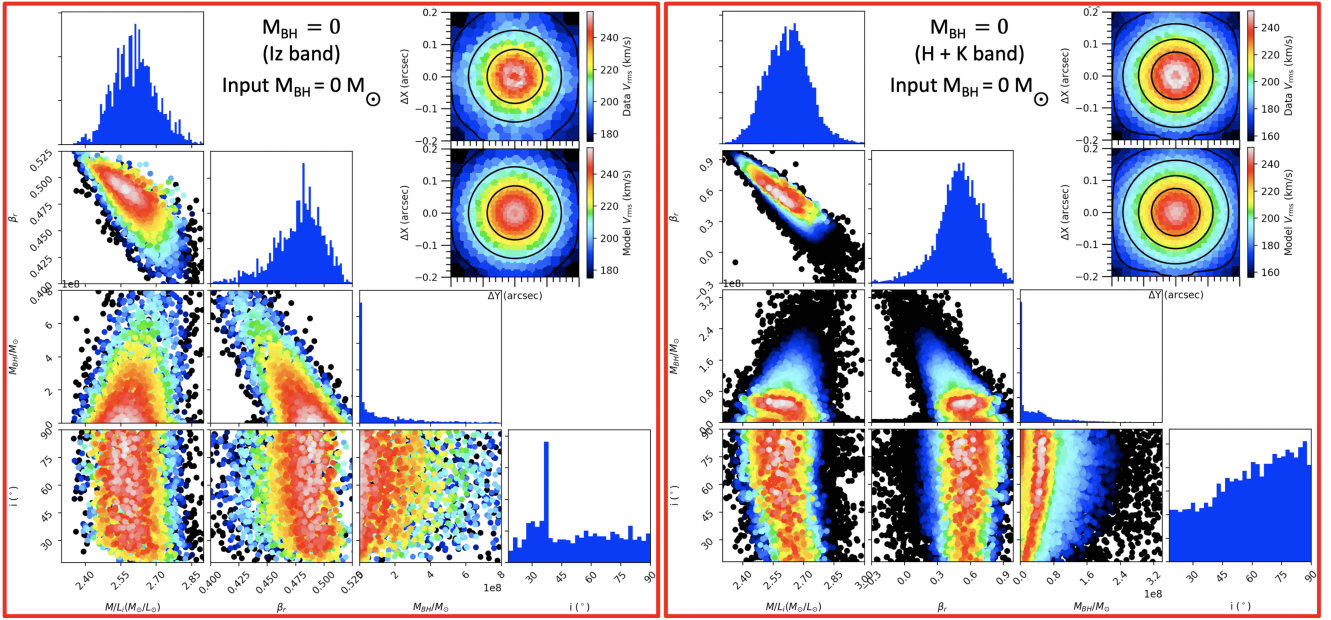

**Figure A8.** Same as Fig. 12 but for the HARMONI simulated kinematics with $M_{BH} = 0$ M$_\odot$ for the galaxy 2MASXJ13080241+0900044. The input parameters, their JAM$_{sph}$ best-fit models, and statistics uncertainties are listed in Table A2.

**Table A2.** Best-fitting JAM$_{sph}$ parameters and their statistical uncertainties for the $I_z$ and $H + K$ band simulated kinematics of 2MASXJ13080241+0900044.

| Parameter name (1) | Search range of parameters (2) | Input value for HSIM (3) | Best-fit value (4) $I_z$ | $1\sigma$ error (16–84%) (5) $I_z$ | $3\sigma$ error (0.14–99.86%) (6) $I_z$ | Best-fit value (7) $H + K$ | $1\sigma$ error (16–84%) (8) $H + K$ | $3\sigma$ error (0.14–99.86%) (9) $H + K$ |
|---|---|---|---|---|---|---|---|---|
| **Assuming no a central SMBH ($M_{BH} = 0$ M$_\odot$)** | | | | | | | | |
| $M_{BH}/M_\odot$ | $(0 \longrightarrow 10^{12})$ | 0 | $2.5 \times 10^5$ | $< 3.5 \times 10^8$ | $< 8.0 \times 10^8$ | $0.5 \times 10^8$ | $\pm 0.3 \times 10^8$ | $< 2.0 \times 10^8$ |
| $M/L_i$ (M$_\odot$/L$_\odot$) | $(0 \longrightarrow 10)$ | 2.5 | 2.5 | $\pm 0.1$ | $\pm 0.4$ | 2.6 | $\pm 0.2$ | $\pm 0.5$ |
| $i$ (°) | $(18 \longrightarrow 90)$ | 60 | 50 | $\pm 17$ | $\pm 38$ | 70 | $\pm 16$ | $\pm 23$ |
| $\beta_r$ | $(-15 \longrightarrow 1)$ | $\pm 0.2$ | 0.50 | $\pm 0.03$ | $\pm 0.10$ | 0.61 | $\pm 0.11$ | $\pm 0.34$ |
| **Assuming a central SMBH with mass $M_{BH} = 2 \times 10^8$ M$_\odot$, derived from the $M_{BH}$–$\sigma_\star$ relation (equation (2) from K18)** | | | | | | | | |
| $M_{BH}/M_\odot$ | $(0 \longrightarrow 10^{12})$ | $2.0 \times 10^8$ | $3.0 \times 10^8$ | $\pm 1.0 \times 10^8$ | $\pm 2.0 \times 10^8$ | $2.1 \times 10^8$ | $\pm 0.2 \times 10^8$ | $\pm 0.6 \times 10^8$ |
| $M/L_i$ (M$_\odot$/L$_\odot$) | $(0 \longrightarrow 10)$ | 2.5 | 2.50 | $\pm 0.05$ | $\pm 0.15$ | 2.50 | $\pm 0.07$ | $\pm 0.20$ |
| $i$ (°) | $(18 \longrightarrow 90)$ | 60 | 55 | $\pm 17$ | $\pm 55$ | 65 | $\pm 16$ | $\pm 25$ |
| $\beta_r$ | $(-15 \longrightarrow 1)$ | $\pm 0.2$ | 0.50 | $\pm 0.10$ | $\pm 0.40$ | 0.55 | $\pm 0.30$ | $\pm 0.55$ |
| **Assuming a central SMBH with mass $M_{BH}$ $3.0 \times 10^9$ M$_\odot$, derived from the $M_{BH}$–$M_\star$ relation (equation (3) from K18)** | | | | | | | | |
| $M_{BH}/M_\odot$ | $(0 \longrightarrow 10^{12})$ | $3.0 \times 10^9$ | $3.0 \times 10^9$ | $\pm 1.5 \times 10^9$ | $< 6.0 \times 10^9$ | $3.1 \times 10^9$ | $\pm 1.6 \times 10^9$ | $< 4.8 \times 10^9$ |
| $M/L_i$ (M$_\odot$/L$_\odot$) | $(0 \longrightarrow 10)$ | 2.5 | 2.50 | $\pm 0.05$ | $\pm 0.15$ | 2.60 | $\pm 0.05$ | $\pm 0.15$ |
| $i$ (°) | $(18 \longrightarrow 90)$ | 60 | 55 | $\pm 16$ | $\pm 35$ | 50 | $\pm 18$ | $\pm 37$ |
| $\beta_r$ | $(-15 \longrightarrow 1)$ | $\pm 0.2$ | 0.45 | $\pm 0.05$ | $\pm 0.15$ | 0.45 | $\pm 0.05$ | $\pm 0.15$ |

*Notes:* Same as Table 6 but are the best-fit JAM$_{sph}$ modellings for the galaxy 2MASXJ13080241+0900044 optimized to its HARMONI simulated kinematics with different $M_{BH}$ which are shown in Figs A6, A7, and A8. Also, see Fig. 16 for a graphically short summary of this Table A2.

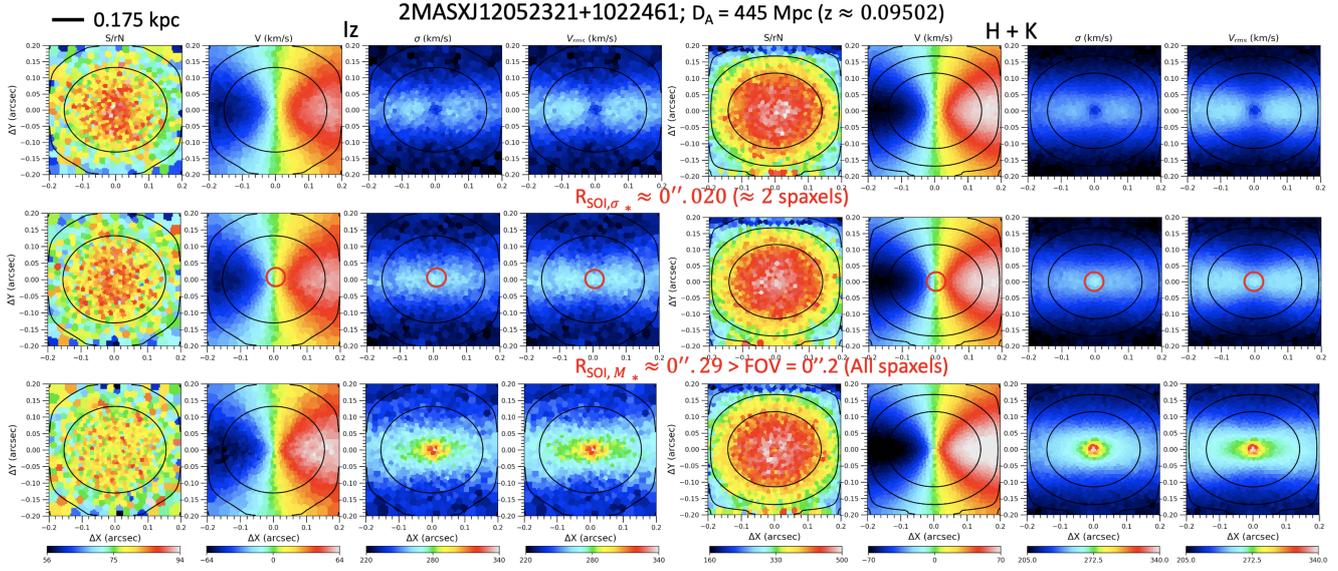

**Figure A9.** Same as Fig. A1, but these stellar-kinematic maps of the galaxy 2MASXJ12052321+1022461 on each row extracted from its mock $I_z$ (the left four-panel plots) and $H + K$ (the right four-panel plots) HSIM IFS cube, respectively.

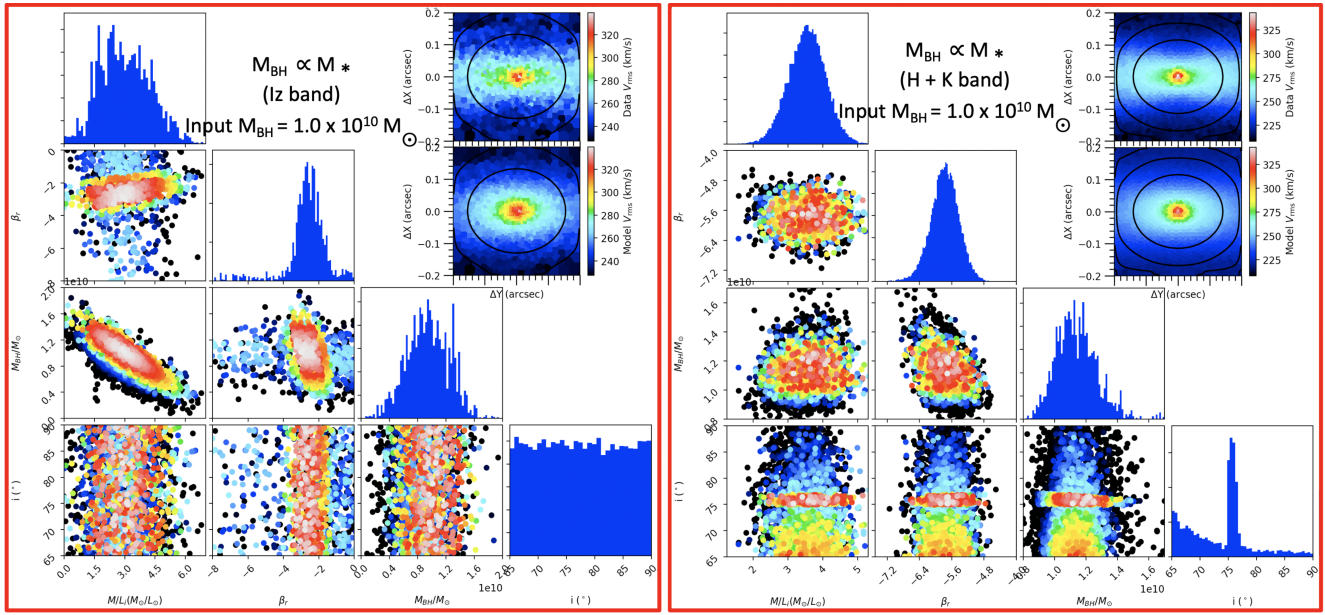

**Figure A10.** Same as Fig. 12 but for the HARMONI simulated kinematics with $M_{BH} = 1.0 \times 10^{10}$ $M_\odot$ that follows the $M_{BH}-M_\star$ relation for galaxies with masses above $M_{crit.}$ predicted by equation (3) of K18 for the galaxy 2MASXJ12052321+1022461. The input parameters, their JAM$_{sph}$ best-fit models, and statistics uncertainties are listed in Table A3 on the next page.

30   *Dieu D. Nguyen et al.*

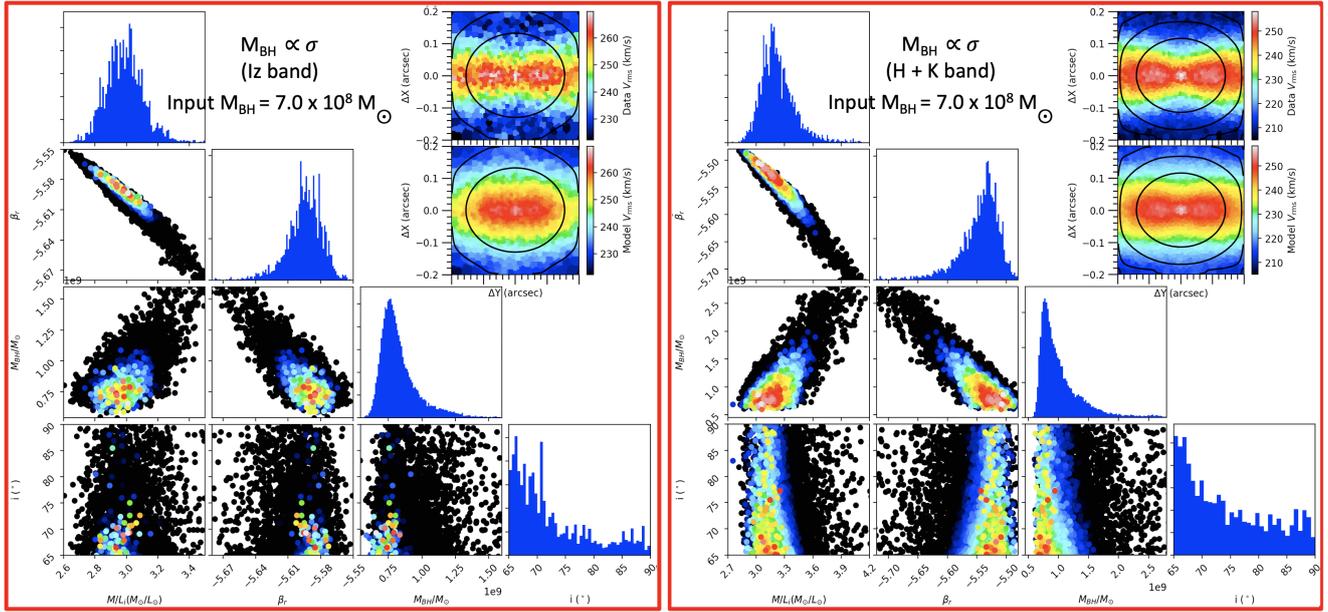

**Figure A11.** Same as Fig. 12 but for the HARMONI simulated kinematics with $M_{BH} = 7.0 \times 10^8\,M_\odot$ that follows the $M_{BH}-\sigma_\star$ relation for galaxies with masses below $M_{crit.}$ predicted by equation (2) of K18 for the galaxy 2MASXJ12052321+1022461. The input parameters, their JAM$_{sph}$ best-fit models, and statistics uncertainties are listed in Table A3 on the next page.

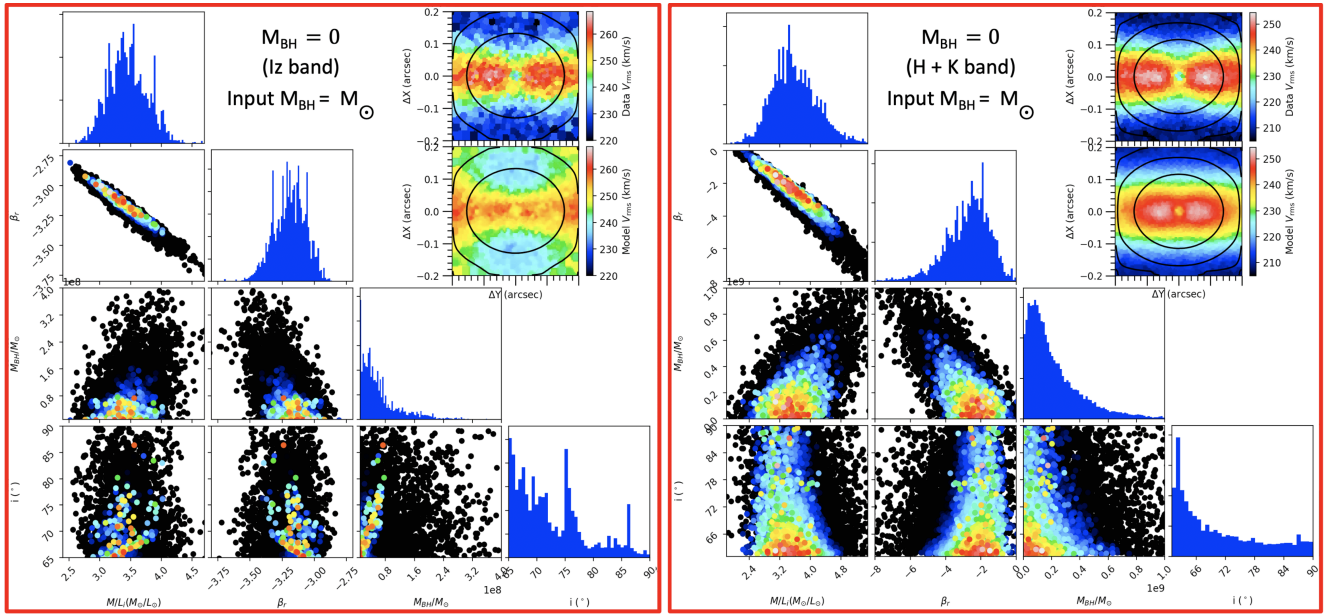

**Figure A12.** Same as Fig. 12 but for the HARMONI simulated kinematics with $M_{BH} = 0\,M_\odot$ for the galaxy 2MASXJ12052321+1022461. The input parameters, their JAM$_{sph}$ best-fit models and statistics uncertainties are listed in Table A3 below.

**Table A3.** Best-fitting JAM$_{sph}$ parameters and their statistical uncertainties for the $I_z$ and $H+K$ band simulated kinematics of 2MASXJ12052321+1022461.

| Parameter name (1) | Search range of parameters (2) | Input value for HSIM (3) | Best-fit value (4) $I_z$ | $1\sigma$ error (16–84%) (5) $I_z$ | $3\sigma$ error (0.14–99.86%) (6) $I_z$ | Best-fit value (7) $H+K$ | $1\sigma$ error (16–84%) (8) $H+K$ | $3\sigma$ error (0.14–99.86%) (9) $H+K$ |
|---|---|---|---|---|---|---|---|---|
| **Assuming no a central SMBH ($M_{BH} = 0\ M_\odot$)** | | | | | | | | |
| $M_{BH}/M_\odot$ | $(0 \longrightarrow 10^{12})$ | 0 | $5.6 \times 10^5$ | $< 3.0 \times 10^7$ | $< 8.0 \times 10^7$ | $6.1 \times 10^5$ | $< 2.0 \times 10^7$ | $< 6.0 \times 10^8$ |
| $M/L_i$ (M$_\odot$/L$_\odot$) | $(0 \longrightarrow 10)$ | 2.5 | 3.3 | $\pm 0.2$ | $\pm 0.5$ | 3.2 | $\pm 0.3$ | $\pm 0.08$ |
| $i$ (°) | $(60 \longrightarrow 90)$ | 60 | 70 | $\pm 5$ | $\pm 10$ | 66 | $\pm 5$ | $\pm 6$ |
| $\beta_r$ | $(-15 \longrightarrow 1)$ | $\pm 0.2$ | $-3.5$ | $\pm 0.2$ | $\pm 0.5$ | $-2.0$ | $\pm 0.3$ | $\pm 1.5$ |
| **Assuming a central SMBH with mass $M_{BH} = 7.0 \times 10^8\ M_\odot$, derived from the $M_{BH}$–$\sigma_\star$ relation (equation (2) from K18)** | | | | | | | | |
| $M_{BH}/M_\odot$ | $(0 \longrightarrow 10^{12})$ | $7.0 \times 10^8$ | $7.5 \times 10^8$ | $\pm 0.1 \times 10^8$ | $\pm 0.2 \times 10^8$ | $7.1 \times 10^8$ | $\pm 0.1 \times 10^8$ | $\pm 0.3 \times 10^8$ |
| $M/L_i$ (M$_\odot$/L$_\odot$) | $(0 \longrightarrow 10)$ | 2.5 | 2.9 | $\pm 0.1$ | $\pm 0.3$ | 3.1 | $\pm 0.1$ | $\pm 0.3$ |
| $i$ (°) | $(60 \longrightarrow 90)$ | 60 | 70 | $\pm 5$ | $\pm 13$ | 73 | $\pm 5$ | $\pm 15$ |
| $\beta_r$ | $(-15 \longrightarrow 1)$ | $\pm 0.2$ | $-5.6$ | $\pm 0.1$ | $\pm 0.4$ | $-5.4$ | $\pm 0.2$ | $\pm 0.5$ |
| **Assuming a central SMBH with mass $M_{BH} = 1.0 \times 10^{10}\ M_\odot$, derived from the $M_{BH}$–$M_\star$ relation (equation (3) from K18)** | | | | | | | | |
| $M_{BH}/M_\odot$ | $(0 \longrightarrow 10^{12})$ | $1.0 \times 10^{10}$ | $1.0 \times 10^{10}$ | $\pm 2.0 \times 10^9$ | $\pm 4.0 \times 10^9$ | $1.1 \times 10^{10}$ | $\pm 2.0 \times 10^9$ | $\pm 4.0 \times 10^9$ |
| $M/L_i$ (M$_\odot$/L$_\odot$) | $(0 \longrightarrow 10)$ | 3.0 | 3.0 | $\pm 0.3$ | $\pm 0.8$ | 3.5 | $\pm 0.5$ | $\pm 1.0$ |
| $i$ (°) | $(60 \longrightarrow 90)$ | 60 | 75 | $\pm 5$ | $\pm 15$ | 76 | $\pm 5$ | $\pm 14$ |
| $\beta_r$ | $(-15 \longrightarrow 1)$ | $\pm 0.2$ | $-3.0$ | $\pm 1.0$ | $\pm 2.5$ | $-5.7$ | $\pm 0.3$ | $\pm 0.8$ |

*Notes:* Same as Table 6 but are the best-fit JAM$_{sph}$ modellings for the galaxy 2MASXJ12052321+1022461 optimized to its HARMONI simulated kinematics with different $M_{BH}$ which are shown in Figs A10, A11, and A12. Also, see Fig. 16 for a graphically short summary of this Table A3.

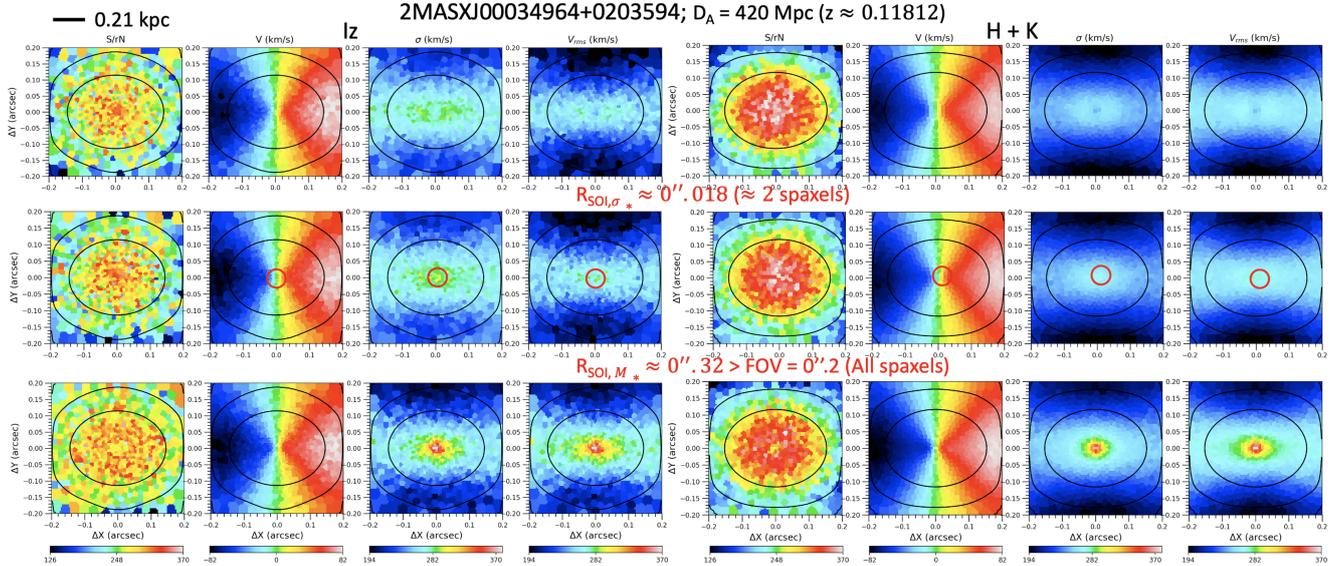

**Figure A13.** Same as Fig. A1, but stellar-kinematic maps of the galaxy 2MASXJ00034964+0203594 on each extracted from its mock $I_z$ (the left four-panel plots) and $H+K$ (the right four-panel plots) HSIM IFS cube, respectively.

32   *Dieu D. Nguyen et al.*

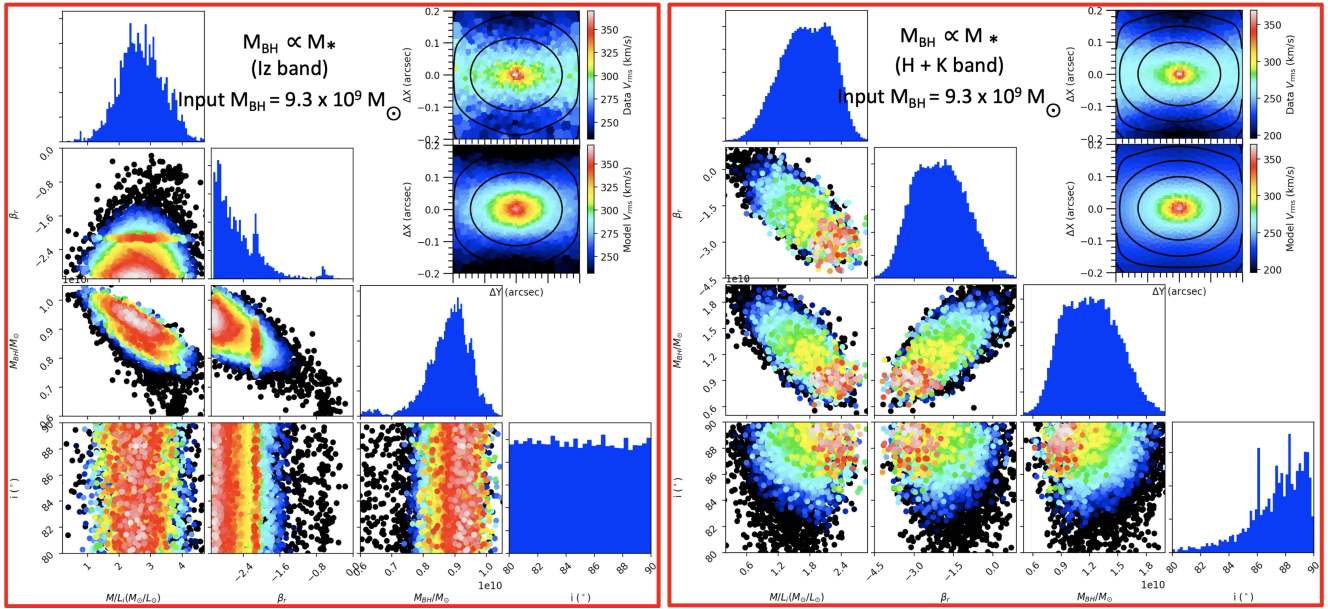

**Figure A14.** Same as Fig. 12 but for the HARMONI simulated kinematics with $M_{BH} = 9.3 \times 10^9$ M$_\odot$ that follows the $M_{BH}$–$M_\star$ relation for galaxies with masses above $M_{crit.}$ predicted by equation (3) of K18 for the galaxy 2MASXJ00034964+0203594. The input parameters, their JAM$_{sph}$ best-fit models, and statistics uncertainties are listed in Table A4 on the next page.

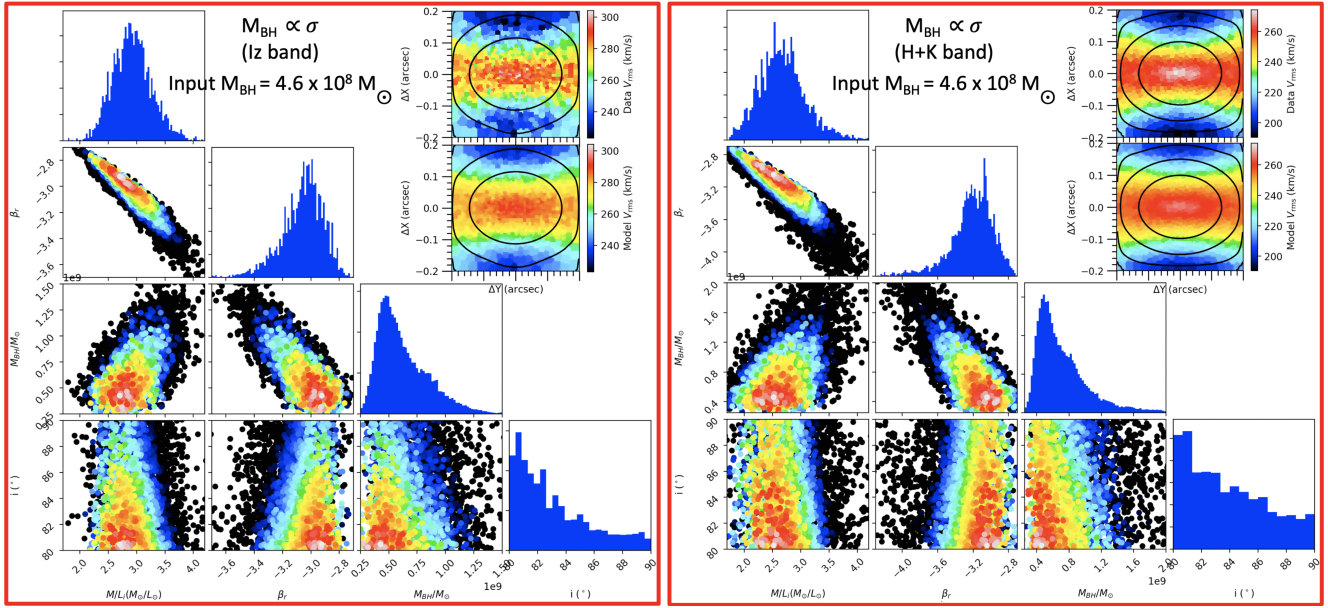

**Figure A15.** Same as Fig. 12 but for the HARMONI simulated kinematics with $M_{BH} = 4.6 \times 10^8$ M$_\odot$ that follows the $M_{BH}$–$\sigma_\star$ relation for galaxies with masses below $M_{crit.}$ predicted by equation (2) of K18 for the galaxy 2MASXJ00034964+0203594. The input parameters, their JAM$_{sph}$ best-fit models, and statistics uncertainties are listed in Table A4 on the next page.

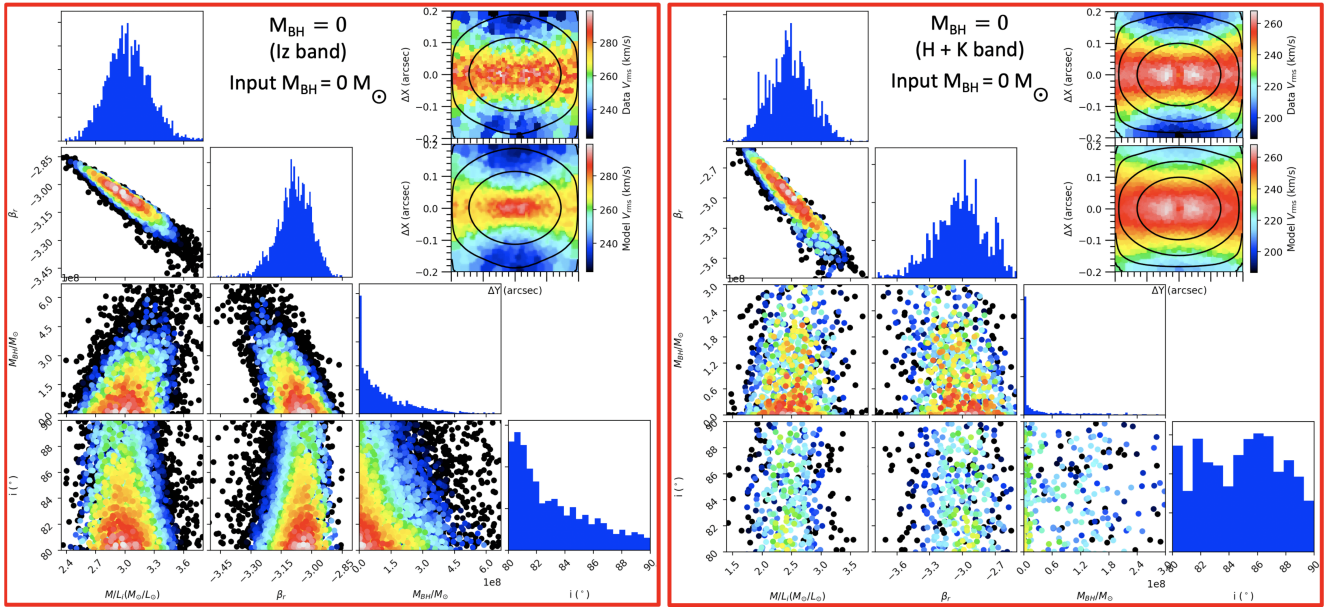

**Figure A16.** Same as Fig. 12 but for the HARMONI simulated kinematics with $M_{BH} = 0$ M$_\odot$ for the galaxy 2MASXJ00034964+0203594. The input parameters, their JAM$_{sph}$ best-fit models, and statistics uncertainties are listed in Table A4 below.

**Table A4.** Best-fitting JAM$_{sph}$ parameters and their statistical uncertainties for the $I_z$ and $H + K$ band simulated kinematics of 2MASXJ00034964+0203594.

| Parameter name (1) | Search range of parameters (2) | Input value for HSIM (3) | Best-fit value (4) $I_z$ | $1\sigma$ error (16–84%) (5) $I_z$ | $3\sigma$ error (0.14–99.86%) (6) $I_z$ | Best-fit value (7) $H + K$ | $1\sigma$ error (16–84%) (8) $H + K$ | $3\sigma$ error (0.14–99.86%) (9) $H + K$ |
|---|---|---|---|---|---|---|---|---|
| **Assuming no a central SMBH ($M_{BH} = 0$ M$_\odot$)** | | | | | | | | |
| $M_{BH}/M_\odot$ | $(0 \longrightarrow 10^{12})$ | 0 | $6.8 \times 10^6$ | $< 1 \times 10^8$ | $< 3 \times 10^8$ | $6.2 \times 10^6$ | $< 6.0 \times 10^7$ | $< 1.8 \times 10^8$ |
| $M/L_i$ (M$_\odot$/L$_\odot$) | $(0 \longrightarrow 10)$ | 2.0 | 3.0 | ±0.2 | ±0.5 | 2.5 | ±0.3 | ±1.0 |
| $i$ (°) | $(80 \longrightarrow 90)$ | 85 | 85 | ±2 | ±5 | 65.0 | ±2 | ±5 |
| $\beta_r$ | $(-15 \longrightarrow 1)$ | ±0.2 | $-3.0$ | ±0.1 | ±0.2 | $-3.0$ | ±0.1 | ±0.3 |
| **Assuming a central SMBH with mass $M_{BH} = 4.6 \times 10^8$ M$_\odot$, derived from the $M_{BH}$–$\sigma_\star$ relation (equation (2) from K18)** | | | | | | | | |
| $M_{BH}/M_\odot$ | $(0 \longrightarrow 10^{12})$ | $4.6 \times 10^8$ | $4.7 \times 10^9$ | $\pm 3.0 \times 10^8$ | $< 5.0 \times 10^9$ | $4.4 \times 10^8$ | $\pm 2.5 \times 10^8$ | $< 1.0 \times 10^9$ |
| $M/L_i$ (M$_\odot$/L$_\odot$) | $(0 \longrightarrow 10)$ | 2.0 | 2.7 | ±0.3 | ±0.8 | 2.5 | ±0.2 | ±0.5 |
| $i$ (°) | $(80 \longrightarrow 90)$ | 85 | 85 | ±2 | ±5 | 71.4 | ±2 | ±5 |
| $\beta_r$ | $(-15 \longrightarrow 1)$ | ±0.2 | $-3.0$ | ±0.2 | ±0.4 | $-3.0$ | ±0.2 | ±0.4 |
| **Assuming a central SMBH with mass $M_{BH} = 9.3 \times 10^9$ M$_\odot$, derived from the $M_{BH}$–$M_\star$ relation (equation (3) from K18)** | | | | | | | | |
| $M_{BH}/M_\odot$ | $(0 \longrightarrow 10^{12})$ | $9.3 \times 10^9$ | $9.3 \times 10^9$ | $\pm 1.0 \times 10^9$ | $\pm 2.0 \times 10^9$ | $9.3 \times 10^9$ | $\pm 3.0 \times 10^8$ | $\pm 9.0 \times 10^9$ |
| $M/L_i$ (M$_\odot$/L$_\odot$) | $(0 \longrightarrow 10)$ | 2.0 | 2.5 | ±0.5 | ±1.5 | 2.0 | ±0.1 | ±0.4 |
| $i$ (°) | $(80 \longrightarrow 90)$ | 85 | 85 | ±2 | ±5 | 85.8 | ±2 | ±5 |
| $\beta_r$ | $(-15 \longrightarrow 1)$ | ±0.2 | $-3.0$ | ±0.5 | ±1.5 | $-3.0$ | ±1.0 | ±2.0 |

*Notes:* Same as Table 6 but are the best-fit JAM$_{sph}$ modellings for the galaxy 2MASXJ00034964+0203594 optimized to its HARMONI simulated kinematics with different $M_{BH}$ which are shown in Figs A14, A15, and A16. Also, see Fig. 16 for a graphically short summary of this Table A4.

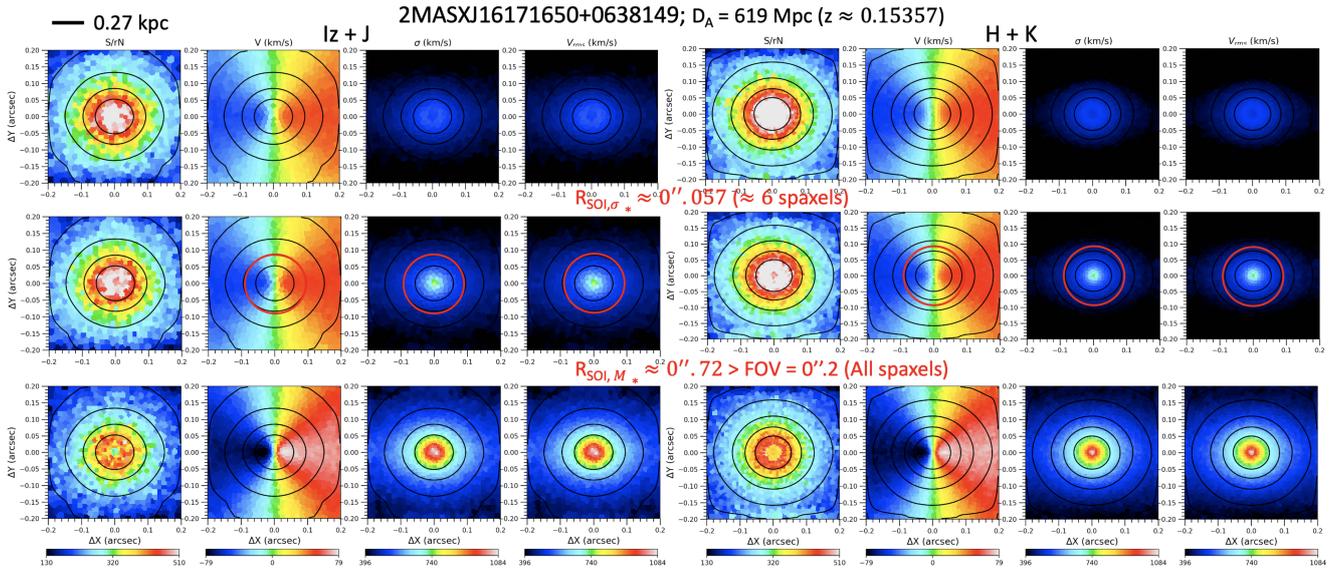

**Figure A17.** Same as Fig. A1, but stellar-kinematic maps of the galaxy 2MASXJ16171650+0638149 on each extracted from its mock $I_z + J$ (the left four-panel plots) and $H + K$ (the right four-panel plots) HSIM IFS cube, respectively.

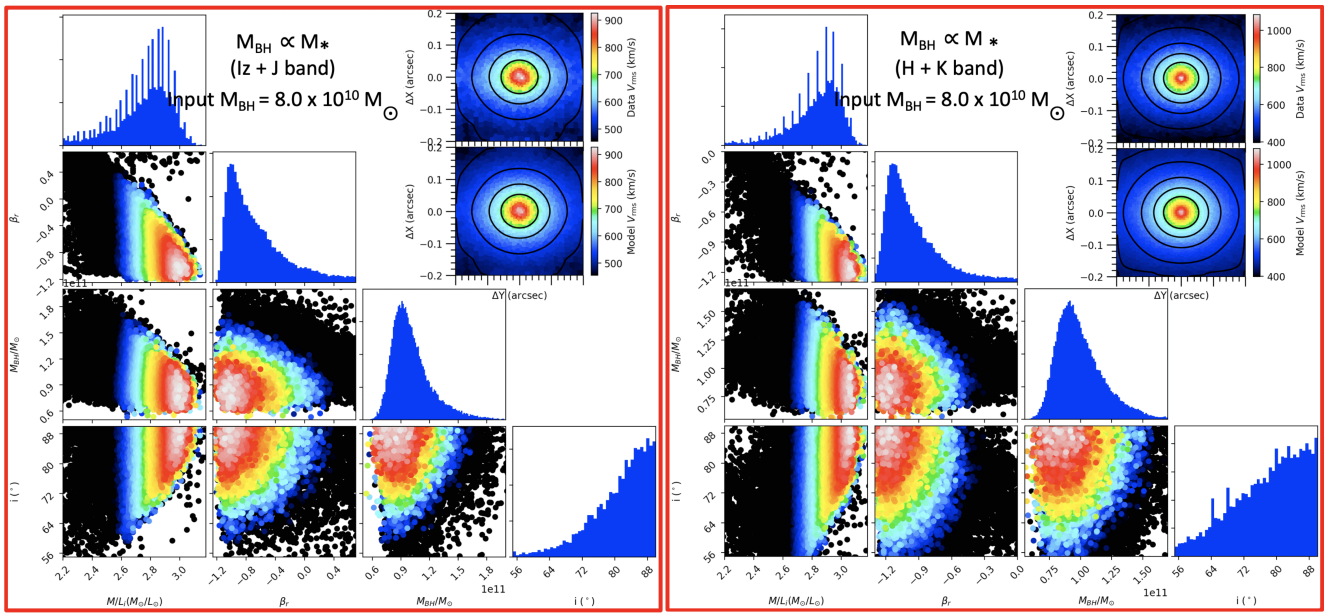

**Figure A18.** Same as Fig. 12 but for the HARMONI simulated kinematics with $M_{BH} = 8.0 \times 10^{10}$ M$_\odot$ that follows the $M_{BH}$–$M_\star$ relation for galaxies with masses above $M_{crit.}$ predicted by equation (3) of K18 for the galaxy 2MASXJ16171650+0638149. The input parameters, their JAM$_{sph}$ best-fit models, and statistics uncertainties are listed in Table A5.

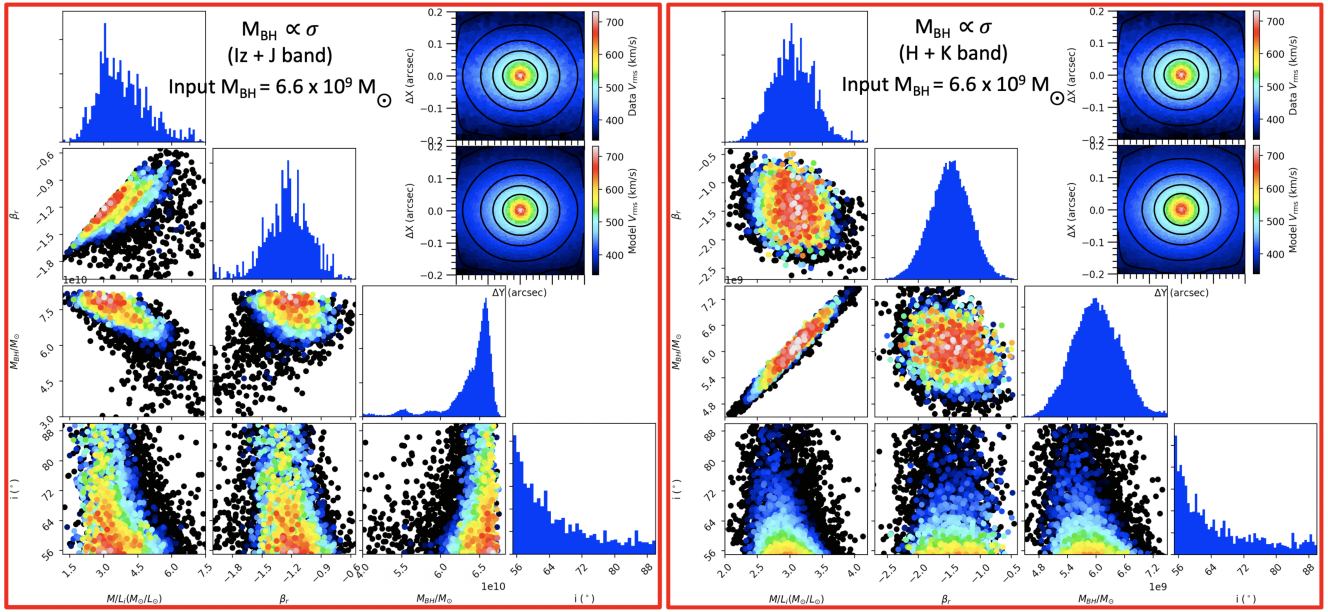

**Figure A19.** Same as Fig. 12 but for the HARMONI simulated kinematics with $M_{BH} = 6.6 \times 10^9$ M$_\odot$ that follows the $M_{BH}-\sigma_\star$ relation for galaxies with masses below $M_{\rm crit.}$ predicted by equation (2) of K18 for the galaxy 2MASXJ16171650+0638149. The input parameters, their JAM$_{\rm sph}$ best-fit models, and statistics uncertainties are listed in Table A5.

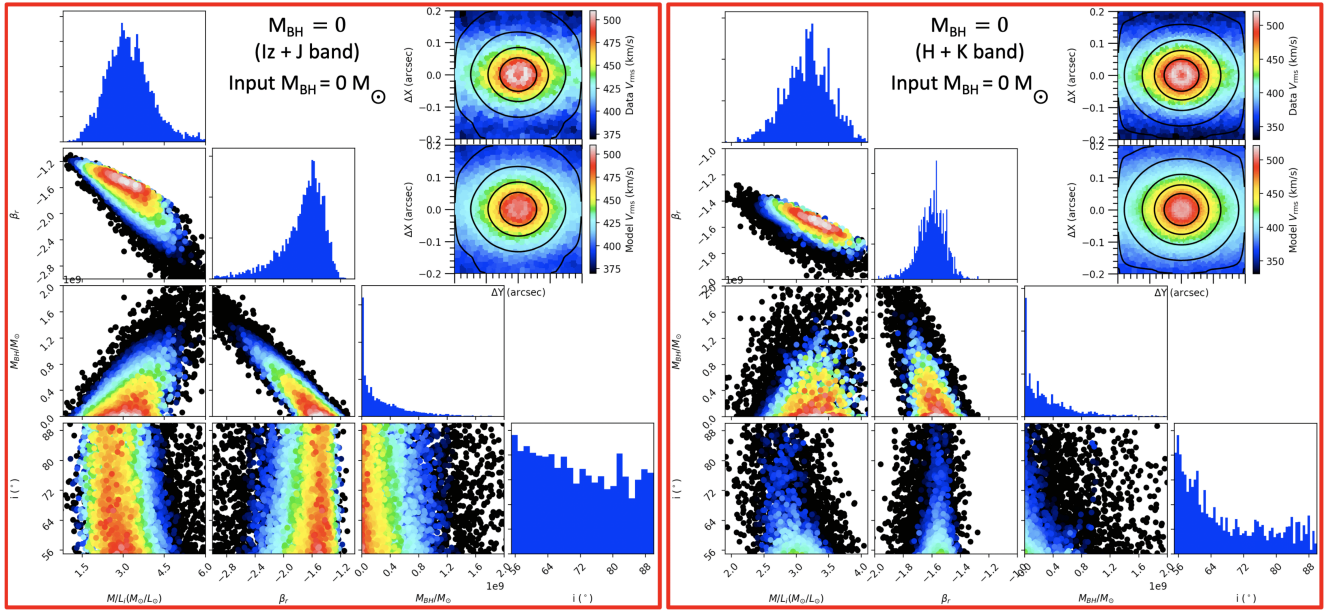

**Figure A20.** Same as Fig. 12 but for the HARMONI simulated kinematics with $M_{BH} = 0$ M$_\odot$ for the galaxy 2MASXJ16171650+0638149. The input parameters, their JAM$_{\rm sph}$ best-fit models, and statistics uncertainties are listed in Table A5.

**Table A5.** Best-fitting JAM$_{sph}$ parameters and their statistical uncertainties for the $I_z+J$ and $H+K$ band simulated kinematics of 2MASXJ16171650+0638149.

| Parameter name (1) | Search range of parameters (2) | Input value for HSIM (3) | Best-fit value (4) $I_z+J$ | $1\sigma$ error (16–84%) (5) $I_z+J$ | $3\sigma$ error (0.14–99.86%) (6) $I_z+J$ | Best-fit value (7) $H+K$ | $1\sigma$ error (16–84%) (8) $H+K$ | $3\sigma$ error (0.14–99.86%) (9) $H+K$ |
|---|---|---|---|---|---|---|---|---|
| **Assuming no a central SMBH ($M_{BH} = 0$ M$_\odot$)** | | | | | | | | |
| $M_{BH}$/M$_\odot$ | ($0 \longrightarrow 10^{12}$) | 0 | $7.5 \times 10^5$ | $< 2.0 \times 10^8$ | $< 1.2 \times 10^9$ | $7.0 \times 10^5$ | $< 1.0 \times 10^8$ | $< 8.0 \times 10^8$ |
| $M/L_i$ (M$_\odot$/L$_\odot$) | ($0 \longrightarrow 10$) | 2.5 | 2.8 | ±0.5 | ±1.5 | 2.9 | ±0.3 | ±1.0 |
| $i$ (°) | ($49 \longrightarrow 90$) | 60 | 70 | ±12 | ±22 | 60 | ±12 | ±25 |
| $\beta_r$ | ($-10 \longrightarrow 1$) | ±0.2 | $-1.5$ | ±0.1 | ±0.3 | $-1.5$ | ±0.1 | ±0.2 |
| **Assuming a central SMBH with mass $M_{BH} = 6.6 \times 10^9$ M$_\odot$, derived from the $M_{BH}-\sigma_\star$ relation (equation (2) from K18)** | | | | | | | | |
| $M_{BH}$/M$_\odot$ | ($0 \longrightarrow 10^{12}$) | $6.6 \times 10^9$ | $6.6 \times 10^9$ | $\pm 0.1 \times 10^9$ | $\pm 0.7 \times 10^9$ | $6.0 \times 10^9$ | $\pm 0.6 \times 10^9$ | $\pm 1.2 \times 10^9$ |
| $M/L_i$ (M$_\odot$/L$_\odot$) | ($0 \longrightarrow 10$) | 2.5 | 3.0 | ±1.1 | ±3.0 | 3.0 | ±0.5 | ±0.8 |
| $i$ (°) | ($49 \longrightarrow 90$) | 60 | 60 | ±10 | ±28 | 60 | ±10 | ±28 |
| $\beta_r$ | ($-10 \longrightarrow 1$) | ±0.2 | $-1.0$ | ±0.3 | ±1.0 | $-1.1$ | ±0.2 | ±0.5 |
| **Assuming a central SMBH with mass $M_{BH} = 8.0 \times 10^{10}$ M$_\odot$, derived from the $M_{BH}-M_\star$ relation (equation (3) from K18)** | | | | | | | | |
| $M_{BH}$/M$_\odot$ | ($0 \longrightarrow 10^{12}$) | $8.0 \times 10^{10}$ | $9.0 \times 10^{10}$ | $\pm 3.0 \times 10^{10}$ | $\pm 6.0 \times 10^{10}$ | $8.5 \times 10^{10}$ | $\pm 3.0 \times 10^{10}$ | $\pm 6.5 \times 10^{10}$ |
| $M/L_i$ (M$_\odot$/L$_\odot$) | ($0 \longrightarrow 10$) | 2.5 | 3.0 | ±0.2 | ±0.4 | 3.0 | ±0.2 | ±0.4 |
| $i$ (°) | ($49 \longrightarrow 90$) | 60 | 80 | ±15 | ±25 | 80 | ±15 | ±25 |
| $\beta_r$ | ($-10 \longrightarrow 1$) | ±0.2 | $-1.2$ | ±0.3 | ±0.6 | $-1.4$ | ±0.4 | ±0.8 |

*Notes:* Same as Table 6 but are the best-fit JAM$_{sph}$ modellings for the galaxy 2MASXJ16171650+0638149 optimized to its HARMONI simulated kinematics with different $M_{BH}$ which are shown in Figs A18, A19, and A20. Also, see Fig. 16 for a graphically short summary of this Table A5.

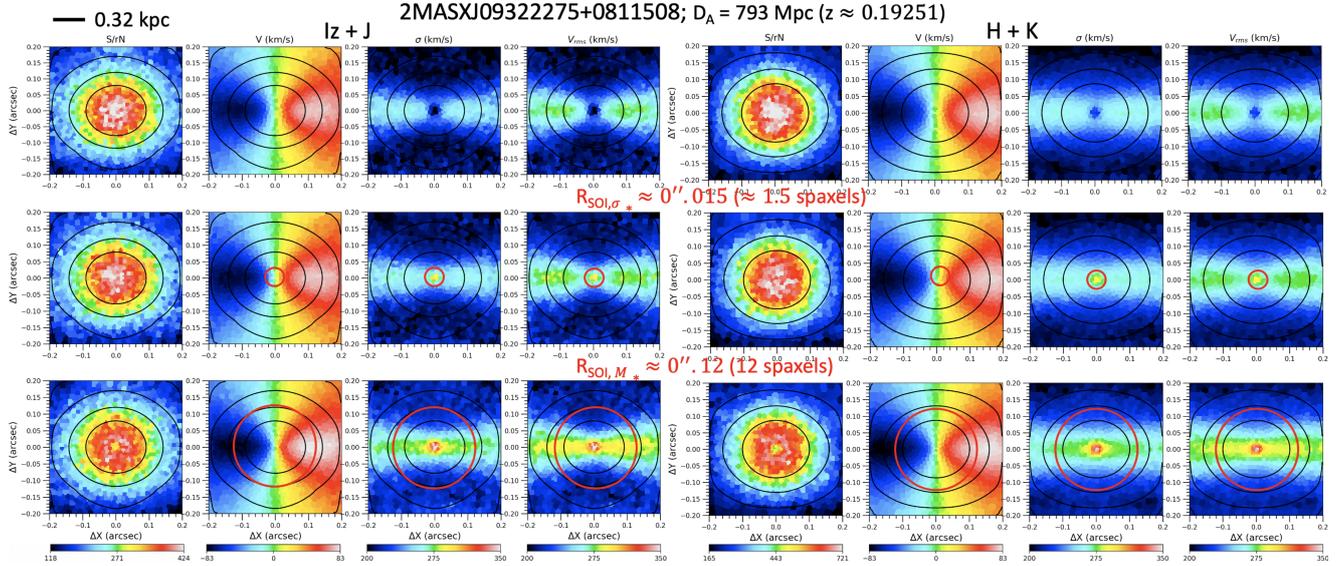

**Figure A21.** Same as Fig. A1, but stellar-kinematic maps of the galaxy 2MASXJ09322275+0811508 on each extracted from its mock $I_z+J$ (the left four-panel plots) and $H+K$ (the right four-panel plots) HSIM IFS cube, respectively.

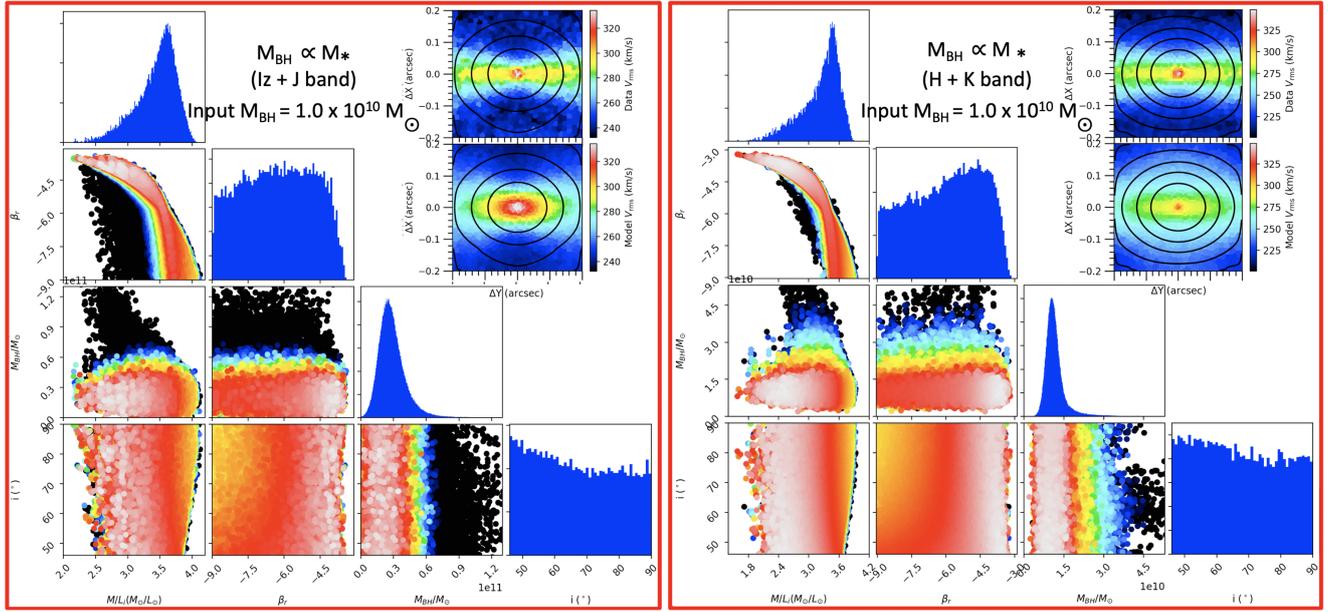

**Figure A22.** Same as Fig. 12 but for the HARMONI simulated kinematics with $M_{BH} = 1.0 \times 10^{10}$ M$_\odot$ that follows the $M_{BH}-M_\star$ relation for galaxies with masses above $M_{\rm crit.}$ predicted by equation (3) of K18 for the galaxy 2MASXJ09322275+0811508. The input parameters, their JAM$_{\rm sph}$ best-fit models, and statistics uncertainties are listed in Table A6 on the next page.

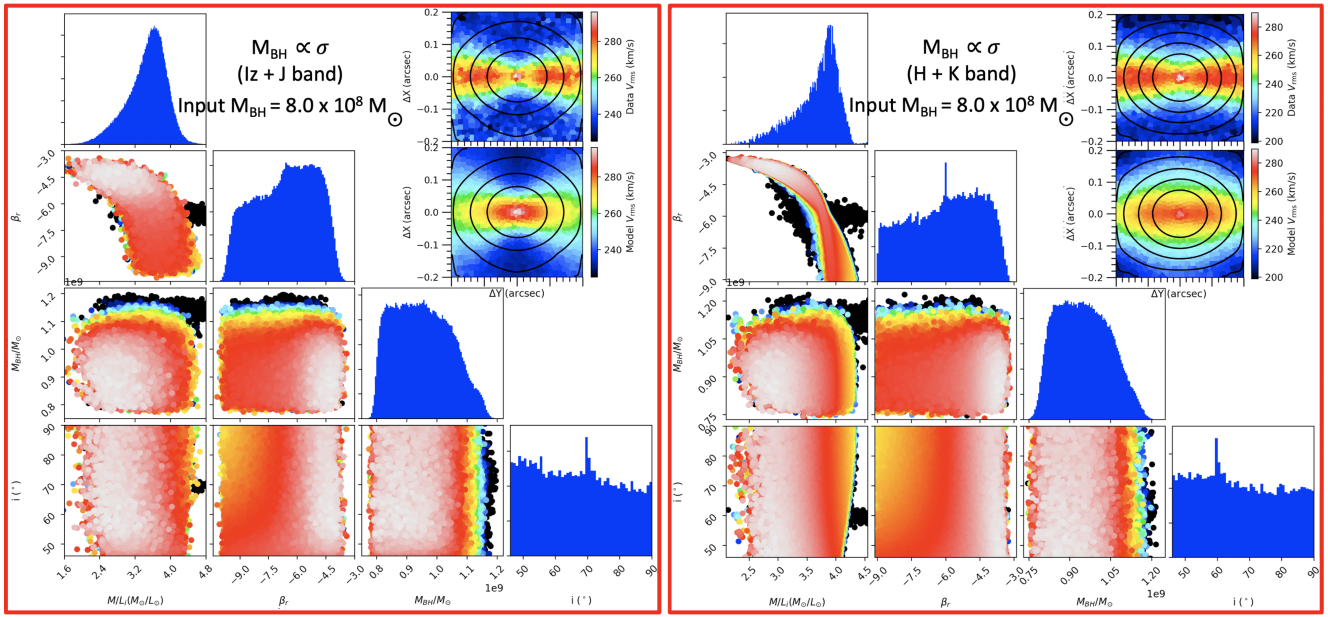

**Figure A23.** Same as Fig. 12 but for the HARMONI simulated kinematics with $M_{BH} = 8.0 \times 10^{8}$ M$_\odot$ that follows the $M_{BH}-\sigma_\star$ relation for galaxies with masses below $M_{\rm crit.}$ predicted by equation (2) of K18 for the galaxy 2MASXJ09322275+0811508. The input parameters, their JAM$_{\rm sph}$ best-fit models, and statistics uncertainties are listed in Table A6 on the next page.

38    *Dieu D. Nguyen et al.*

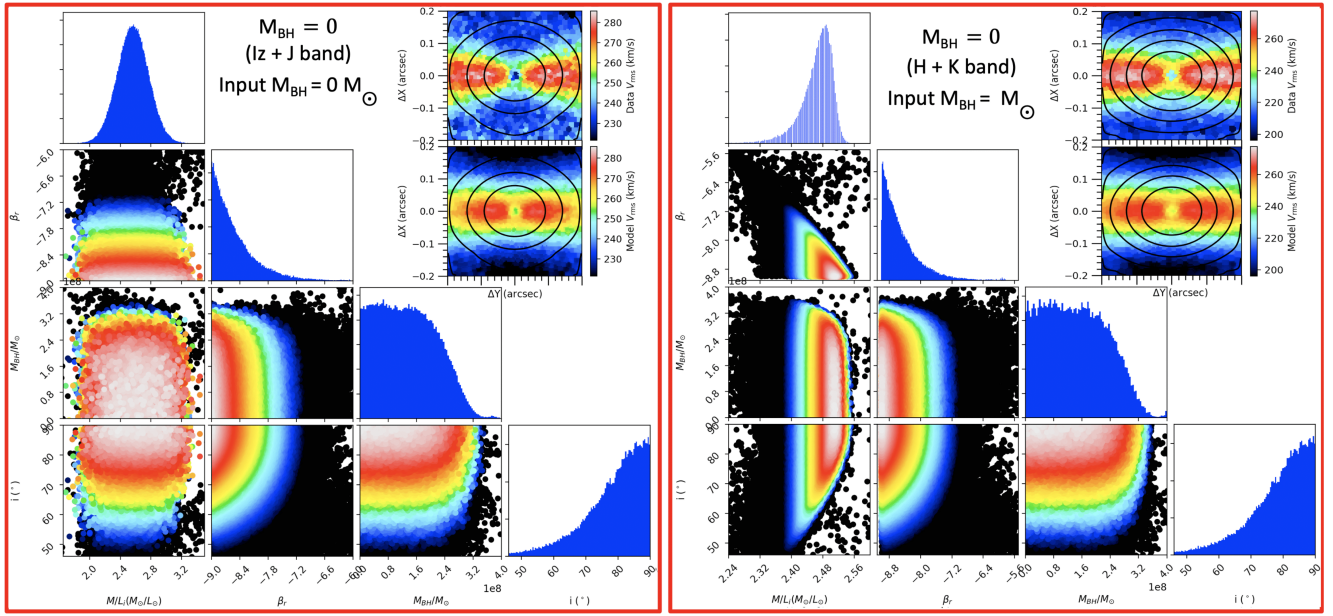

**Figure A24.** Same as Fig. 12 but for the HARMONI simulated kinematics with $M_{BH} = 0\,M_\odot$ for the galaxy 2MASXJ09322275+0811508. The input parameters, their JAM$_{sph}$ best-fit models, and statistics uncertainties are listed in Table A6 below.

**Table A6.** Best-fitting JAM$_{sph}$ parameters and their statistical uncertainties for the $I_z + J$ and $H + K$ band simulated kinematics of 2MASXJ09322275+0811508.

| Parameter name (1) | Search range of parameters (2) | Input value for HSIM (3) | Best-fit value (4) $I_z + J$ | $1\sigma$ error (16–84%) (5) $I_z + J$ | $3\sigma$ error (0.14–99.86%) (6) $I_z + J$ | Best-fit value (7) $H + K$ | $1\sigma$ error (16–84%) (8) $H + K$ | $3\sigma$ error (0.14–99.86%) (9) $H + K$ |
|---|---|---|---|---|---|---|---|---|
| **Assuming no a central SMBH ($M_{BH} = 0\,M_\odot$)** | | | | | | | | |
| $M_{BH}/M_\odot$ | $(0 \longrightarrow 10^{12})$ | 0 | $1.2 \times 10^8$ | $< 1.8 \times 10^8$ | $< 2 \times 10^8$ | $1.2 \times 10^8$ | $< 1.8 \times 10^8$ | $< 2.0 \times 10^8$ |
| $M/L_i$ (M$_\odot$/L$_\odot$) | $(0 \longrightarrow 10)$ | 2.5 | 2.5 | $\pm 0.2$ | $\pm 0.6$ | 2.5 | $\pm 0.2$ | $\pm 0.6$ |
| $i$ (°) | $(45 \longrightarrow 90)$ | 60 | 85 | $\pm 15$ | $\pm 40$ | 85 | $\pm 15$ | $\pm 40$ |
| $\beta_r$ | $(-15 \longrightarrow 1)$ | $\pm 0.2$ | $-9.0$ | $< 1.2$ | $< 3.5$ | $-9.0$ | $< 1.2$ | $< 3.5$ |
| **Assuming a central SMBH with mass $M_{BH} = 8 \times 10^8\,M_\odot$, derived from the $M_{BH}$–$\sigma_\star$ relation (equation (2) from K18)** | | | | | | | | |
| $M_{BH}/M_\odot$ | $(0 \longrightarrow 10^{12})$ | $8.0 \times 10^8$ | $9.5 \times 10^8$ | $\pm 0.5 \times 10^8$ | $\pm 1.5 \times 10^8$ | $9.6 \times 10^8$ | $\pm 0.6 \times 10^8$ | $\pm 1.6 \times 10^8$ |
| $M/L_i$ (M$_\odot$/L$_\odot$) | $(0 \longrightarrow 10)$ | 2.5 | 2.8 | $\pm 0.5$ | $\pm 1.0$ | 2.7 | $\pm 0.5$ | $\pm 1.5$ |
| $i$ (°) | $(45 \longrightarrow 90)$ | 60 | 70 | $\pm 10$ | $\pm 20$ | 70 | $\pm 10$ | $\pm 20$ |
| $\beta_r$ | $(-15 \longrightarrow 1)$ | $\pm 0.2$ | $-0.45$ | $\pm 0.15$ | $\pm 0.50$ | $-0.44$ | $\pm 0.16$ | $\pm 0.46$ |
| **Assuming a central SMBH with mass $M_{BH} = 1.0 \times 10^{10}\,M_\odot$, derived from the $M_{BH}$–$M_\star$ relation (equation (3) from K18)** | | | | | | | | |
| $M_{BH}/M_\odot$ | $(0 \longrightarrow 10^{12})$ | $1.0 \times 10^{10}$ | $2.0 \times 10^{10}$ | $\pm 0.7 \times 10^{10}$ | $\pm 2.0 \times 10^{10}$ | $1.6 \times 10^{10}$ | $\pm 0.5 \times 10^{10}$ | $\pm 1.5 \times 10^{10}$ |
| $M/L_i$ (M$_\odot$/L$_\odot$) | $(0 \longrightarrow 10)$ | 2.5 | 2.8 | $\pm 0.5$ | $\pm 1.5$ | 2.7 | $\pm 0.6$ | $\pm 1.5$ |
| $i$ (°) | $(45 \longrightarrow 90)$ | 60 | 70 | $\pm 10$ | $\pm 20$ | 70 | $\pm 10$ | $\pm 20$ |
| $\beta_r$ | $(-15 \longrightarrow 1)$ | $\pm 0.2$ | $-0.45$ | $\pm 0.15$ | $\pm 0.45$ | $-0.44$ | $\pm 0.16$ | $\pm 0.46$ |

*Notes:* Same as Table 6 but are the best-fit JAM$_{sph}$ modellings for the galaxy 2MASXJ09322275+0811508 optimized to its HARMONI simulated kinematics with different $M_{BH}$ which are shown in Figs A22, A23, and A24. Also, see Fig. 16 for a graphically short summary of this Table A6.

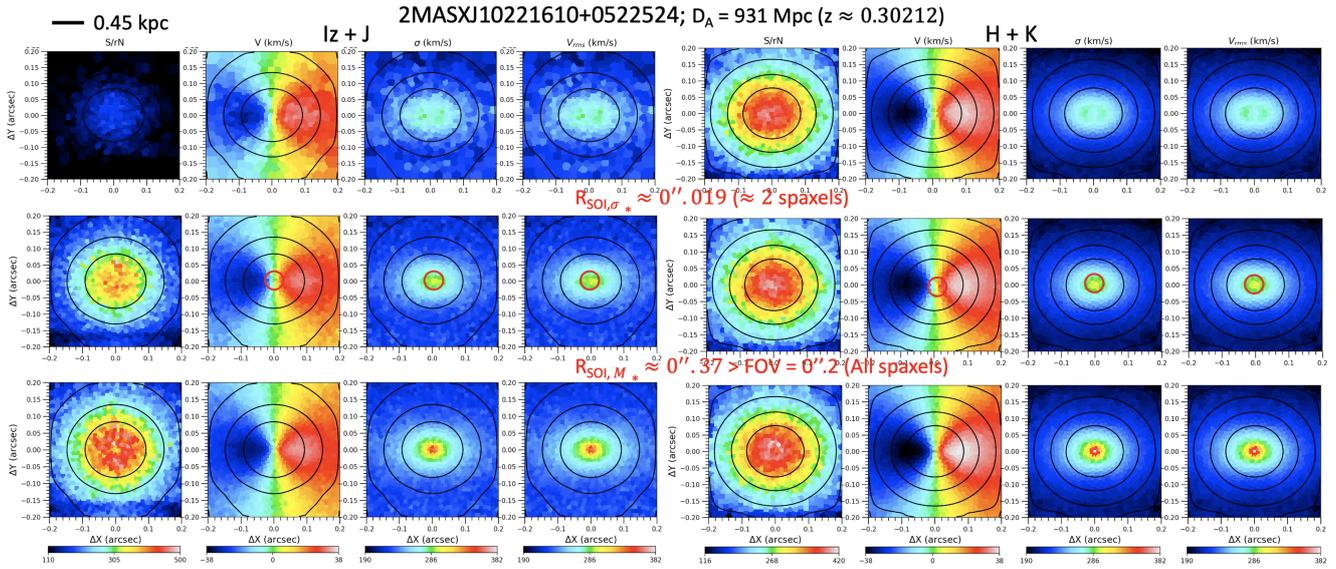

**Figure A25.** Same as Fig. A1, but stellar-kinematic maps of the galaxy 2MASXJ10221610+0522524 on each extracted from its mock $I_z + J$ (the left four-panel plots) and $H + K$ (the right four-panel plots) HSIM IFS cube, respectively.

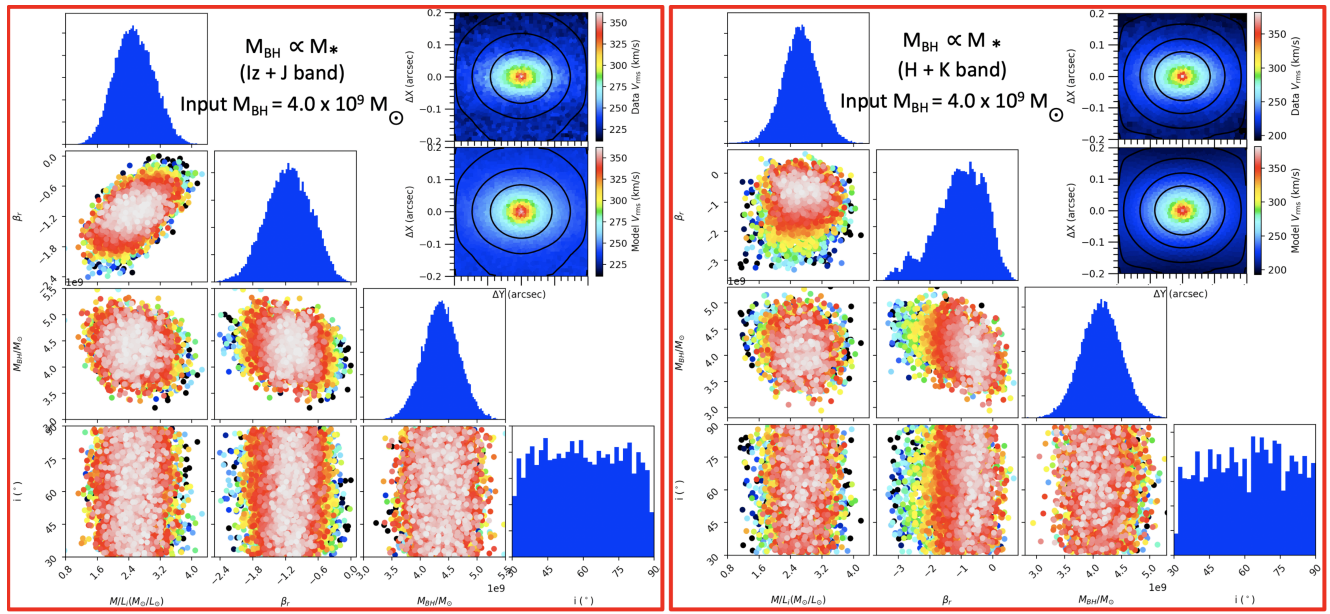

**Figure A26.** Same as Fig. 12 but for the HARMONI simulated kinematics with $M_{BH} = 4.0 \times 10^9$ $M_\odot$ that follows the $M_{BH}$–$M_\star$ relation for galaxies with masses above $M_{crit.}$ predicted by equation (3) of K18 for the galaxy 2MASXJ10221610+0522524. The input parameters, their JAM$_{sph}$ best-fit models and statistics uncertainties are listed in Table A7.

40  *Dieu D. Nguyen et al.*

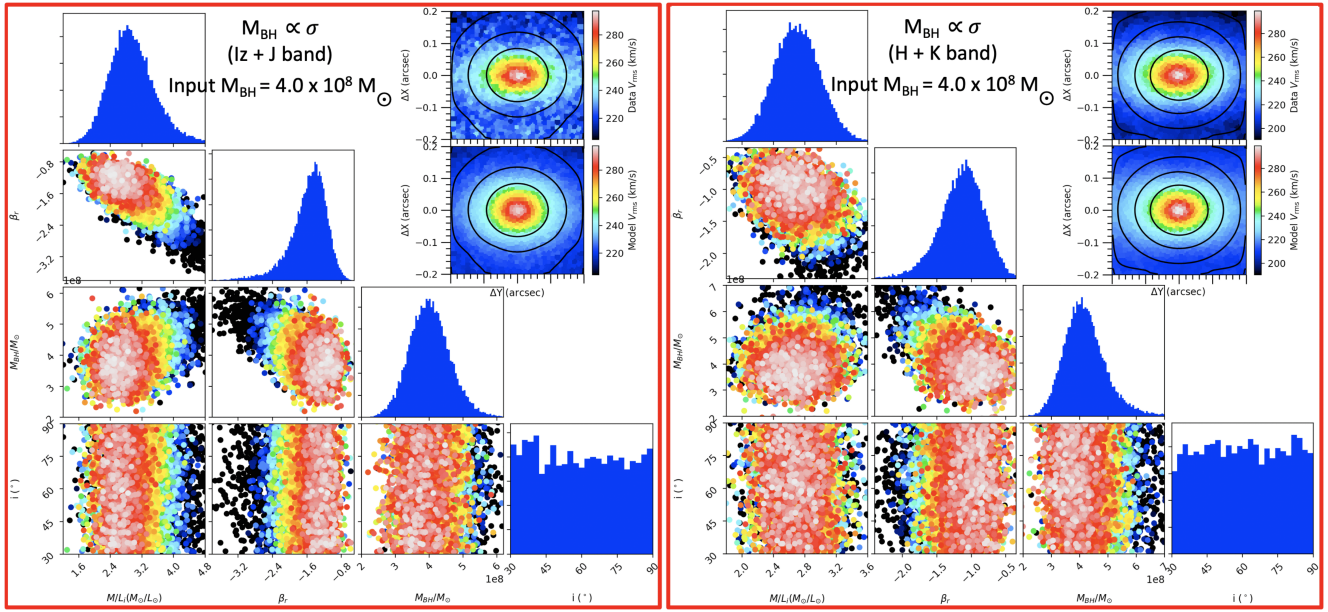

**Figure A27.** Same as Fig. 12 but for the HARMONI simulated kinematics with $M_{BH} = 4.0 \times 10^8$ M$_\odot$ that follows the $M_{BH}-\sigma_\star$ relation for galaxies with masses below $M_{crit.}$ predicted by equation (2) of K18 for the galaxy 2MASXJ10221610+0522524. The input parameters, their JAM$_{sph}$ best-fit models and statistics uncertainties are listed in Table A7.

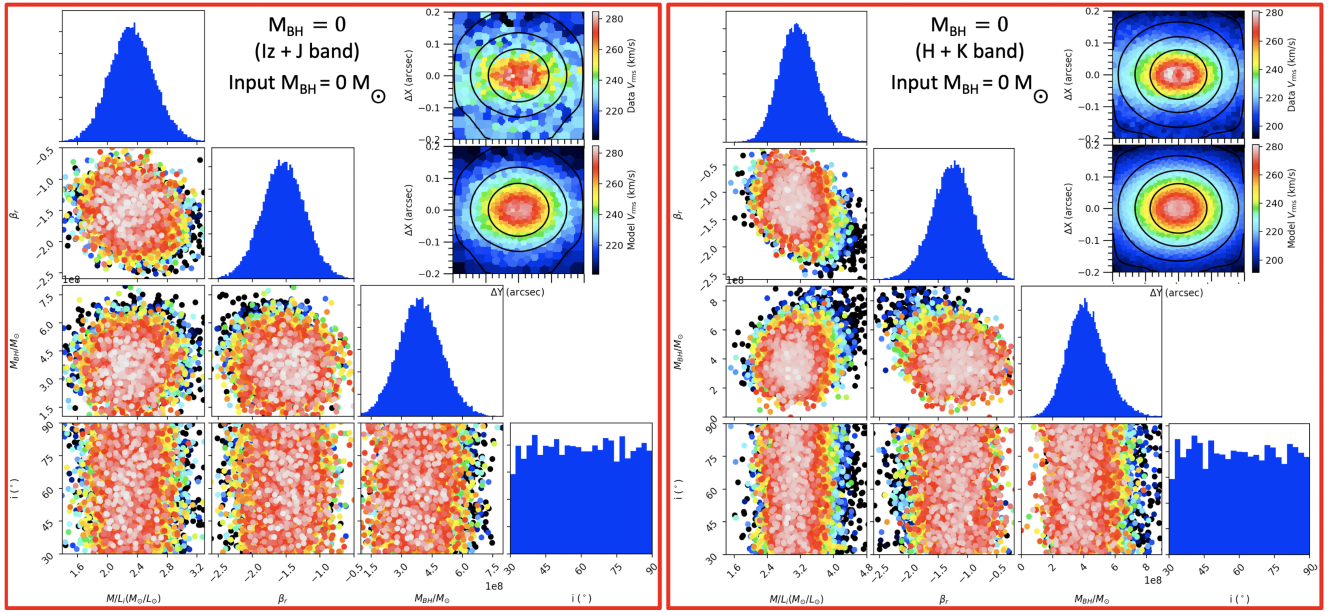

**Figure A28.** Same as Fig. 12 but for the HARMONI simulated kinematics with $M_{BH} = 0$ M$_\odot$ for the galaxy 2MASXJ10221610+0522524. The input parameters, their JAM$_{sph}$ best-fit models and statistics uncertainties are listed in Table A7.

**Table A7.** Best-fitting JAM$_{sph}$ parameters and their statistical uncertainties for the $I_z + J$ and $H + K$ band simulated kinematics of 2MASXJ10221610+0522524.

| Parameter name (1) | Search range of parameters (2) | Input value for HSIM (3) | Best-fit value (4) $I_z + J$ | $1\sigma$ error (16–84%) (5) $I_z + J$ | $3\sigma$ error (0.14–99.86%) (6) $I_z + J$ | Best-fit value (7) $H + K$ | $1\sigma$ error (16–84%) (8) $H + K$ | $3\sigma$ error (0.14–99.86%) (9) $H + K$ |
|---|---|---|---|---|---|---|---|---|
| **Assuming no a central SMBH ($M_{BH} = 0$ M$_\odot$)** | | | | | | | | |
| $M_{BH}/M_\odot$ | $(0 \longrightarrow 10^{12})$ | 0 | $4.2 \times 10^8$ | $\pm 1.0 \times 10^8$ | $\pm 2.8 \times 10^8$ | $4.0 \times 10^8$ | $\pm 1.5 \times 10^8$ | $\pm 3.5 \times 10^8$ |
| $M/L_i$ (M$_\odot$/L$_\odot$) | $(0 \longrightarrow 10)$ | 2.5 | 2.3 | $\pm 0.3$ | $\pm 0.8$ | 2.5 | $\pm 0.6$ | $\pm 2.0$ |
| $i$ (°) | $(30 \longrightarrow 90)$ | 60 | 60 | $\pm 18$ | $\pm 28$ | 60 | $\pm 18$ | $\pm 28$ |
| $\beta_r$ | $(-15 \longrightarrow 1)$ | $\pm 0.2$ | $-1.5$ | $\pm 0.2$ | $\pm 0.7$ | $-1.2$ | $\pm 0.4$ | $\pm 1.1$ |
| **Assuming a central SMBH with mass $M_{BH} = 4.0 \times 10^8$ M$_\odot$, derived from the $M_{BH}$–$\sigma_\star$ relation (equation (2) from K18)** | | | | | | | | |
| $M_{BH}/M_\odot$ | $(0 \longrightarrow 10^{12})$ | $4.0 \times 10^8$ | $4.0 \times 10^8$ | $\pm 1.0 \times 10^8$ | $\pm 3.0 \times 10^8$ | $4.0 \times 10^9$ | $\pm 1.0 \times 10^8$ | $\pm 3.0 \times 10^8$ |
| $M/L_i$ (M$_\odot$/L$_\odot$) | $(0 \longrightarrow 10)$ | 2.5 | 2.4 | $\pm 0.8$ | $\pm 2.4$ | 2.5 | $\pm 0.6$ | $\pm 1.8$ |
| $i$ (°) | $(30 \longrightarrow 90)$ | 60 | 60 | $\pm 18$ | $\pm 28$ | 60 | $\pm 18$ | $\pm 28$ |
| $\beta_r$ | $(-15 \longrightarrow 1)$ | $\pm 0.2$ | $-1.2$ | $\pm 0.3$ | $\pm 1.2$ | $-1.0$ | $\pm 0.4$ | $\pm 1.5$ |
| **Assuming a central SMBH with mass $M_{BH} = 4.0 \times 10^9$ M$_\odot$, derived from the $M_{BH}$–$M_\star$ relation (equation (3) from K18)** | | | | | | | | |
| $M_{BH}/M_\odot$ | $(0 \longrightarrow 10^{12})$ | $4.0 \times 10^9$ | $4.2 \times 10^9$ | $\pm 3.0 \times 10^8$ | $\pm 1.0 \times 10^9$ | $4.2 \times 10^9$ | $\pm 3 \times 10^8$ | $\pm 1.0 \times 10^9$ |
| $M/L_i$ (M$_\odot$/L$_\odot$) | $(0 \longrightarrow 10)$ | 2.5 | 2.5 | $\pm 0.3$ | $\pm 1.0$ | 2.5 | $\pm 0.3$ | $\pm 1.0$ |
| $i$ (°) | $(30 \longrightarrow 90)$ | 60 | 60 | $\pm 18$ | $\pm 28$ | 60 | $\pm 18$ | $\pm 28$ |
| $\beta_r$ | $(-15 \longrightarrow 1)$ | $\pm 0.2$ | $-1.2$ | $\pm 0.3$ | $\pm 1.0$ | $-1.0$ | $\pm 0.3$ | $\pm 1.0$ |

*Notes:* Same as Table 6 but are the best-fit JAM$_{sph}$ modellings for the galaxy 2MASXJ10221610+0522524 optimized to its HARMONI simulated kinematics with different $M_{BH}$ which are shown in Figs A26, A27, and A28. Also, see Fig. 16 for a graphically short summary of this Table A7.

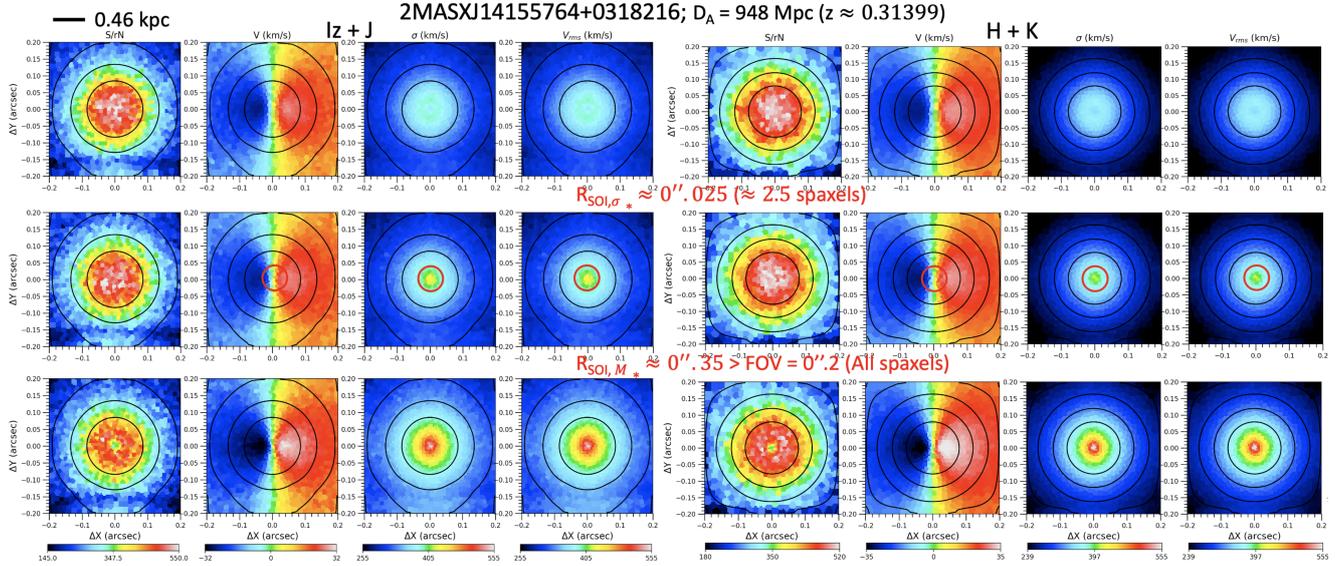

**Figure A29.** Same as Fig. A1, but stellar-kinematic maps of the galaxy 2MASXJ14155764+0318216 on each extracted from its mock $I_z + J$ (the left four-panel plots) and $H + K$ (the right four-panel plots) HSIM IFS cube, respectively.

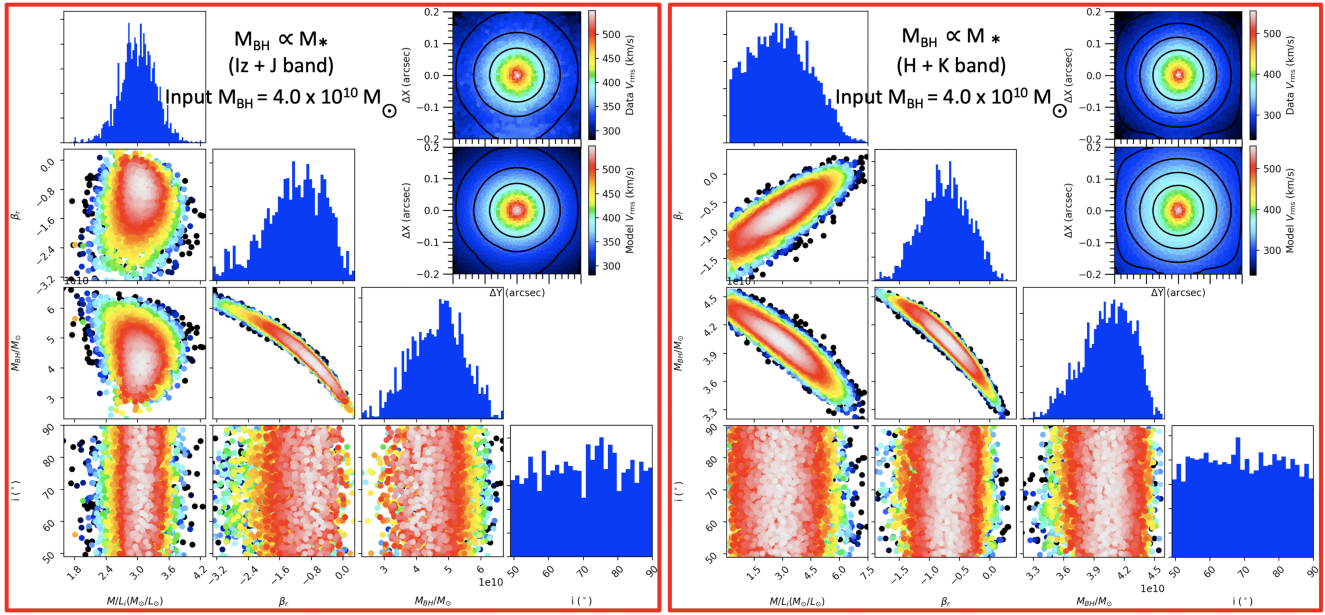

**Figure A30.** Same as Fig. 12 but for the HARMONI simulated kinematics with $M_{\rm BH} = 4.0 \times 10^{10}$ M$_\odot$ that follows the $M_{\rm BH}-M_\star$ relation for galaxies with masses above $M_{\rm crit.}$ predicted by equation (3) of K18 for the galaxy 2MASXJ14155764+0318216. The input parameters, their JAM$_{\rm sph}$ best-fit models, and statistics uncertainties are listed in Table A8.

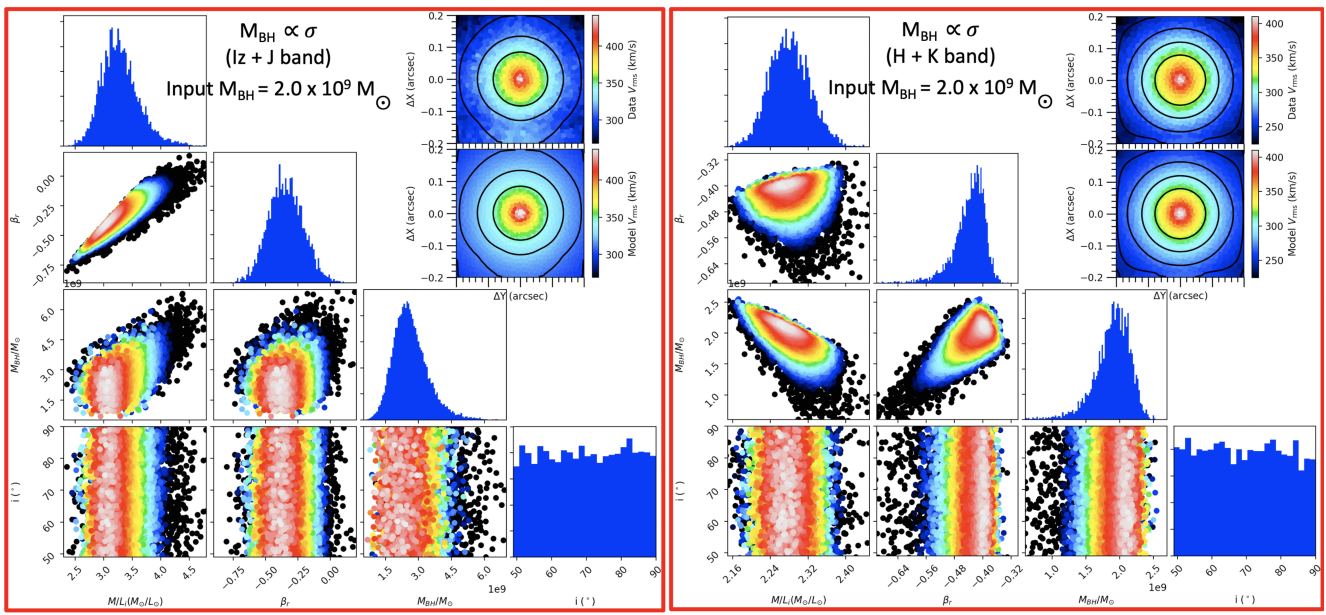

**Figure A31.** Same as Fig. 12 but for the HARMONI simulated kinematics with $M_{\rm BH} = 2.0 \times 10^9$ M$_\odot$ that follows the $M_{\rm BH}-\sigma_\star$ relation for galaxies with masses below $M_{\rm crit.}$ predicted by equation (2) of K18 for the galaxy 2MASXJ14155764+0318216. The input parameters, their JAM$_{\rm sph}$ best-fit models, and statistics uncertainties are listed in Table A8.

*Simulating SMBH mass measurements with HARMONI IFS* 43

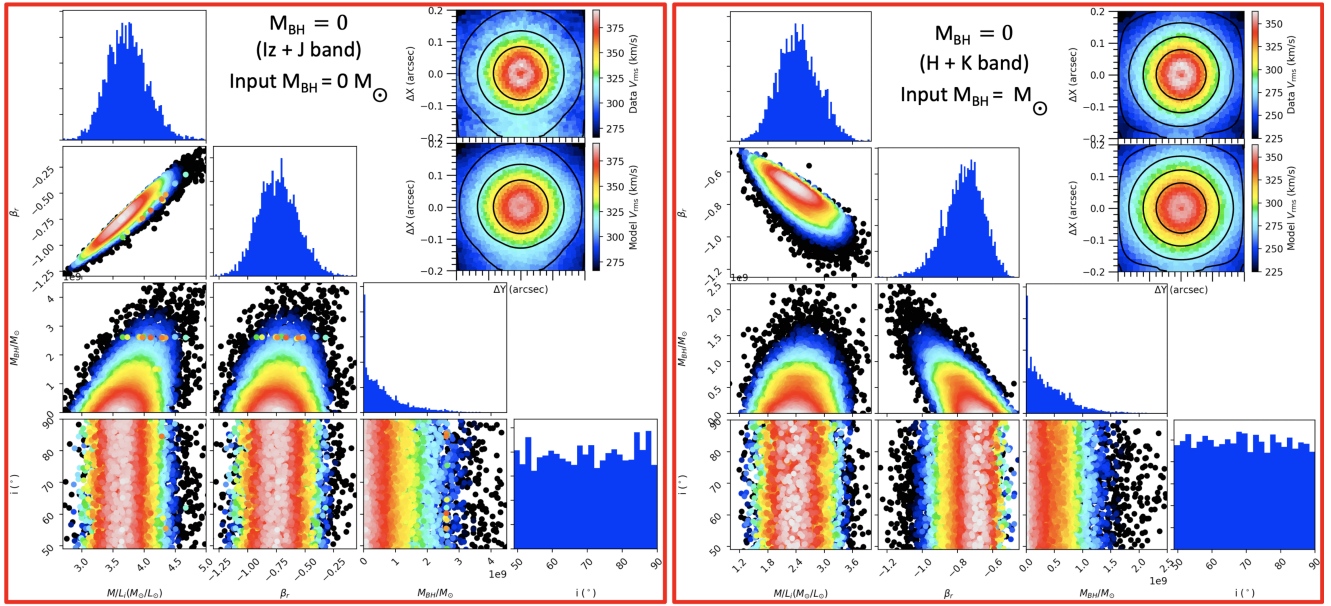

**Figure A32.** Same as Fig. 12 but for the HARMONI simulated kinematics with $M_{BH} = 0$ M$_\odot$ for the galaxy 2MASXJ14155764+0318216. The input parameters, their JAM$_{sph}$ best-fit models, and statistics uncertainties are listed in Table A8.

**Table A8.** Best-fitting JAM$_{sph}$ parameters and their statistical uncertainties for the $I_z + J$ and $H + K$ band simulated kinematics of 2MASXJ14155764+0318216.

| Parameter name (1) | Search range of parameters (2) | Input value for HSIM (3) | Best-fit value (4) $I_z + J$ | $1\sigma$ error (16–84%) (5) $I_z + J$ | $3\sigma$ error (0.14–99.86%) (6) $I_z + J$ | Best-fit value (7) $H + K$ | $1\sigma$ error (16–84%) (8) $H + K$ | $3\sigma$ error (0.14–99.86%) (9) $H + K$ |
|---|---|---|---|---|---|---|---|---|
| **Assuming no a central SMBH ($M_{BH} = 0$ M$_\odot$)** | | | | | | | | |
| $M_{BH}/M_\odot$ | $(0 \longrightarrow 10^{12})$ | 0 | $6.2 \times 10^7$ | $< 7.1 \times 10^8$ | $< 2.2 \times 10^9$ | $4.4 \times 10^5$ | $< 3.4 \times 10^8$ | $< 1.3 \times 10^9$ |
| $M/L_i$ (M$_\odot$/L$_\odot$) | $(0 \longrightarrow 10)$ | 3 | 3.6 | $\pm 0.3$ | $\pm 1.0$ | 2.5 | $\pm 0.5$ | $\pm 1.3$ |
| $i$ (°) | $(49 \longrightarrow 90)$ | 60 | 80 | $\pm 18$ | $\pm 25$ | 55 | $\pm 18$ | $\pm 25$ |
| $\beta_r$ | $(-15 \longrightarrow 1)$ | $\pm 0.2$ | $-0.74$ | $\pm 0.19$ | $\pm 0.54$ | $-0.69$ | $\pm 0.12$ | $\pm 0.37$ |
| **Assuming a central SMBH with mass $M_{BH} = 2.0 \times 10^9$ M$_\odot$, derived from the $M_{BH}$–$\sigma_\star$ relation (equation (2) from K18)** | | | | | | | | |
| $M_{BH}/M_\odot$ | $(0 \longrightarrow 10^{12})$ | $2.0 \times 10^9$ | $2.0 \times 10^9$ | $\pm 8.2 \times 10^8$ | $< 2.5 \times 10^9$ | $2.1 \times 10^9$ | $\pm 2.7 \times 10^8$ | $\pm 1.0 \times 10^9$ |
| $M/L_i$ (M$_\odot$/L$_\odot$) | $(0 \longrightarrow 10)$ | 3.0 | 3.0 | $\pm 0.4$ | $\pm 1.2$ | 2.3 | $\pm 0.1$ | $\pm 0.2$ |
| $i$ (°) | $(49 \longrightarrow 90)$ | 60 | 60 | $\pm 18$ | $\pm 25$ | 61 | $\pm 18$ | $\pm 25$ |
| $\beta_r$ | $(-15 \longrightarrow 1)$ | $\pm 0.2$ | $-0.4$ | $\pm 0.2$ | $\pm 0.4$ | $-0.4$ | $\pm 0.1$ | $\pm 0.2$ |
| **Assuming a central SMBH with mass $M_{BH} = 4.0 \times 10^{10}$ M$_\odot$, derived from the $M_{BH}$–$M_\star$ relation (equation (3) from K18)** | | | | | | | | |
| $M_{BH}/M_\odot$ | $(0 \longrightarrow 10^{12})$ | $4.0 \times 10^{10}$ | $4.0 \times 10^{10}$ | $\pm 9.1 \times 10^9$ | $\pm 2.0 \times 10^{10}$ | $4.0 \times 10^{10}$ | $\pm 2.8 \times 10^9$ | $\pm 6.4 \times 10^9$ |
| $M/L_i$ (M$_\odot$/L$_\odot$) | $(0 \longrightarrow 10)$ | 3.0 | 3.0 | $\pm 0.4$ | $\pm 1.4$ | 3.1 | $\pm 1.8$ | $\pm 3.5$ |
| $i$ (°) | $(49 \longrightarrow 90)$ | 60 | 86 | $\pm 14$ | $\pm 20$ | 53 | $\pm 14$ | $\pm 20$ |
| $\beta_r$ | $(-15 \longrightarrow 1)$ | $\pm 0.2$ | $-0.64$ | $\pm 0.85$ | $\pm 1.92$ | $-0.65$ | $\pm 0.45$ | $\pm 1.1$ |

*Notes:* Same as Table 6 but are the best-fit JAM$_{sph}$ modellings for the galaxy 2MASXJ14155764+0318216 optimized to its HARMONI simulated kinematics with different $M_{BH}$ which are shown in Figs A30, A31, and A32. Also, see Fig. 16 for a graphically short summary of this Table A8.